\documentclass[aps,prx,10pt,twocolumn,superscriptaddress,floatfix,amsmath,amssymb]{revtex4-2}

\usepackage{amsmath}
\usepackage{amssymb}
\usepackage{graphicx}
\usepackage{hyperref}

\newcommand{\omc}{\omega_{\rm c}}
\newcommand{\fz}{\zeta}
\newcommand{\thmax}{\theta_{\rm max}}
\newcommand{\wmax}{\pi^2/\lambda^2}
\newcommand{\HGKP}{\hat{H}_{\rm GKP}}
\newcommand{\HRWA}{\hat H_{\rm RWA}}
\newcommand{\Heff}{\hat H_{\rm eff}}
\newcommand{\cM}{\hat{\mathcal{M}}}
\newcommand{\Vz}{\bar {\mathcal V}_z }
\newcommand{\Tr}{{\rm Tr}}
\newcommand{\sgn}{{\rm sgn}}
\newcommand{\ii}{i}
\newcommand{\be}{\begin{equation}}
\newcommand{\ee}{\end{equation}}

\begin{document}
\title{
Self-correcting qubit in a strongly driven LC resonator }
\author{Max Geier}
\affiliation{Department of Physics, Massachusetts Institute of Technology, Cambridge, MA 02139, USA}
\author{Frederik Nathan}
\affiliation{%
NNF Quantum Computing Programme,
Niels Bohr Institute, University of Copenhagen, 2100 Copenhagen, Denmark}

\date{\today}

\begin{abstract}
We propose a device to realize quantum error correction by dissipation.
The device generates Gottesman-Kitaev-Preskill (GKP) states as steady states in a single driven LC resonator. The resonator should have
impedance above $h/2e^2\approx 12.9\,{\rm k}\Omega$, and should be driven across a Josephson junction with a high-amplitude voltage tone at frequency near-commensurate
with the resonator frequency.
In this regime, we show that GKP states emerge as steady-states of the resonator; these states encode a qubit protected against phase-space local noise.
Low-lying excitations
preserve the quantum information and increase the energy proportionally to a measure of the error degree. Thus, quantum error correction of both bit and phase errors can be achieved with
dissipative cooling.
Our calculations indicate an achievable coherence time of $ 1000 \times$ the resonator photon lifetime at operating temperatures above $100 \, {\rm mK}$.
Beyond steady-state quantum error correction, the device supports protected single-qubit Clifford gates and readout with exponential accuracy corrected by the same dissipative mechanism, as well as magic-state preparation above the distillation threshold.
\end{abstract}

\maketitle

\section{Introduction}

Harnessing dissipation for quantum error correction---or {\it dissipative quantum error correction} (DQEC) ---offers
a route to quantum computing that could address some of its key challenges, by reducing, or completely eliminating, the need for measurement, data processing, and feedback control for stabilizing qubits~\cite{barnes_automatic_2000,verstraete_quantum_2009,bombin_self-correcting_2013,brown_quantum_2016,reiter_dissipative_2017,SellemarXiv2023Apr,Nathan2024May,kastoryano_little_2024}.

First put forth as an idea in the 2000s~\cite{barnes_automatic_2000}, proposals for DQEC for a long time remained elusive~\cite{brown_quantum_2016}.
This situation changed recently, with proposals emerging for circuit-QED realizations of DQEC that protects against both dephasing and bit flip~\cite{SellemarXiv2023Apr,Nathan2024May}.
These approaches use Gottesman-Kitaev-Preskill (GKP) encoding to encode a qubit in charge and phase variables of an electromagnetic resonator~\cite{GottesmanPhysRevA2001Jun,Campagne-Ibarcq2020Aug,LachanceQuirion2024Apr}. The qubits are dissipatively stabilized by coupling the resonator to a thermal reservoir and modulating an attached Josephson junction with a periodic driving protocol, either a pulse comb~\cite{SellemarXiv2023Apr} or a stepwise activation protocol~\cite{Nathan2024May}.
These works indicate a potential for significant improvements in qubit stability.
At the same time, these proposals assume exotic elements beyond experimental reach, such as a high-contrast rapidly-switchable Josephson junction, or a high-bandwidth pulse comb. As a result, DQEC has remained beyond experimental reach.

\begin{figure}[h!]\includegraphics[width=1\columnwidth]{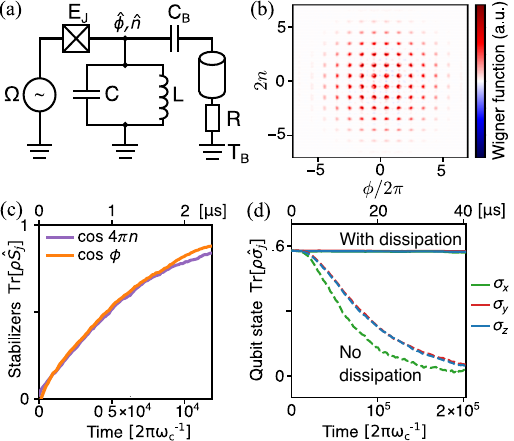}
\caption{{\bf Self-correcting GKP qubit.}
(a) An LC resonator
is coupled to an oscillating voltage bias with fundamental angular frequency $\Omega$ through a Josephson junction, and to a filtered dissipative environment (cylinder) with spectral support near the LC resonance frequency, $\omc=1/\sqrt{LC}$. We show that the circuit realizes a dissipatively stabilized GKP qubit for a broad class of configurations of the drive and circuit---e.g., when $\sqrt{L/C}\approx h/2e^2\approx 12.9\,{\rm k}\Omega$ and $\Omega \approx 2\omc$.
(b) Simulated Wigner function of the system resulting at time $t=2.4\,\mu{\rm s}$ after initialization in a high-temperature state (see Sec.~\ref{sec:numerical} for parameters and details). (c) Evolution of GKP stabilizers
for the same simulation.
(d) Evolution of encoded information after initializing in a GKP state with an equal superposition of $\hat\sigma_x$, $\hat \sigma_y$, and $\hat \sigma_z$ expectation values with (solid) and without (dashed lines) coupling to the environment, demonstrating dissipative quantum error correction. The simulations include photon loss from the resonator with a loss rate of $5\,{\rm kHz}$.}
\label{fig:1}
\end{figure}

Here, we propose an architecture for a dissipatively error corrected GKP qubit based on components
within experimental reach.
The device is shown in Fig.~\ref{fig:1}(a): it consists of an electromagnetic resonator with impedance above $h/2e^2\approx 12.91\,{\rm k}\Omega$,
connected through a Josephson junction to
an oscillating voltage bias.
When the voltage tone is suitably configured, we show that the GKP stabilizer Hamiltonian emerges as the effective Floquet Hamiltonian of the circuit. The low-energy eigenstates of this Hamiltonian are logical GKP states that encode a protected qubit.
The circuit can both realize the stabilizer Hamiltonian of conventional square-lattice GKP qubit, and those of a broader family of oblique and hexagonal encodings; additionally, it can realize qudit encodings. 

 DQEC can be realized
by
coupling the resonator to a suitably filtered environment, e.g. via a detuned lossy resonator or Purcell
filter~\cite{ferreira_collapse_2020,Putterman_2022}.
We show that such dissipation effectively cools the resonator down to the low-energy subspace of the effective Hamiltonian, which coincides with the GKP codespace.
In this codespace the encoded quantum information remains uncorrupted even in the presence of finite constant noise, when dissipation is turned on. The qubit can thus remain in a pure state while the resonator is in a mixed, noisy state.
Our simulations [Fig.~\ref{fig:1}(b,c)] suggest that dissipation can error-correct the GKP qubit this way at bath temperatures below $200\,{\rm mK}$ for NbN-based resonators with frequencies $ \sim 5 \, {\rm GHz}$.

We show that protected Clifford gates can be implemented in the qubit via appropriate modulation of the oscillating bias, while readout/initialization can be realized via measurement of the DC supercurrent in an ancilla Josephson junction.
Finally, magic states can be generated with sufficient fidelity for distillation~\cite{Bravyi2005Feb,Baragiola2019Nov} by applying the protocol starting from the resonator ground state.

This paper is structured as follows:
In Sec.~\ref{sec:gkp_encoding} we introduce the GKP encoding considered here.
In Sec.~\ref{sec:dec} we show that a dissipatively corrected GKP qubit with exponentially long lifetime can be realized by coupling the GKP stabilizer Hamiltonian to a generic dissipative environment.
In Sec.~\ref{sec:device} we show that the GKP stabilizer Hamiltonian emerges as the effective Hamiltonian of the circuit in Fig.~\ref{fig:1}(a) for particular configurations of drive and circuit parameters.
In Sec.~\ref{sec:lindblad} we show that the GKP code subspace can be stabilized and dissipatively error corrected in the circuit by coupling it to a suitably filtered environment.
In Sec.~\ref{sec:numerical}
we provide data from numerical simulations.
In Secs.~\ref{sec:gates}, \ref{sec:magic_state_initialization}, and~\ref{sec:readout}
 we propose dissipatively corrected Clifford gates, magic state preparation, and readout/initialization protocols.
 We conclude with a discussion in Sec.~\ref{sec:discussion}, where we discuss possible challenges and open questions of the approach, along with potential benefits.

\section{ GKP encoding}
\label{sec:gkp_encoding}
\begin{figure*}
    \centering
    \includegraphics[width=0.99\linewidth]{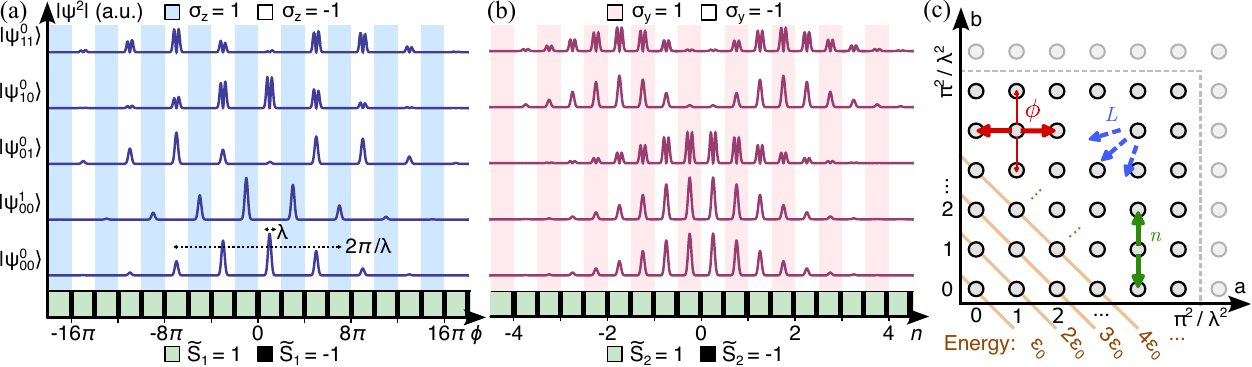}
    \caption{{\bf Lattice picture of dissipative quantum error correction}. (a,b) Probability distributions for $\hat \phi$ (a) and $\hat n$ (b) in the codespace lattice states $|\psi_{ab}^s\rangle $ [Eq.~\eqref{eq:lattice_basis_wf}], for various values of $a,b,$ and $s$, with the squeezing parameter $\lambda$ indicated. Colors in top indicate intervals corresponding to $\hat \sigma_z$ (a) and $\hat \sigma_y$ (b) eigenvalues $1$ and $-1$, while colors in bottom indicate intervals corresponding to eigenvalues $1$ and $-1$ of the canonical GKP stabilizers $\tilde S_1$ and $\tilde S_2$ [see Eq.~\eqref{eq:can_stab}]. (c) The states $|\psi_{ab}^s\rangle$ can be visualized as sites on a semi-infinite square lattice with two orbitals per cell, with site coordinates $(a,b)$, and orbital index $s$.
    States with $a,b\ll \wmax$ [see Eq.~\eqref{eq:lambdadef}] are orthonormal and confined in the GKP codespace up to $\mathcal O(e^{-\wmax})$ corrections; for these states $s$ moreover encodes the quantum information up to similar $\mathcal O(e^{-\wmax})$ corrections.
    The quadratures $\hat \phi$ and $\hat n$ can be represented by local coupling terms in the lattice (blue and green). They, and thereby any physical quantity, formally leave $s$ invariant.
    The GKP stabilizer Hamiltonian is approximately diagonal in the codespace lattice basis, and assigns an energy $\varepsilon_0(a+b)$ to site $(a,b)$.
    When coupling a phase-space local observable in the system to an environment, the dissipative processes, here represented by $L$ (blue), therefore cause local flow towards the lower-left corner of the lattice that formally leaves $s$ invariant. This confines the system in the GKP codespace without corrupting the encoded information. }
    \label{fig:gkp_states}
\end{figure*}

The qubit is redundantly encoded in the Fock space of a superconducting resonator mode, formed by the LC part of the circuit in Fig.~\ref{fig:1}(a).
We describe the mode via its canonically conjugate quadratures, the dimensionless phase $\hat \phi$ and charge (Cooper pair number) $\hat n=-i\partial_{\hat \phi}$.

The qubit is encoded via a Gottesman-Kitaev-Preskill (GKP) type encoding~\cite{GottesmanPhysRevA2001Jun}.
While our architecture can realize a broad class of GKP encodings (see below), for concreteness, we first focus on the square-lattice GKP qubit encoding where the logical Pauli matrices of the qubit are given by
\be
\hat \sigma_z = \Xi(\hat \phi/2\pi),\quad \hat \sigma_y = \Xi(2\hat n),\quad \hat \sigma_x = -i\hat \sigma_z\hat \sigma_y.
\label{eq:GKP_encoding}
\ee
Here $\Xi(x) \equiv (-1)^{{\rm round}(x+1/2)}$ and ${\rm round}(x)$ returns the integer nearest to $x$~\cite{Pantealoni_2020,Nathan2024May,Pauliproperty,LogicalChoice}. In Fig.~\ref{fig:gkp_states}(a,b), we indicate the eigenvalues of $\hat \sigma_{z}$ and $\hat \sigma_y$ as a function of $\hat \phi$ and $\hat n$: essentially, $\hat \sigma_z$ gives the parity of the integer closest to $\hat\phi/2\pi -1/2$, while $\hat \sigma_y$ gives the parity of the integer closest to $2\hat n-1/2$.
Nontrivially, the Pauli operators $\hat \sigma_x$, $\hat \sigma_y$, and $\hat \sigma_z$ satisfy the Pauli commutation relations, and hence define a valid qubit encoding~\cite{Pantealoni_2020,SellemarXiv2023Apr,Nathan2024May}.
The encoding above allows multiple physical resonator states to encode the same logical qubit state. Indeed, a {\it mixed} resonator state can encode a {\it pure} logical state~\cite{Pantealoni_2020,Nathan2024May}.

The GKP encoding enjoys significant protection against a broad class of perturbations.
To see this, consider a perturbation of the resonator state $|\psi\rangle$, $|\psi\rangle \to\hat{ \mathcal O}|\psi\rangle$, where $\hat{\mathcal O}$ is a unitary or Kraus operator representing the perturbation, e.g. resulting from noise, partial measurement, or dephasing.
From the definitions
above, we see that $\langle \hat \sigma_z\rangle$ can only change if the perturbation shifts the system's $\hat \phi$-probability support between the blue and white regions in Fig.~\ref{fig:gkp_states}(a).
Likewise, $\langle \hat \sigma_y\rangle$ can only change if the $\hat n$-probability support shifts between the red and white regions in Fig.~\ref{fig:gkp_states}(b).
Finally, $\langle \hat \sigma_x\rangle$ can only change if either $\hat n$ support or $\hat \phi$ support shifts between regions of different color in Fig.~\ref{fig:gkp_states}(a,b)~\footnote{This follows by noting that $\hat{\mathcal O}\hat \sigma_{z,y}|\psi\rangle =\hat \sigma_{z,y} \hat{\mathcal O}|\psi\rangle$ if $\hat{\mathcal O}$ does not shift $\hat n$ and $\hat \phi$ probability support between regions of different colors in Fig.~\ref{fig:gkp_states}(a,b)}.
Thus when the system's $\hat \phi$- and $\hat n$-probability distributions remain confined a finite distance away from the ``phase boundaries'' of $\hat \sigma_z$ and $\hat \sigma_y$, $\hat \phi \in 2\pi \mathbb Z$ and $\hat n\in \mathbb Z/2$, the qubit's Pauli expectation values $\{\langle \hat \sigma_i\rangle \}$ are unaffected under infinitesimal perturbations that generate continuous flow of $\hat \phi$ and $\hat n$ distributions. This includes any perturbation generated by finite-order polynomials of $\hat \phi$ and $\hat n$, such as, e.g., charge noise, flux noise, or photon loss.
We refer to such perturbations
as {\it phase-space local} perturbations.

To quantify the distance from the system's $\hat \phi$ and $\hat n$ supports to the phase boundaries of the logical operators, $\hat \phi \in 2\pi \mathbb Z$ and $\hat n\in \mathbb Z/2$, we introduce the two {\it GKP stabilizers}
\be
\hat S_1 \equiv \cos(\hat \phi),\quad \hat S_2 \equiv \cos(4 \pi \hat n).\label{eq:stabilizers}
\ee
Since the GKP stabilizers commute, they can be taken to share eigenstates ~\footnote{The commutativity can be seen by noting that $e^{\pm 4\pi i n}$ translates $\phi$ by $\pm 4\pi$, and thus commutes with $\cos(\hat \phi)$}.
Thus, any state $|\psi\rangle$ in the span of the mutual eigenstates of $\hat S_1$ and $\hat S_2$ with eigenvalues smaller than $1-\epsilon$ for some arbitrary $\epsilon>0$ have $\hat \phi$ and $\hat n$ support in the bulk of the colored regions of Fig.~\ref{fig:gkp_states}(a,b).
We refer to such a state as a {\it GKP state} in this work.
Common for GKP states is that their Wigner functions exhibit a characteristic grid structure, as in Fig.~\ref{fig:1}(b).
Examples of GKP states include Zak states~\cite{zak_finite_1967,shaw_stabilizer_2024}, and some of their linear combinations, including normalizable, finitely-squeezed, GKP states.

We term the subspace of all GKP states as the {\it GKP codespace}. Viewing the encoding above as a stabilizer code, the GKP codespace is given by the joint eigenvalue-$1$ eigenspace of the two {\it canonical} stabilizers $\tilde S_{1}$ and $\tilde S_2$, defined as \begin{equation}
    \tilde S_{1}\equiv \sgn(1-\epsilon-\hat S_{1}), \qquad \tilde S_{2}\equiv \sgn(1-\epsilon-\hat S_{2}).
    \label{eq:can_stab}
\end{equation}
These operators are illustrated in the bottom panels of Fig.~\ref{fig:gkp_states}(a,b). For convenience, we use the term {\it stabilizer} to refer to $\hat S_1$ and $\hat S_2$ below.

The square-lattice GKP encoding above
can be generalized to a
broader class of {\it oblique} GKP-type qubit encodings. Oblique encodings are obtained by substituting $\hat n$ with $\hat n+\alpha\hat \phi$ in the definitions of the logical operators and stabilizers above, where $\alpha$ can be any real number~\footnote{Note that the commutation relations between the stabilizers and logical operators are unaffected by this substitution, and hence they continue to define a valid qubit encoding.}.
Oblique encodings include the {\it hexagonal} GKP encoding~\cite{GottesmanPhysRevA2001Jun}, which arises when $\alpha=1/2$.

\section{Dissipative quantum error correction}
\label{sec:dec}
Here we show that the GKP encoded logical operators $\{\hat \sigma_i\}$ above can be error corrected by {\it dissipation}, without any readout, data processing or feedback control~\cite{Nathan2024May}.
Specifically, we will show this can be achieved when the resonator mode is described by the following Hamiltonian,
\be
\hat H_{\rm GKP} \equiv \frac{\hbar \delta \omega}{2} \Big( \frac{\hat \phi^2}{4\pi}+4\pi \hat n^2 \Big)\!+\tilde{E}_J\! \left[\cos\hat \phi +\cos 4\pi \hat n\right]
\label{eq:gkp_stab_hamil}\ee
when $\tilde E_J\gg \hbar \delta \omega$.
We term this the {\it GKP stabilizer Hamiltonian.}
In the next section we show that $\HGKP$ can be realized with the driven circuit in Fig.~\ref{fig:1}(a).

Below, we show that the low-energy eigenstates of $\HGKP$ are doubly degenerate and reside in the GKP codespace (up to exponentially suppressed corrections) of the square-lattice GKP qubit we introduced in the previous section. Subsequently, we show that coupling a generic phase-space local observable to a generic dissipative environment will confine the resonator in this codespace without corrupting the encoded information. The confinement persists in the presence of constant external noise, and hence this coupling to the environment realizes DQEC.
Our analysis exploits that we work in the limit $\tilde E_J\gg \hbar \delta \omega$, and will involve expansion in the following small parameter:
\be
\lambda \equiv \left(\frac{4\pi \hbar \delta \omega}{\tilde E_J}\right)^{1/4},
\label{eq:lambdadef}
\ee
which we identify below as a squeezing parameter for the stabilized GKP states.

To see how DQEC emerges, we first consider the low-energy spectrum and eigenstates of $\HGKP$.
In Appendix~\ref{app:diagonalization} we show that, to leading order in $\lambda$, these are given by a particular basis of states $\{|\psi_{ab}^s\rangle\}$ for $a,b=0,1,2,\ldots $ and $s=0,1$, which we term the {\it lattice} basis. Here the $\hat \phi$-wavefunction of the lattice basis state $|\psi_{ab}^s\rangle$ is given by
\be
\psi_{ab}^s(\phi)= \sum_{k=-\infty}^\infty \chi_{a}\left(\frac{\phi-\phi_{ks}}{\lambda}\right)\chi_b\left(\frac{\lambda\phi}{4\pi}\right)(-i)^{b+2k},
\label{eq:lattice_basis_wf}
\ee
where $\phi_{ks} \equiv 2\pi (2k+s) +\pi $,
 and $\chi_a (\phi)$ denotes the $a$th harmonic oscillator eigenfunction with unit width~\footnote{Specifically, $\chi_a(\phi) \equiv {H_a\left(\phi\right)e^{-\frac{\phi^2}{2}} }/{\sqrt{2^a a!\sqrt{\pi}}}$ with
$H_a(x)$ denoting the $a$th Hermite polynomial.}.
Specifically, for $a,b\ll \pi^2/\lambda^2$, $|\psi_{ab}^{0,1}\rangle $ are eigenstates of $\HGKP $ up to $\mathcal O(\lambda^2) $ corrections. The associated eigenvalues, $E_{ab}^{0}$ and $E_{ab}^1$, are {\it exponentially close} to each other in $\wmax$, and given by
\begin{equation}
   E_{ab}^{s}
   =(a+b+1)\varepsilon_0%
   + \mathcal O(\lambda^2 \varepsilon_0),
   \label{eq:energies}
\end{equation}
for $s=0,1$, with $\varepsilon_0 = \lambda^2 \tilde E_J$.
We plot the $\hat \phi$- and $\hat n$-distributions for the first few lattice basis states in Fig.~\ref{fig:gkp_states}(a,b).
Heuristically, $|\psi_{00}^s\rangle$ corresponds to a canonical GKP state encoding a logical $|s\rangle $ state, with finite squeezing $\lambda$. For nonzero $a,b$, $|\psi_{ab}^s\rangle$ corresponds to a photon-dressed GKP state, also encoding logical $|s\rangle$ state: its $\phi$- and $\hat n$-distributions
consist of well-separated peaks with Gaussian tails, confined near odd multiples of $\pi$ and $1/2$, respectively.

The lattice basis states have several useful features: first, they reside in the GKP codespace for $a,b\ll \pi^2/\lambda^2$.
Specifically,
\begin{align}
\tilde S_1|\psi_{ab}^s\rangle & =|\psi_{ab}^s\rangle +\mathcal O(e^{-\wmax}), \label{eq:s1action}\\
\tilde S_2|\psi_{ab}^s\rangle & = |\psi_{ab}^s\rangle +\mathcal O(e^{-\wmax})\label{eq:s2action}
\end{align}
for $a,b\ll \wmax$. These results are evident for the illustrated states in Fig.~\ref{fig:gkp_states}(a,b).
The first result follows straightforwardly from Eq.~\eqref{eq:lattice_basis_wf}, while the second result is shown in Appendix~\ref{app:lattice_basis}.
The second useful property of the lattice basis states is that, while not strictly orthonormal, their deviations from orthonormality are exponentially suppressed in the codespace region $a,b\ll \wmax$. Specifically, we show in Appendix~\ref{app:lattice_basis} that
\begin{equation}
    \langle \psi_{a_1b_1}^{s_1}|\psi_{a_2b_2}^{s_2}\rangle =\delta_{a_1a_2}\delta_{b_1b_2}\delta_{s_1s_2}+\mathcal O (e^{-\wmax}).
    \label{eq:near_orth}
\end{equation}
with $\delta_{ij}$ denoting the Kronecker delta.
The third useful feature is that, for $a,b\ll \wmax$, the index $s$ encodes the logical state of the qubit, up to similar corrections:
\begin{align}
\hat \sigma_z|\psi_{ab}^s\rangle
&=(-1)^s|\psi_{ab}^s\rangle +\mathcal O(e^{-\wmax}),\label{eq:zl} \\
\hat \sigma_y| \psi_{ab}^s \rangle
&=(-i)^s|\psi_{ab}^{1-s}\rangle+\mathcal O(e^{-\wmax}).\label{eq:yl}
\end{align}
The action of $\hat \sigma_x$ follows from the two relations above.

It is useful to visualize the states $|\psi_{ab}^s\rangle$ as sites on a semi-infinite square lattice with two orbitals per unit cell as in Fig.~\ref{fig:gkp_states}(c), with lattice coordinates ($a,b$) and orbital index $s$.
In this picture, the lower-left region of the lattice, $a,b\ll \wmax$ correspond to the GKP codespace, where the $s$ index encodes the qubit.
Crucially, the quadratures can be represented as local operations in the lattice. Specifically~\footnote{The lattice basis states form an overcomplete basis. Hence the right-hand side below is not the only way to represent the action of the quadratures on $|\psi_{ab}^s\rangle$ as linear combination of lattice basis states.},
\begin{align}
\hat \phi|\psi_{ab}^s\rangle&=%
 \frac{4 \pi}{\sqrt{2}i\lambda}(\sqrt{b+1}|\psi_{a,b+1}^s\rangle -\sqrt{b}|\psi_{a,b-1}^s\rangle ) ,
\\
\hat n|\psi_{ab}^s\rangle &=
\frac{1}{\lambda\sqrt{2}i}(\sqrt{a+1}|\psi_{a+1,b}^s \rangle -\sqrt{a}|\psi_{a-1,b}^s\rangle)\notag\\ &%
+\frac{\lambda}{4\pi\sqrt{2}}(\sqrt{b+1}|\psi_{a,b+1}^s\rangle + \sqrt{b}|\psi_{a,b-1}^s\rangle).
\end{align}
Thus, $\hat \phi$ and $\hat n$ take lattice basis states to their adjacent sites on the lattice and formally leave $s$ invariant.
More generally, a phase-space local observable, in the form of a $k$th order polynomial of the quadratures, can only couple sites separated by a Manhattan distance $k$ and leaves $s$ invariant.
The ``ideal'' GKP state $|\psi_{00}^s\rangle$ can thus tolerate $\mathcal O( \wmax)$ applications of $\hat \phi$ or $\hat n$ without leaving codespace, and the encoded information being corrupted.
Motivated by this, we can interpret
$
    w\equiv a+b
$
as a syndrome weight, and $\wmax$ as a code distance.

To understand how the GKP qubit is protected, note that any observable, including the Hamiltonian of the resonator, can be written as a function of $\hat \phi$ and $\hat n$. Thus $s$ can be taken to be formally invariant under the evolution of the system.
Since $s$ encodes the quantum information in the codespace region $a,b\ll \wmax$, quantum information can only be corrupted when the state of the system acquires support on lattice basis states where $(a+b)\gtrsim \wmax$
(See Appendix~\ref{app:diagonalization} for further details).
Since $\HGKP$ assigns an energy penalty $(a+b+1)\varepsilon_0$ to site $(a,b)$ in the codespace lattice, up to $\lambda^2\varepsilon_0$ corrections, the system thus has to overcome an energy barrier of height $\varepsilon_0 \wmax \sim \tilde E_J$ to leave the codespace region $a,b\ll\wmax$ and corrupt the encoded information.

The above features imply that DQEC emerges when coupling $\HGKP$ to a generic thermodynamic environment at temperature $T_{\rm B}\ll \tilde E_J$ through a phase-space local coupling:
In the lattice picture above, such coupling will thermalize the system to a Gibbs-state with relative probability weight $\sim e^{-(a+b)\varepsilon /k_{\rm B} T}$ on site $(a,b)$. When $T_{\rm B}\ll \varepsilon_0 \wmax \sim \tilde E_J$, this will confine the system to the GKP codespace region $(a+b)\ll \wmax$, implying the encoded information remains uncorrupted.
The protection persists even when there is extrinsic phase-space local noise or perturbation applied to the system: the noise will simply induce a quasilocal dressing of the steady-state in the lattice picture, and renormalize the effective temperature of the Gibbs state, $T_{\rm eff}$, to account for the competition between energy outflux to the environment and energy influx to the system.
As a result, the support leakage rate out of the codespace region $(a+b)\ll \wmax $ must follow an Arrhenius law, scaling as $ \frac{\varepsilon_0}{\hbar} e^{-\tilde E_J/k_{\rm B}T_{\rm eff}}$~\cite{ankerhold_dissipative_1995},
\begin{equation}
t_{\rm coherence} \sim \frac{\hbar}{\varepsilon_0} \exp\left[\frac{\tilde E_J}{k_{\rm B}T_{\rm eff}}\right]\label{eq:arrhenius}
\end{equation}%
Thus the lifetime of the qubit can remain exponentially large, even in the presence of noise. We provide further details of this argument in Appendix~\ref{app:diagonalization} and also refer to Ref.~\cite{cheung2026}.

\section{Realizing $H_{\rm GKP}$ with Floquet engineering}
\label{sec:device}

\begin{figure}
    \includegraphics[width=0.999\columnwidth]{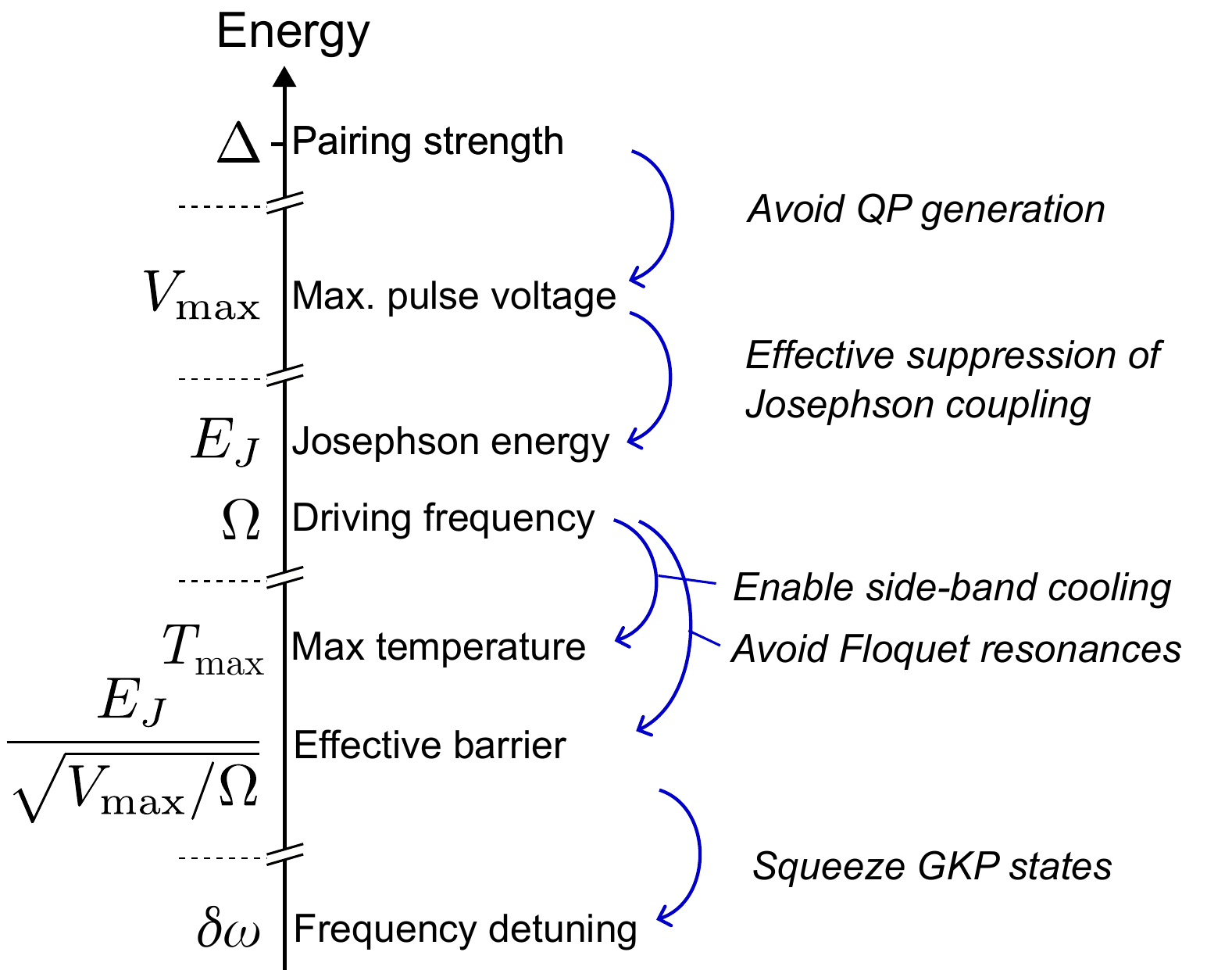}
    \caption{{\bf Hierarchy of energy scales required for the qubit.} Italic text indicates the origin of the scale-separation requirements. Dashed lines indicate a separation of scales (typically a factor $5-10$, but see text for details). Here $V_{\rm max} $ denotes the maximum amplitude of the bias voltage $V(t)$, $T_{\rm max}$ denotes the maximal value of the bath temperature $T_{\rm B}$ allowed for operation of the qubit, while $\delta\omega=\omc-\tilde \Omega$, with $\tilde \Omega$ a rational multiple of $\Omega$. Quantities above are represented in units where $\hbar = 2 e=1$.}
    \label{fig:hierarchy}
\end{figure}
As one of our main results, we now show that the GKP stabilizer Hamiltonian $\hat H_{\rm GKP}$
can be realized in the driven superconducting circuit in Fig.~\ref{fig:1}(a).
The circuit consists of an electromagnetic resonator with inductance $L$ and capacitance $C$, connected through a Josephson junction
to an oscillating voltage bias $V(t) $
with fundamental angular frequency $\Omega = 2\pi/T$.
The Hamiltonian of the driven circuit reads
\be
\hat H(t) =\frac{\hbar\omc}{2}\Big[\frac{\hat \phi^2}{2\pi \fz}+2\pi \fz \hat n^2 \Big] - E_J \cos[\hat \phi-
\Phi(t)] .
\label{eq:hamiltonian_definition}\ee
where $E_J$ denotes the Josephson energy of the junction, and $
    \fz \equiv {\sqrt{L/C}}/Z_Q
$
gives the circuit impedance in units of the superconducting quantum resistance $Z_Q\equiv h/4e^2\approx 6.45\,{\rm k}\Omega$. Finally $\Phi(t)$ denotes the phase induced by the voltage drive: $\partial_t \Phi(t) =\frac{2e}{\hbar} V(t)$~\footnote{We work in a gauge where the time-average of $\Phi(t)$
vanishes.}.

The circuit above can realize the stabilizer Hamiltonian of square-lattice, oblique, and hexagonal GKP encodings for a broad class of waveforms $V(t)$ and circuit parameters, that we detail in Sec.~\ref{sec:extensions_to_more_general} below.
For concreteness, we first show that the square-lattice GKP stabilizer Hamiltonian $\HGKP$
can be realized using a harmonic drive $V(t) = V_{\rm max} \sin \Omega t $ with $2eV_{\rm max}\gg E_J,\hbar \Omega$ and $\Omega\approx 2 \omc$, and picking $\fz = 2$.

The GKP stabilizer Hamiltonian is realized
in the sense that the evolution of the circuit sampled at integer multiples of $2T$, here termed the {\it stroboscopic evolution}, is approximately equivalent to the dynamics generated by $\hat H_{\rm GKP}$. 
Specifically, the Floquet theorem~\cite{floquet_sur_1883,oka_floquet_2019} states that the stroboscopic evolution is generated by the {\it effective Hamiltonian} \begin{equation}
    \Heff \equiv \frac{i\hbar}{2 T} \log \hat U(2 T).
    \label{eq:heff}
\end{equation}
where $\hat U(t)\equiv \mathcal T e^{-i\int_0^t dt' \hat H(t')/\hbar }$ and $\mathcal T$ the time-ordering symbol~\footnote{The logarithm above is defined with branch cut along the negative real axis, such that $\Heff$ has all its eigenvalues in the interval $[-\hbar \tilde \Omega/2,\hbar \tilde \Omega /2)$.}.
In this section, we shall show that $\Heff \approx \HGKP$.

We first transform to a comoving frame of phase space that rotates with angular frequency $\tilde \Omega \equiv \Omega/2$. In this frame, the evolution operator is given by $\tilde U(t)\equiv e^{i\hat H_{\rm r}t/\hbar}\hat U(t)$, where $H_{\rm r} = \frac{\hbar\tilde \Omega}{2}({\hat \phi^2}/{4\pi }+4\pi \hat n^2)$.
The Hamiltonian in the comoving frame---i.e., the generator of $\tilde U(t)$---is given by
\begin{align}
    \tilde H\!(t) =& \frac{\hbar\delta \omega }{2} \left(\frac{\hat\phi^2}{4\pi }+4\pi \hat n ^2\right) \label{eq:tildehdef}%
    \\ &-E_{\!J} \cos\!\left[ \hat\phi \cos \tilde{\Omega}t - 4\pi \hat n \sin \tilde{\Omega}t \!- \thmax \!\cos \Omega t\right]\! ,\notag
\end{align}
where $\thmax=\frac{2eV_{\rm max}}{ \hbar \Omega }$ denotes the phase oscillation amplitude induced by drive,
and $\delta \omega \equiv \omc - \tilde \Omega $ denotes the detuning of the resonator frequency from half the drive frequency~\cite{Schon_2013,Zhang_2017,NathanPhysRevRes2020Dec}.
Importantly,
$\tilde H(t)$ is explicitly time-periodic with a {\it doubled period}, $\tilde T\equiv 2T$ relative to that of $\hat H(t)$.
Henceforth, we shall use tilde ($\tilde{\cdot}$) to indicate frequencies and timescales associated with this doubled period.
Since the eigenvalues of $\hat H_r$ are integer multiples of $\hbar \tilde \Omega$, the lab and comoving frames coincide at integer multiples of $\tilde T$: $\tilde U(2 T) = \hat U(2T)$. Hence {the stroboscopic dynamics generated by $\tilde H(t)$ are equivalent to those generated by $\Heff$.}

We now exploit the separation of energy scales we assumed above, $2e V_{\rm max}\gg E_J, \hbar \tilde \Omega$, which implies $\Omega \thmax \gg E_J/\hbar,\delta \omega$. This separation of frequency scales allows us to obtain $\Heff$ via a combination of rotating wave (RWA) and stationary phase (SPA) approximations~\cite{Zhang_2017,NathanPhysRevRes2020Dec}. We provide the derivation in App.~\ref{app:SPA}, but provide an outline here: first, for $\thmax \Omega\gg E_J/\hbar,\delta \omega $ the time-dependent phase term $\Phi(t)=\thmax \cos\Omega t$ in Eq.~\eqref{eq:tildehdef} causes an oscillation of the second term in $\tilde H(t)$ much faster than any other scale of $\tilde H(t)$; thus we can use the RWA to approximate $\Heff$ by the time average of $\tilde H(t)$ over one extended period $\tilde T$. Due to the same rapid oscillation, this time-average only gets contributions from the stationary points of $\Phi(t)$, which are located at $t= n\tilde T/4$ for $n\in\mathbb Z$---this is the SPA. Thus, we can effectively obtain $\Heff$ from a weighted average of $\tilde H(t)$ sampled at $t\in \mathbb Z\tilde T/4$.
 \begin{figure*}
    \centering
    \includegraphics[width=0.78 \linewidth]{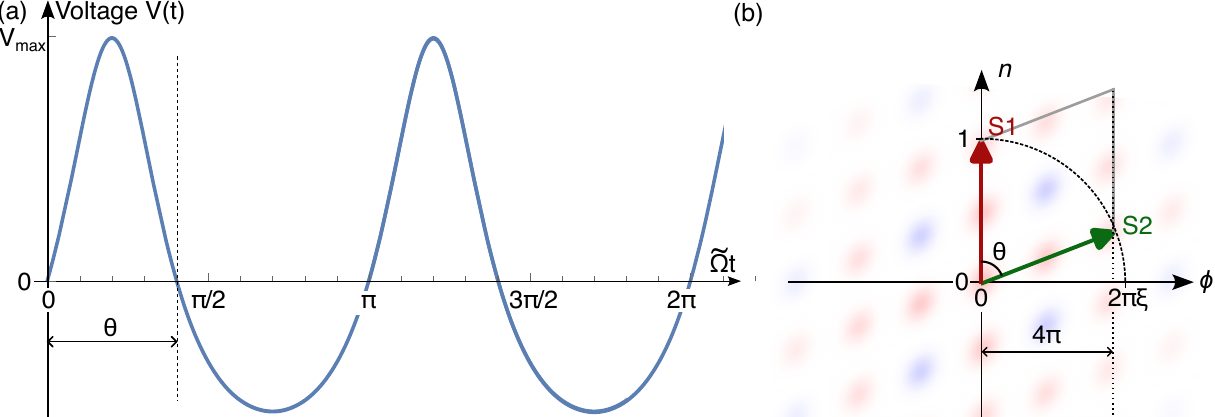}
    \caption{{\bf Oblique GKP encoding via generalized waveforms}. In (a), we show a possible waveform that can be used to generate the stabilizer Hamiltonian of an oblique GKP encoding. The stabilizers S1 and S2 of the encoding are illustrated in (b) via the phase-space displacement they generate. The oblique-GKP encoded qubit emerges when the area of the parallelogram spanned by S1 and S2 equals $4\pi $. Hence such a Hamiltonian can be realized for any $\fz>2$ by choosing the node location $\theta$ of the voltage tone in (a) such that $2\pi\fz \sin \theta=4\pi $.}
    \label{fig:oblique}
\end{figure*}
After these approximations, we obtain
\begin{equation}
     \hat {H}_{\rm eff}\! \approx \frac{\hbar \delta \omega }{2} \Big( \frac{\hat\phi^2}{4\pi }+4\pi \hat n^2 \Big)\!-\tilde{E}_J\! \left[\cos \hat \phi +\cos 4\pi \hat n\right],
     \label{eq:H_eff_GKP}
\end{equation}
where
\begin{align}
\tilde{E}_J = \!\frac{E_J \cos\left(\thmax-\pi/4\right)}{ \sqrt{ 2\pi \thmax }} .
\label{eq:tildej}
\end{align}
We thus see that $\Heff \approx \hat H_{\rm GKP}$, as we sought to show.

Physically, the approximations above reflect that, for $2eV_{\rm max}\gg E_J$, Cooper pair tunneling through the Josephson junction is effectively blocked away from the nodes of $V(t)$ [the stationary points of $\Phi(t)$]; hence we may effectively set the second term in Eq.~\eqref{eq:tildehdef} to zero except near these nodes. The RWA is then justified through the arguments in Refs.~\cite{conrad_2021_GKP,shaw_stabilizer_2024}.

We finally note that the oscillatory dependence of $\tilde E_J$ on $\thmax$ is a feature of the particular example we considered here. Below, we identify a drive configuration that results in the same stabilizer Hamiltonian as above, but where $\tilde E_J$ is not an oscillatory function of $\thmax$ [see Eq.~\eqref{eq:no_oscillation}].

\subsection{Extension to more general waveforms and impedances}
\label{sec:extensions_to_more_general}

We now generalize the example above to a broad family of drive/circuit configurations that realize the stabilizer Hamiltonians of square-, hexagonal-, or oblique-lattice GKP encodings.

For a GKP stabilizer Hamiltonian to emerge as the effective Hamiltonian of the circuit, we require the circuit and drive parameters satisfy the following conditions, summarized in Fig.~\ref{fig:hierarchy}: first, the maximal amplitude of $V(t)$, $V_{\rm max}$, must be smaller than the superconducting gap of the Josephson junction, $\Delta$, to avoid generation of quasiparticles, which would otherwise limit the qubit's performance (see Sec.~\ref{sec:challenges}). Second, we require the voltage tone $V(t)$ to have an amplitude away from its nodes which is much larger than $E_J/2e$ and $\hbar \Omega/2e$. The voltage must climb to this magnitude from its nodes on a timescale much shorter than $1/\Omega$. This is required to ensure the applicability of the RWA and SPA---or, equivalently, to effectively block Cooper pairs from tunnelling across the junction away from the nodes of $V(t)$.
Thirdly, we require $\Omega$ to be close to a rational multiple $q/p$ of the resonator frequency $\omc$, such that the detuning $\delta \omega=\omc-p\Omega /q$ is much smaller than both $E_J/\hbar$ and $\Omega$.
Once these conditions are satisfied, one can employ the RWA and SPA to obtain the effective Floquet Hamiltonian $\Heff$ for the circuit, through the same steps we made above. This derivation, detailed in Appendix~\ref{app:SPA}, results in an effective Hamiltonian given by
\begin{align}
\Heff &\approx\frac{\hbar \delta \omega}{2} \left( \frac{\hat \phi^2}{2\pi \fz}+2\pi \fz \hat n^2 \right)
\label{eq:hrwa_general}\\
&-\sum_{i=1}^N \!\frac{E_J\cos[\hat\phi \cos\tilde \Omega t_i \! - \! 2\pi\fz \hat n\sin\tilde \Omega t_i\!-\! \Phi (t_i)\!-\!\frac{\pi}{4}\mu_i] }{qT\sqrt{2 e |\partial_t V(t_i)|/h}} .\notag
\end{align}
Here $\tilde \Omega = p\Omega/q$, while $t_1,\ldots t_N$ denotes the nodes of $V(t)$ within an extended period $\tilde T = qT$, and $\mu_i = \sgn[\partial_t V(t_i)]$.

The stabilizer Hamiltonian of a GKP encoding emerges on the right-hand side of Eq.~\eqref{eq:hrwa_general} when the nodes $\{t_i\}$ are located such that the terms in the sum commute; i.e., when all pairs of the phase-space displacements sampled at the nodes, $\{{\mathbf D}_i=(\cos\tilde \Omega t_i,2\pi \fz \sin \tilde \Omega t_i)\}$, span parallelograms that are integer multiples of $2\pi$: $|\mathbf D_i\times \mathbf D_j|\in 2\pi \mathbb Z$.
In this case each term in the sum in Eq.~\eqref{eq:hrwa_general} will be proportional to a stabilizer of a generalized (square, hexagonal, or oblique) GKP encoding.
For instance, the square-lattice GKP stabilizer Hamiltonian $\hat H_{\rm GKP}$ in Eq.~\eqref{eq:gkp_stab_hamil} can be realized with $\fz \approx 2$ for any voltage tone $V(t)$ whose nodes are uniformly spaced by $\tilde T/4$.

A broad family of waveforms and circuit configurations satisfy the conditions above, and thus lead to the realization of a GKP stabilizer Hamiltonian.
For instance, for any $\fz\geq 2$, the stabilizer Hamiltonian of an oblique GKP encoding can be realized for any voltage tone with fundamental angular frequency $\Omega\approx 2\omc $ whose nodes within a period are located at times $0,2\theta/ \Omega$, where $\theta =\arcsin(2/\fz)$, as in Fig.~\ref{fig:oblique}(a). For this setting, Eq.~\eqref{eq:hrwa_general} leads to the stabilizer Hamiltonian of the oblique-GKP encoding with commuting stabilizers $ \cos(\hat \phi)$ and $\cos(\cos(\theta)\hat\phi +4\pi \hat n) $ [see Fig.~\ref{fig:oblique}(b)].
Likewise, the stabilizer Hamiltonian for the hexagonal GKP encoding emerges when $\fz\approx 4/\sqrt{3}$ and $\Omega\approx 3\omc$, and the nodes of the voltage tone are evenly spaced by period $T/2$; we demonstrate this analytically and numerically in Appendix~\ref{app:hex}.
We finally also note that the circuit can implement a $d$-dimensional GKP encoded qubit for $\fz \geq d$ when $\theta =\arcsin(d/\fz)$.

We finally note that the prefactor of the stabilizers in the effective Hamiltonian does not have to be an oscillatory function of the phase amplitude $\thmax$, as in Eq.~\eqref{eq:tildej}: for instance, picking $V(t)$ to have fundamental angular frequency $\Omega\approx \omc$ and two nodes within each period, located at $t_0=0$ and $t_1=T/4$, will lead to~\footnote{Here
we assumed $\partial_t V(t_0)\geq0$ without loss of generality} \begin{align}
    \hat H_{\rm eff}\approx& \frac{E_J\big[\cos(\hat \phi\!-\!\Phi(t_0)
    \!-\!\pi/4)+\cos(4\pi \hat n\!-\!\Phi(t_1)\!+\!\pi/4)\big]}{T\sqrt{2e\partial_t V(t_0)/h}}
  \notag \\ &+\frac{\hbar \delta \omega}{2} \left( \frac{\hat \phi^2}{4\pi }+4\pi \hat n^2 \right) .\label{eq:no_oscillation}%
\end{align}
Above, a shift of phase amplitudes $\{\Phi(t_i)\}$ will not cause a modulation of $\tilde E_{0,1}$, but rather generate a displacement of the stabilizers and GKP states in phase space.

\subsection{Corrections to the RWA and SPA}
\label{sec:finite_freq}

Our derivation above was based on the RWA and SPA. These approximations are valid in the limit $2e V_{\rm max}/E_J,2eV_{\rm max}/\hbar \Omega \gg 1$.
Realistically, however, the ratios will be finite. As a result, $\Heff$, while still well-defined and numerically tractable, will acquire weak finite-frequency corrections away from $\HGKP$~\cite{eckardt_high-frequency_2015}. Here we discuss the implications of these corrections.

The finite-frequency corrections above will in general be phase-space nonlocal due to the phase-space nonlocality of the time-dependent terms in $\tilde H(t)$~\footnote{Specifically, the corrections can be obtained by perturbative expansion in $E_J/2eV_{\rm max}$ and will, in general, contain powers of the Fourier coefficients of $\tilde H(t)$~\cite{eckardt_high-frequency_2015}. These in turn can be obtained with the SPA. The $z$th component will thus be given by terms of the form $\sim \sum_i \tilde E_z^i \cos(\cos[\tilde \Omega t^i_z] \hat \phi + \sin[\tilde \Omega t^i_z] 2\pi \fz\hat n)$, where $\tilde E_z^i $ are some energy prefactors, and for a given $z$, $\{t^i _z\}$ denote the values of $t$ for which $2eV(t)=\hbar z \Omega$.}. They can compromise the protection of the qubit by lifting the degeneracy of the eigenvalues of $\Heff$ for the doublets of its eigenstates that correspond to distinct logical GKP states, $E_{ab}^1$ and $E_{ab}^0$ [see Eq.~\eqref{eq:energies}]. This will lead to a slow, but finite, precession of the logical state on the Bloch sphere, whose axis and rate will in general depend on the indices $a,b$; thus this precession will ultimately dephase the qubit.
The adverse effects of these finite-frequency corrections can be partly mitigated using encodings such as Eq.~\eqref{eq:GKP_encoding}, where eigenstates of $\HGKP$ have $\hat \phi$ support near odd multiples of $\pi$, and thus the near-degenerate levels correspond to combinations of these states that are even and odd under inversion symmetry.
Since the finite-frequency corrections also respect the inversion symmetry, they cannot lift the near-degeneracy between the near-degenerate doublets by coupling them. This leads to a significant reduction in the dephasing rate.

A second potential impact of finite-frequency corrections is drive-induced state transitions (DUST)~\cite{dumas_measurement-induced_2024}, where low-lying eigenstates of $\hat H_{\rm GKP}$---in codespace---are coupled to high-lying eigenstates outside codespace when they accidentally differ in energy by an integer multiple of $\hbar\tilde \Omega$.
We expect such resonances to be rare, however: since $\hbar \tilde \Omega\gg \tilde E _J $ in the parameter regime we consider (see Fig.~\ref{fig:hierarchy}), high-energy eigenstates of $\hat H_{\rm GKP}$ that are resonant with low-lying eigenstates in codespace must have energy at least $\sim \hbar \tilde \Omega $. Moreover, since again $\tilde E_J \ll \hbar \tilde \Omega$, such resonant states must have their energy dominated by the quadratic part of $\hat H_{\rm GKP} $, i.e., $\frac{\hbar \delta \omega }{2} \Big( \frac{\hat\phi^2}{2\pi \fz}+2\pi \fz \hat n^2 \Big)\sim \hbar \tilde \Omega $. As a result their phase space
support must be concentrated at a distance $\sim r_{\rm res}=\sqrt{2\pi \fz \Omega/\delta \omega}$ from the origin, with phase-space distance $r$ defined as $ r^2=\hat \phi^2+(2\pi \fz\hat n)^2$. On the other hand, the phase-space support of low-lying GKP states in Eq.~\eqref{eq:lattice_basis_wf} only extends up to a distance $ \sim \lambda^{-1}\sim ({\tilde E_J/\delta \omega})^{1/4}$ from the origin, which is much smaller than $r_{\rm res}$ in the parameter regime we consider (Fig.~\ref{fig:hierarchy}). Thus, DUST resonances may only emerge through high-order resonant processes that generate phase-space displacement by a distance of order $r_{\rm res}-r_{\rm GKP}\sim r_{\rm res}$. Since finite-frequency corrections to $\Heff$ arise from the nonlinear term $\cos(\hat \phi\cos\Omega t+2\pi\fz \hat n \sin\Omega t+\Phi(t)) $, which generates phase-space displacements of order $2\pi \fz $, we heuristically expect that the DUST resonances will emerge as a resonant process of order $\gtrsim r_{\rm res}/2\pi\fz \sim \sqrt{\Omega/2\pi\fz \delta \omega}$, which is much greater than $1$. Thus, we expect DUST resonances to be rare and slow.

\begin{figure}
    \centering
    \includegraphics[width=\linewidth]{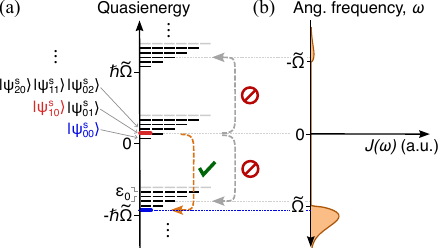}
    \caption{\textbf{Quantum error correction via sideband-cooling.}
    (a) Low-lying quasienergy spectrum of the driven circuit in the regime of GKP stabilization. Each line represents a degenerate doublet of Floquet eigenstates, which are approximately given by the lattice basis states $\{|\psi_{ab}^s\rangle\}$, as indicated by the arrows, with quasienergies $\varepsilon_0(a+b+1)$ [Eq.~\eqref{eq:energies}]. As described in Sec.~\ref{sec:dec},
    $w=a+b$ provides a notion of syndrome weight, with GKP codespace reached at smaller values of $w$.
    (b) Illustration of dissipative error correction via sideband-cooling. Orange curve plots an example of a suitable PSD $J(\omega)$---specifically, the one we simulate in Sec.~\ref{sec:numerical}: see Table~\ref{tab:parameters} for parameters. Note that the vertical axis has the same scale as in (a), but has its direction flipped and is shifted to have origin aligned with the energy of the ``perturbed'' $w=1$ state $|\psi_{10}^s\rangle$ [red in (a)]. This state may relax into the syndrome-free state $|\psi_{00}^s\rangle$ [blue in (a)] by emitting a quantum of energy $\hbar\tilde \Omega +\varepsilon_0$ to the environment, as indicated by the orange arrow---a process that realizes DQEC. The reverse process,
    where the system {\it absorbs} a quantum of energy $\hbar\tilde \Omega +\varepsilon_0$ (upper gray arrow), is thermodynamically suppressed by a factor of $e^{-2\hbar \tilde \Omega/k_{\rm B} T_{\rm B}}$. Furthermore, the $w$-increasing process where the bath absorbs energy $\hbar \tilde \Omega - \varepsilon_0$ (lower gray arrow), is forbidden due to the constrained spectrum of the environment.}
    \label{fig:2}
\end{figure}

\section{Dissipative error correction from side-band cooling}
\label{sec:lindblad}
We now show how DQEC can be implemented.
In Sec.~\ref{sec:dec} we showed that DQEC can be achieved by coupling a resonator with Hamiltonian $\hat H_{\rm GKP}$ to a generic thermodynamic environment at temperature smaller than $\tilde E_J/k_{\rm B}$.
However, since the circuit from Sec.~\ref{sec:device} realizes $\hat H_{\rm GKP}$ through Floquet engineering, dissipation will in general introduce drive-assisted incoherent processes~\cite{Zhang_2024}.
Here we analyze the impact of these drive-induced incoherent processes on DQEC. We find they in fact can {\it enhance} performance with suitably structured environments, via a sideband-cooling mechanism that enables DQEC at significantly higher temperatures than $\tilde E_J/k_{\rm B}$.

DQEC can be achieved with a broad class of configurations, as in Sec.~\ref{sec:extensions_to_more_general}, and in principle by coupling any phase-space local observable to the filtered environment. However, for simplicity, here we focus on the example that we also considered in the previous section, where $V(t) = V_{\rm max} \sin\Omega t$ and $\fz=2$.
Moreover, we consider a capacitive coupling to the environment, which can be efficiently implemented in superconducting circuits.

We model the capacitively coupled environment by adding the term
$\hat H_{\rm RC} = \frac{2e \hat n \hat q_ {\rm B}(t)}{C_ {\rm B}}$ to the Hamiltonian in Eq.~\eqref{eq:hamiltonian_definition},
where $\hat q_{\rm B}(t)$ denotes the fluctuating charge on the bath-side of the coupler, and $C_{\rm B}$ the capacitance of the coupler.
Here $\hat q_B(t)$ is a {Gaussian} quantum noise variable with {zero mean} and power spectral density (PSD) $J(\omega )$, such that $\langle \hat q_{\rm B}(t)\hat q_{\rm B}(t')\rangle = \int d\omega e^{-i\omega t} J( \omega)$.
For an environment in thermal equilibrium at temperature $T_{\rm B}$, $J(\omega) = S(|\omega|)\big[\Theta(\omega )+n_{\rm B}(|\hbar \omega|)\big]$, where $n_{\rm B}(E)=\frac{1}{e^{ E/k_{\rm B}T_{\rm B}}-1}$ denotes {the Bose-Einstein distribution at temperature $T_{\rm B}$}, $\Theta(x)$ denotes the Heaviside step function, and $S(\omega )$ the spectral density of $\hat q_{\rm B}(t)$.
For DQEC to emerge, $S(\omega)$ should have its support effectively confined in the region $\hbar \tilde \Omega<\omega<\frac{3}{2}\hbar \tilde \Omega $, as in Fig.~\ref{fig:2}(b), while $T_{\rm B}$ should be much smaller than $\hbar \tilde \Omega /k_{\rm B}$, as indicated in Fig.~\ref{fig:hierarchy}. In Appendix~\ref{app:chebyshev} we show that such constrained spectral densities $S(\omega)$, can e.g., be achieved via a single lossy resonator or an array of these~\cite{krantz_quantum_2019,ferreira_collapse_2020}.

We model the dissipative dynamics of the circuit through the universal Lindblad equation (ULE) from Refs.~\cite{Kirsanskas_2018,Phd_thesis,Davidovic2020completelypositive,NathanPRB2020_ULE,NathanarXiv2022Jun}. The ULE is valid when the bath-induced loss rate is much smaller than the inverse decay time of $J(t)=\langle \hat q_{\rm B}(t)\hat q_B(0)\rangle$~\cite{NathanPRB2020_ULE,NathanarXiv2022Jun}.
In App.~\ref{app:dissipative-dynamics}, we show that, in this limit, and for $ \hbar \Omega \gg \tilde E_J , \hbar \delta \omega $, the stroboscopic dynamics within the code subspace becomes approximately equivalent to those generated by the {\it time-independent} master equation
\begin{align}
\partial_{t}\hat \rho(t)& = \frac{i}{\hbar}[\hat \rho(t),\Heff] \, +\!\!\!\!\sum_{z=-\infty}^\infty\!\!\!\! \Big(\hat L_z\hat \rho(t)\hat L_z^\dagger-\frac{1}{2}\{{ \hat L_z^{\dagger} \hat L_z},\hat \rho(t)\}\Big).
\label{eq:Lindblad_eq_HighFrequency}
\end{align}
Here
$ \hat L_z $ is constructed from the eigenvalues~\footnote{Recall from the text below Eq.~\eqref{eq:tildehdef} that the eigenvalues of $\Heff$ are uniquely defined, and located in the interval $[-\hbar \pi/\tilde T,\hbar \pi/\tilde T]$.} $\{\varepsilon_a\}$ and eigenstates $\{|\Psi_a\rangle\}$ of $\Heff$ through
\begin{align}
\hat L_z &= \sum_{ab}\sqrt{2\pi J(\varepsilon_b/\hbar -\varepsilon_a/\hbar + z \tilde \Omega)}|\Psi_a\rangle\langle \Psi_a|\bar {\mathcal V} _z |\Psi_b\rangle \langle\Psi_b| ,\notag \\
\bar {\mathcal V}_z &\!\equiv\! \frac{2e}{\tilde {T}C_{\rm B}}\int_0^{\tilde T} \!dt \cM^ \dagger(t) [\hat n\cos\tilde \Omega t \!+ \!\frac{\hat \phi}{2\pi\fz } \sin\tilde \Omega t] \cM(t) e^{- iz \tilde \Omega t}
\label{eq:jump_operator}
\end{align}
with $\cM(t) \equiv \tilde U(t) e^{i\Heff t/\hbar} $ denoting the time-periodic micromotion operator of the system, which satisfies $\cM(t)= \cM(t+\tilde T)$. Heuristically, $\hat L_z$ represents drive-assisted relaxation processes involving the absorption of $z$ drive photons.

Importantly, $\cM(t)$
approaches the identity in the rapid-driving limit we consider~\cite{eckardt_high-frequency_2015}.
In this case $\Vz$ is most significant for $z=\pm 1$, where it takes value $\bar {\mathcal V}_{\pm 1}\approx \frac{e}{2C_{\rm B}}(\hat n \pm i\hat \phi/2\pi \fz) $. Thus, relaxation is dominated by the side-band cooling and heating processes $\hat L_{\pm 1}$.
Moreover, since $|\varepsilon_a |\leq \tilde \Omega /2$ for low-energy eigenstates of $\Heff$, and $J(\omega)$ has its support confined between $\tilde \Omega$ and $3\tilde \Omega/2$, only the terms with $z\in \{0,1,2\}$ contribute.
Hence the dissipative processes are dominated by the sideband cooling term $\hat L_1$.
When keeping only the $z=1$ term in
the sum in Eq.~\eqref{eq:Lindblad_eq_HighFrequency}, the master equation is equivalent to that of a time-independent system described by $\Heff\approx \hat H_{\rm GKP}$ coupled to a bath with PSD $J(\omega + \tilde \Omega)$ via the observable $\bar{\mathcal V}_1\approx \frac{e}{2C_{\rm B}} (\hat n + i\hat \phi/2\pi \fz)$~\cite{NathanPRB2020_ULE}.
In Sec.~\ref{sec:dec} we found that this realizes DQEC provided the effective temperature of the environment $T_{\rm eff}$ is much smaller than $\tilde E_J/k_{\rm B}$.

To estimate $T_{\rm eff}$ we note that the bath is not in thermal equilibrium; the ratio between upward and downward transition probabilities between levels separated by energy $\hbar\omega $ is given by $ \frac{J( \tilde \Omega +\omega )}{J( \tilde \Omega -\omega)}$, and thus cannot be expressed via a Boltzmann factor. Instead, there is a transition-frequency dependent temperature $\tilde T_{\rm B}(\omega)=\frac{\hbar \omega}{k_{\rm B}\log\left[{J( \tilde \Omega +\omega )}/{J( \tilde \Omega -\omega)}\right]}$.
However, in the low-energy eigenspace of $\Heff$ (coinciding with the GKP codespace), $\hat {\mathcal V}_1\approx \frac{e}{2C_{\rm B}} (\hat n + i\hat \phi/2\pi \fz)$ dominantly couples adjacent sites in the GKP lattice, and thus increases or decreases the energy by $\varepsilon_0$.
Thus the sum defining the jump operator $\hat L_1$ in Eq.~\eqref{eq:jump_operator} is dominated by terms with transition frequencies $\pm \varepsilon_0/\hbar$, and the effective temperature experienced by the system is
\be
T_{\rm eff} \approx \frac{\varepsilon_0 /k_{\rm B}}{ \log S( \tilde \Omega+\varepsilon_0/\hbar )-\log S(\tilde \Omega-\varepsilon_0/\hbar)}
\ee
where we used that $J(\omega )\approx S(\omega)$ for $\hbar \omega \gg k_{\rm B}T_{\rm B}$.
Note that $T_{\rm eff}$ depends exclusively on the structure of the bath spectrum: if $S( \tilde \Omega - \varepsilon_0/\hbar)\ll S(\tilde \Omega + \varepsilon_0/\hbar) $, $T_{\rm eff}$ can be much smaller than $\tilde E_J$.
In this case, the bath effectively cools the resonator into the low-energy subspace of $\Heff$---i.e., the code subspace---, achieving dissipative quantum error correction [see Fig.~\ref{fig:2}(b)].

As in Sec.~\ref{sec:device}, there are finite-frequency corrections to the picture above: for finite $E_J/2eV_{\rm max}$, $\cM(t)$ will deviate from $1$ by phase-space nonlocal corrections, by the same considerations that we outlined in Sec.~\ref{sec:finite_freq}. As a result, each $\hat L_z$ in Eq.~\eqref{eq:jump_operator} will contain phase-space nonlocal finite-frequency corrections that give it a finite, but small, probability for coupling GKP states encoding distinct logical states. Hence, the error-correcting quantum jumps $\hat L_z$ will have a small, but finite probability of inducing logical errors due to finite-frequency corrections. This feature may provide an eventual limit on the qubit's performance, at least within the ULE approximation. At the same time, we emphasize the small error probability still allows for significant lifetime improvement by DQEC, as we observe in our simulations in the next section.

\begin{figure}
\includegraphics[width=0.8\columnwidth]{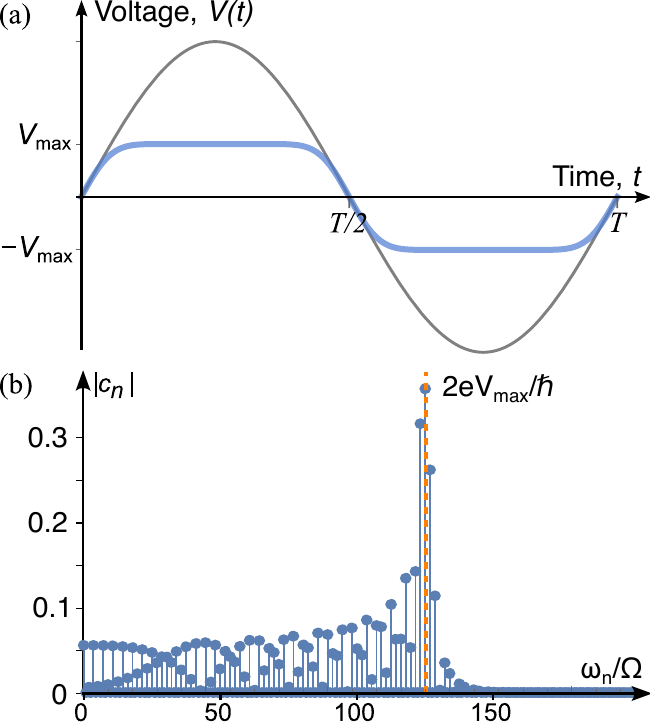}
\caption{{\bf Capped voltage tone}. (a) Blue: smoothly capped voltage tone used in our simulations that approximates a sinusoidal drive (gray) outside the cap. See Appendix~\ref{app:pulse} for explicit expression. (b) Absolute value of Fourier coefficients of $e^{\pm i\Phi(t)}$, which form the time-dependent term in the circuit Hamiltonian. The orange line indicates the frequency corresponding to the voltage amplitude, $2eV_{\rm max}/\hbar$.}
\label{fig:pulse}
\end{figure}

\section{Numerical simulations}
\label{sec:numerical}

\begin{figure*}[t]
    \centering
    \includegraphics[width=\linewidth]{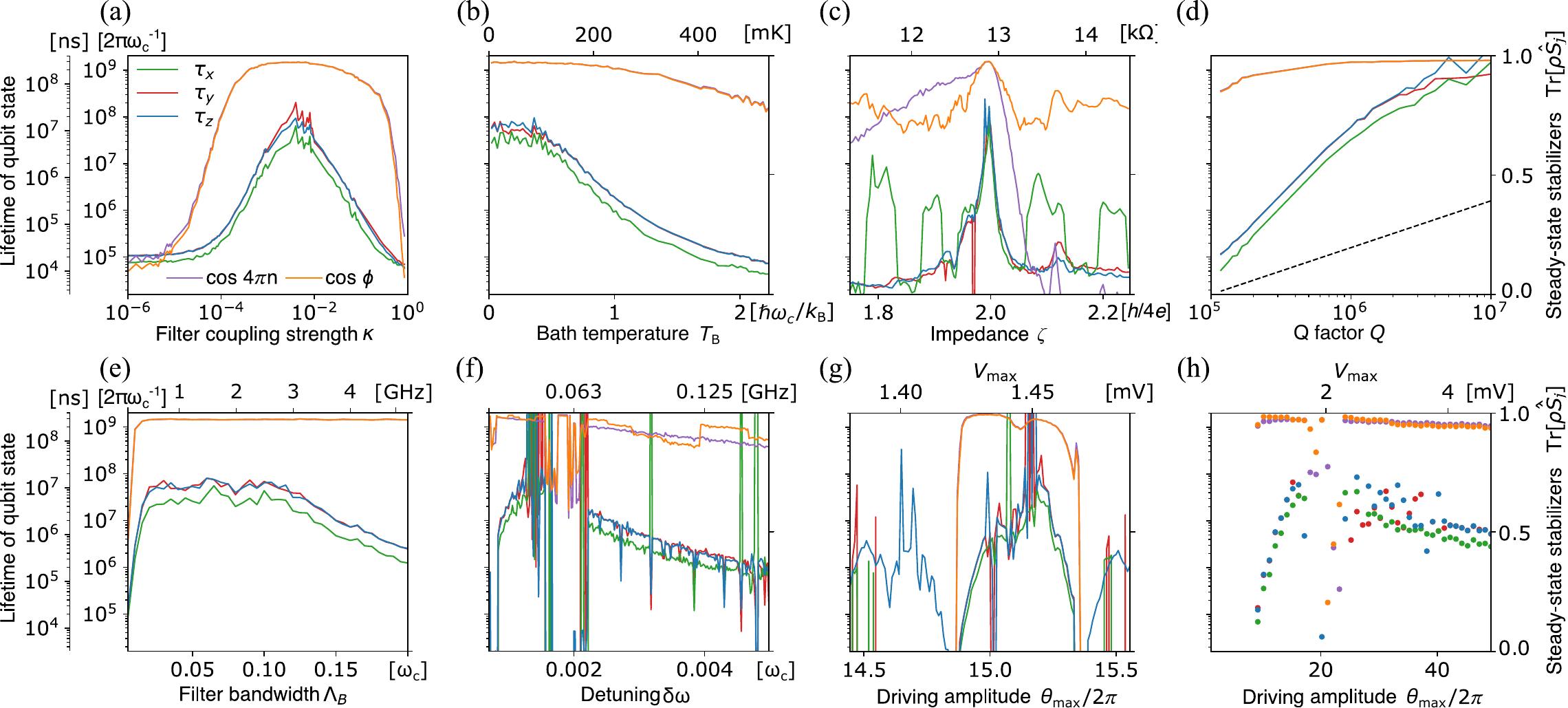}
    \caption{\textbf{Dependence of lifetime and stabilizers on key parameters.} Lifetimes of information $\tau_j,\, j = x,y,z$ stored in the expectation values of the logical operators $\langle \hat{\sigma}_j \rangle,\, j = x,y,z$ (green, red, and blue; left y-axes) and steady-state stabilizer expectation values (yellow and orange; right y-axes) as a function of
    (a) dimensionless coupling to the filtered bath,
    (b) bath temperature
    (c) impedance,
    (d) $Q$ factor, $\omc /\Gamma_{\hat a}$,
    (e) filter bandwidth, and
    (f) detuning.
    Panels (g) and (h) display logical lifetime and stabilizers against the driving amplitude $V_{\rm max}$ ($\thmax$), for an interval corresponding to $14.5<\thmax/2\pi\leq 15.5$ (g), and sampled discretely at values $\thmax /2\pi = z + 0.2$ for integer $z$, where we expect the logical stabilizers and lifetime to approximately attain its local maxima.
    We keep all non-varied parameters fixed at the default values in Table~\ref{tab:parameters}.
    The dashed line in (d) denotes the photon lifetime.
    }
    \label{fig:lifetimes}
\end{figure*}

To verify our predictions, we simulate the driven circuit, in a way which does not rely on the SPA and RWA approximations we made in Sec.~\ref{sec:device}.

As above, we consider the configuration resulting in a square-lattice GKP qubit that we also considered in the above sections, where $\fz = 2$ and $\Omega \approx 2\omc$.
Throughout our simulations we use the parameters from Table~\ref{tab:parameters}, unless stated otherwise.
We take the voltage tone $V(t)$ to be the smoothly capped sinusoidal driving signal shown in Fig.~\ref{fig:pulse}(a)---see Appendix~\ref{app:pulse} for explicit expression.
The pulse shape is chosen to balance the first and second requirements in Sec.~\ref{sec:extensions_to_more_general}, namely to minimize the time-window where $|V(t)|\sim E_J/2e$, to satisfy the conditions for the SPA and RWA (see Sec.~\ref{sec:device}), while minimizing higher harmonics in the pulse shape and the maximum voltage $V_{\rm max}$, to avoid quasiparticle generation.
With this pulse shape, $V_{\rm max}$ remains significantly lower than the superconducting gap of NbN ($\Delta/e \approx 3\,{\rm mV}$), while the modulation remains adiabatic with respect to the quasiparticle excitation gap. Thus, we expect our parameters are compatible with NbN.
Supporting this, Fig.~\ref{fig:pulse}(b) shows the Fourier coefficients of $e^{\pm i\Phi(t)}$ resulting from this voltage tone, which constitute the driving terms of $\hat H(t)$. Evidently, these are effectively limited by the frequency associated with the voltage amplitude, $2eV_{\rm max}/\hbar$, meaning the drive-induced sidebands do not resonantly excite quasiparticles.
The parameters and pulse shape we pick result in $\tilde{E}_{J}/h \approx {0.11} \, \tilde\Omega/2\pi \approx 0.55\,{\rm GHz}$ (see Appendix~\ref{app:SPA} for expression).
We emphasize that the pulse shape is not essential for our simulation, but
only chosen to be consistent with the gap parameters of NbN.
For instance, in Appendix~\ref{app:hex} we
simulate a realization of the hexagonal GKP stabilizer Hamiltonian using a pure sinusoidal tone.

We use a bandpass filter with PSD given by
\begin{align}
    J(\omega) = \left\{\begin{array}{ll}\hbar \kappa \theta(\omega - \tilde \Omega)(\omega - \tilde \Omega) e^{-\frac{(\omega - \tilde \Omega)^2}{2 \Lambda_{\rm B}^2}}, \ &\omega \geq 0\vspace{1mm} \\
    e^{\frac{\hbar\omega}{k_{\rm B} T_{\rm B}}} J(-\omega),
&\omega<0\end{array}\right.
    \label{eq:J(E)}
\end{align}
with $\kappa$ a dimensionless number parameterizing the system bath coupling.
In Fig.~\ref{fig:2}(b) we plot $J(\omega)$ with our simulated parameters, $T_{\rm B} = 120\,{\rm mK}$ and $\Lambda_B = 0.07 \omc$.
The PSD is chosen to have its support confined within the interval $\tilde\Omega \leq |\omega|\leq 3\tilde \Omega /2$, but we emphasize that the qubit is not sensitive to the details of the PSD as long as it has this property.
In Appendix~\ref{app:chebyshev} we demonstrate that the device can also be operated with more general filters with the same features, consisting of a single lossy LC resonator, or arrays of lossy LC resonators \cite{Sete2015BandpassPurcell,Cattaneo2021EngineeringDissipation,Thorbeck2024PassiveLeakageReset}.

To demonstrate the qubit's resilience to noise, we also include extrinsic photon loss in the simulation, by adding an additional jump operator to Eq.~\eqref{eq:Lindblad_eq_HighFrequency} given by the annihilation operator $\hat a = \sqrt{\pi\Gamma_ a \fz} (\phi - i n/2\pi \fz)$,
which, e.g., can result from uncontrolled capacitive coupling of the resonator to a bosonic environment (see App.~\ref{app:dissipative-dynamics} for details). Here $\Gamma_{\hat a}$ denotes the loss rate, and corresponds to a resonator $Q$ factor $Q=\omega_c/\Gamma_a$. We emphasize we consider this noise channel only as an example: as we show in Appendices~\ref{app:voltage_noise}-\ref{app:low_frequency_noise}, the qubit is also resilient towards other sources of noise, such as noise in the voltage bias, or low frequency charge and flux noise.

\begin{table}
\begin{tabular}{|c|c|c|c|c|c|c|c|}
\hline
       $\omc/2\pi$%
     & $\delta \omega/2\pi$
     & $E_J/h$
     & $Q$
     & $\kappa$ &
     $\Lambda_{\rm B}$
     & $T_{\rm B}$
     & $V_{\rm max}$
     \\
     \hline\hline

      $5\, {\rm GHz}$ & $6\,{\rm MHz}$
     & $ 20\,{\rm GHz}$ & $160\, 000$ & $.002$ &$ .35\:{\rm GHz}$ &$.05\,{\rm K} $& {$1.45\, {\rm mV}$}\\
     \hline
         $1/2\pi$& $.0075$
      & $25 $ & $ 160\, 000$ & $.002$ & $.44$ & $.63$%
      & $430 $ \\
     \hline
\end{tabular}
\caption{{\it Row 1}: Default parameters for simulations.
Here, $\omc$ denotes the resonator angular frequency, $\delta \omega$ the detuning of half the angular drive frequency from $\omc$, $E_J$ the junction Josephson energy, $Q=\omega_c/\Gamma_{\rm \hat a}$ the $Q$ factor from the extrinsic photon loss, $\kappa$ the dimensionless system-bath coupling, $\Lambda_{\rm B}$ filter bandwidth, $T_{\rm B}$ bath temperature, and $V_{\rm max}$ voltage amplitude (see Fig.~\ref{fig:pulse} and Appendix~\ref{app:pulse} for explicit waveform).
The resonator frequency corresponds to inductance $L = 410 \,{\rm nH}$ and capacitance $C=2.5 \,{\rm fF}$.
{\it Row 2:} parameter values in units where $\omc=1$ and $h=e=k_{\rm B}=1$.
 }
\label{tab:parameters}
\end{table}

Using the parameters above, we obtain $\Heff$ exactly from $\frac{i\hbar}{\tilde T}\log\mathcal T e^{-i\int_0^{\tilde T} \hat H(t) dt/\hbar }$, discretizing the Hilbert space via a regularly-spaced $\phi$ grid,
 and compute the jump operators $\hat L_{-1},\hat L_0,\hat L_1$ via Eq.~\eqref{eq:jump_operator}; the remaining jump operators are neglected since $J(\omega)$ is exponentially suppressed outside this interval (we only include $\hat L_{\pm 2}$ explicitly for calculations where we vary the filter width $\Lambda_B$).
See Appendix~\ref{app:numerics} for further description of the simulation.
We then numerically integrate Eq.~\eqref{eq:Lindblad_eq_HighFrequency} without any further approximation via the stochastic Schr\"odinger equation~\cite{dalibard_wave-function_1992,carmichael_open_1993}, using 1000 realizations per parameter set. Thus our simulation is agnostic to the assumptions and approximations we made to obtain $\HGKP$ in Sec.~\ref{sec:device}.

We first demonstrate the dissipative stabilization of GKP states in the circuit.
To this end, we initialize the resonator in an ensemble of random high-energy states whose components in the discretized $\phi$ basis are given by $\psi_j = (a_j + i b_j) / \mathcal{N}$ with $a_j, b_j$ randomly drawn from $ [-1, 1]$, and $\mathcal{N}$ a normalization prefactor.
We then evolve the state over a time-window $12000\times 2\pi /\omc \approx 2.4 \,\mu{\rm s}$.
In Fig.~\ref{fig:1}(b), we show the Wigner function of the final state, while Fig.~\ref{fig:1}(c) shows the evolution of GKP stabilizers over the simulated time-window.
Evidently, the system relaxes to the GKP codespace within $\sim 2\,\mu{\rm s}$, demonstrating that dissipation stabilizes the resonator in codespace.

We next demonstrate that DQEC is achieved in the circuit.
To this end, we obtain the evolution of the logical operators $\{\langle \hat{\sigma}_i\rangle \}$ defined in Eq.~\eqref{eq:GKP_encoding}, in the presence and absence of dissipation, starting from a state in the GKP code subspace with $\langle \hat{\sigma}_i\rangle = 1/\sqrt{3}$ for $i=1,2,3$.
The data are shown in Fig.~\ref{fig:1}(d).
Evidently, the dissipation leads to a significant improvement of the qubit lifetime.
The halved lifetime of $\hat{\sigma}_x$ relative to those of $\hat{\sigma}_{y,z}$ is consistent with the dominant error processes being caused by transitions across the potential energy barriers of $H_{\rm GKP}$---located at
{$\hat \phi \in 2\pi \mathbb Z$ and $\hat n\in \frac{1}{2} \mathbb Z$}.
These transitions result in error processes generated by $\hat{\sigma}_y$ and $\hat{\sigma}_z$ at equal rates.

We next obtain the qubit lifetimes and final-state stabilizer expectation values as a function of the most relevant device parameters.
The lifetimes $\tau_j,\, j=x,y,z$ were computed by a linear fit to the logarithm of the expectation values of the logical operators ${\rm Tr}(\rho(t) \hat{\sigma}_j)$ for the time-evolved density matrix. The results are plotted in Fig.~\ref{fig:lifetimes}.

Fig.~\ref{fig:lifetimes}(a) plots the qubit lifetime and stabilizer expectation values versus system-bath coupling $\kappa $. Both quantities increase with $\kappa$ for $\kappa\gtrsim 10^{-4}$, demonstrating the onset of DQEC. This value of $\kappa$ marks the point where the stabilization time of GKP states matches the photon loss time.
For $\kappa\gtrsim 10^{-2}$, lifetime begins to decrease again; we expect this to be due to the finite-frequency corrections to the jump operators discussed at the end of Sec.~\ref{sec:lindblad}.

Fig.~\ref{fig:lifetimes}(b) plots lifetimes and stabilizers versus bath temperature. The data shows that stabilization of GKP states occurs for bath temperatures
$ T_{\rm B} \lesssim 4 \, \hbar \omc/k_{\rm B}\approx 1000 \,{\rm mK}$, while DQEC emerges for
$T_{\rm B} \lesssim \hbar \omc/k_{\rm B}\approx 250\,{\rm mK}$.
As $T_{\rm B}$ is decreased below this threshold, lifetime appears to increase exponentially, consistent with the Arrhenius law we obtained in Eq.~\eqref{eq:arrhenius}.

In Fig.~\ref{fig:lifetimes}(c) we plot lifetime and stabilizers against the resonator impedance $\fz$. The data indicate that DQEC has a narrow relative tolerance of $\approx 0.5 \%$ for deviations of $\fz$ away from $2$,
while stabilization of GKP states has a significantly larger tolerance for deviations of the impedance, of order $\approx 10\%$. Note that DQEC can be restored for any positive deviation of $\fz$ by adjusting the drive, as detailed in Sec.~\ref{sec:extensions_to_more_general} (see also Fig.~\ref{fig:oblique}).

Fig.~\ref{fig:lifetimes}(d) plots lifetime and stabilizers versus the $Q$ factor. The data indicates that stabilization of GKP states sets in for $Q\gtrsim 10^5$; there the time between two photon loss events $Q/\omc$ becomes longer than the time we observe it takes to stabilize the GKP states, $t_{\rm stab}$; for the parameters we use (see Table~\ref{tab:parameters}), $t_{\rm stab} \approx 10^4 \times 2\pi \omc^{-1}$ [compare Fig.~\ref{fig:1}(c)].
Lifetime grows superlinearly with $Q$ above this threshold, and appears to saturate for $Q\gtrsim 10^{6}$.

We next probe the role of the spectral structure of the environment. Fig.~\ref{fig:lifetimes}(e) plots the lifetimes and stabilizers as a function of the upper cutoff $\Lambda_B$ for our
 PSD in Eq.~\eqref{eq:J(E)}. Evidently, the lifetime achieves a plateau as long as $\Lambda_B$
 is sufficiently small to prohibit absorption at frequency $2 \tilde \Omega-\varepsilon_0/\hbar $, while admitting absorption at frequency $ \tilde \Omega +\varepsilon_0/\hbar $, consistent with our discussion in Sec.~\ref{sec:lindblad} (see in particular Fig.~\ref{fig:2}).
 The plateau demonstrates that the results do not depend on the precise shape of the filter spectral function, as long as these main qualitative features are preserved.

We next consider the role of frequency detuning $\delta \omega$: Fig.~\ref{fig:lifetimes}(f) shows that lifetime is maximized at
$\delta \omega_{\rm opt} \approx .0012 \omc$.
Larger $\delta \omega$ leads to lifetime decreasing, because the resulting GKP squeezing parameter $\lambda \propto (\delta \omega/\tilde E_J)^{1/4}$ increases, enhancing the error rate due to interwell-tunneling~\cite{Nathan2024May}.
Decreasing $\delta \omega$ from $\delta \omega_{\rm opt}$ leads to lifetime decrease. We expect this is due to the growth of the phase space extent of the stabilized GKP states, which is proportional to $ \lambda^{-1}\sim \delta \omega^{-1/4}$, as this makes them more susceptible to errors from the photon loss~\cite{Nathan2024May}.
Varying $\delta \omega $ takes the system through narrow DUST-like resonances, where stabilized GKP states undergo avoided crossings with high-lying states outside codespace (see Sec.~\ref{sec:finite_freq}). The observed high density of resonances may be artifacts of our simulation, due to low-lying GKP states becoming resonant with unphysical states with support near, or beyond the boundaries of the phase-space region we cover with our finite $\hat \phi$ grid.

We finally consider the roles of the waveform parameters.
In Figs.~\ref{fig:lifetimes}(g,h) we plot lifetime and stabilizer expectation value against the phase bias amplitude {$\theta_{\rm max}$, which is proportional to the voltage bias amplitude $ V_{\rm max}$}. Fig.~\ref{fig:lifetimes}(g) plots these quantities in a window $14.5\leq \thmax/2\pi \leq 15.5$.
Evidently, both lifetime and stabilizers depend strongly on {$\theta_{\rm max}$} in this window:
{The lifetime peaks for $\thmax$ close to $2\pi z + \pi/4,\, z \in \mathbb{N}$ where the GKP barrier height of the effective Hamiltonian, $\tilde E_J \propto\cos(\thmax-\pi/4)/\sqrt{\thmax}$, [see Eq.~\eqref{eq:tildej}] attains its maximum.}
The shown pattern is repeated outside the plotted window, consistent with the effective barrier height being approximately $2\pi$-periodic in the phase oscillation amplitude, $\thmax$.
As described in Sec.~\ref{sec:extensions_to_more_general}, this observed oscillatory behavior can be avoided using other tones than the one we picked here.

In Fig.~\ref{fig:lifetimes}(h), we plot the stabilizer expectation values and lifetimes for a broader range of $\thmax$, picked at values $2\pi (z + 0.2),\, z\in\mathbb Z$, close to the local optimum over the oscillation.
Evidently, local-maximum lifetime decreases with $\thmax$, starting from a minimal threshold
$\thmax \gtrsim 20 \pi$ required for the effective Hamiltonian description in Eq.~\eqref{eq:H_eff_GKP} to be valid.
This is consistent with our expectation that
the effective GKP barrier height $\tilde E_J$ scales as {$\thmax^{-1/2}$}.

\section{Protected single-qubit Clifford gates}
\label{sec:gates}
\begin{figure}
    \centering
    \includegraphics[width=1.0\linewidth]{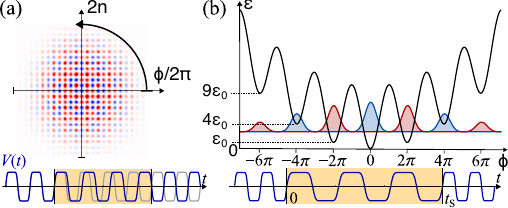}
    \caption{\textbf{Protected single-qubit Clifford gates.} (a) A protected $e^{-i\frac{\pi}{4}\hat \sigma_x}$ gate, expressed as a $\pi/2$ rotation in phase space, is performed by a $\pi$ phase shift of the voltage pulse (blue) relative to normal operation (gray). (b) A protected $e^{-i\frac{\pi}{4}\hat \sigma_z}$-gate is performed
    by switching to a driving frequency close to the resonance angular frequency $\omc$ and then letting even and odd wells acquire a relative $\pi/2$ phase shift from dynamic phases $k^2 \varepsilon_1 t/\hbar$ of well $k$ with $\varepsilon_1 = \frac{\pi}{2} \hbar \delta \omega'$.
    }
    \label{fig:gates}
\end{figure}
We now describe how protected single-qubit Clifford gates can be implemented in the circuit via modulation of the driving tone. As above, we consider the square-lattice GKP qubit configuration described in Sec.~\ref{sec:device}.

A protected
$e^{-i\frac{\pi}{4}\hat \sigma_x}$
gate can be generated by a $\pi/2$ phase space rotation: this operation effectively maps $\hat \phi\to 2\pi \fz \hat n$ and $\hat n \to - \hat \phi/2\pi\fz$ and hence $\hat \sigma_z$ to $\hat \sigma_y$ and $\hat \sigma_y\to -\hat \sigma_z$, consistent with a $e^{-i\frac{\pi}{4}\hat \sigma_x}$ gate [See Eq.~\eqref{eq:GKP_encoding}]. Such an operation can be implemented in the circuit by adiabatically shifting the phase of the voltage tone by $\pi$, see Fig.~\ref{fig:gates}(a).
Provided the phase shift is sufficiently slow, the system will remain confined in the low-energy subspace of the $\Heff$,
implying strong error suppression.

A
protected $e^{-i\frac{\pi}{4}\hat{\sigma}_z}$ gate can be implemented using the phase-revival mechanism described in Ref.~\cite{Nathan2024May}.
The protocol involves the following steps [see Fig.~\ref{fig:gates}(b) and App.~\ref{app:S_T_gates} for details]: first, adiabatically turn on an additional driving tone at $\tilde{\Omega}$
so that $\Phi(t) \to \Phi(t) + \pi \cos \tilde{\Omega}t$. This adiabatically shifts the minima of $\Heff$ by $\Delta \phi =\pi$, so that $|0\rangle$ and $|1\rangle$ GKP logical states have support centered around $\phi \in 4\pi \mathbb Z$ and $\phi \in 4\pi(\mathbb Z + 1/2) $, respectively.
Second, abruptly shift to a driving frequency at $\Omega' = \omc + \delta\omega'$ and voltage amplitude so that $\thmax = 2\pi k - 3\pi/4$ and the minima of the potential remain at $\hat \phi \in 2\pi \mathbb Z$. This turns off the term $\propto \cos (2\pi\fz \hat n)$ in $\Heff$ stabilizing inter-well phase coherence; allowing for a relative phase accumulation between the two logical sectors. Then, by waiting a time $\Delta t_{\rm S} = \frac{1}{\delta \omega'}$, states with support near $\hat \phi = 2\pi k$ have acquired a phase factor $e^{-i k^2 \pi \delta \omega ' t_S /2} =(-i)^{k^2}$; this is equivalent to an $e^{-i\frac{\pi}{4}\hat{\sigma}_z}$ gate.
Finally, quenching back to the resonantly modulated driving tone and adiabatically turning off the modulation at frequency $\tilde \Omega$, reactivates the interwell-coherence stabilizing term $\propto \cos (2\pi\fz \hat n)$, which will correct small relative phase errors from a non-perfect waiting time $t_{\rm S}$.
\section{Magic state preparation}
\label{sec:magic_state_initialization}
We now discuss a native protocol for generating magic states.
Our approach utilizes the results from Ref.~\cite{Baragiola2019Nov}, which show that a magic $H$-type state, $|H_+\rangle \equiv \frac{1}{\sqrt{2}}(|X\rangle+|Y\rangle)$, can be created by applying a GKP error correction scheme to the ground state of the resonator mode.
This mechanism can be achieved in our protocol by (i) initializing the resonator in its ground state while the Josephson junction is decoupled (e.g., via a SQUID or a voltage bias across the junction) (ii) suddenly switching on the Josephson coupling and resonant voltage drive that activates the dissipative error correction. The amplitude of the voltage drive needs to be chosen such that $\tilde E_J>0$ such that the effective GKP potential has a well at the origin [see Eqs.~\eqref{eq:H_eff_GKP}-\eqref{eq:tildej}].

\begin{figure}
    \centering
    \includegraphics[width=\linewidth]{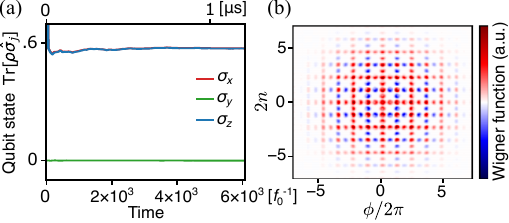}
    \caption{\textbf{Magic state preparation.} (a) Logical information ${\rm Tr}[\rho(t) \hat{\sigma}_j]$ after preparing the system in the resonator ground state and suddenly switching on the driven Josephson coupling at $t = 0$.
    (b) Wigner function of the resulting density matrix at $t = 6000 \times 2\pi \omc^{-1}$. This Wigner function represents a GKP magic state \cite{GrimsmoPRXQuantum2021Jun}. Parameters as in Table~\ref{tab:parameters} but with driving amplitude $V_{\rm max} = 1.40 \, {\rm mV}$, bath temperature $T_{\rm B} = 0.5 \hbar \omc /k_{\rm B}= 120 \, {\rm mK}$, and increased coupling $\kappa = 0.01$ to the filtered bath for faster relaxation to the magic state.}
    \label{fig:magic_state_preparation}
\end{figure}

In Fig.~\ref{fig:magic_state_preparation} we show a numerical demonstration of the magic state preparation protocol. Within a few thousand periods after switching on the driven Josephson coupling, the resonator state settles in a magic $|H_+\rangle = (|X\rangle + |Y\rangle)/\sqrt{2}$ state with fidelity $F = (1 + {\rm Tr[}\rho (\hat{\sigma}_y + \hat{\sigma}_z)]/\sqrt{2})/2 \approx 0.90$. This fidelity is above the threshold for magic state distillation \cite{Bravyi2005Feb,Baragiola2019Nov}.
The infidelity arises because the steady state of the magic state generation protocol is an incoherent mixture of opposite $H$ states, $|H_\pm\rangle = (|X\rangle \pm |Y\rangle)/\sqrt{2}$, with a $\sim 90\%$ probability for a $|H_+\rangle$ state and a $\sim 10\%$ probability for a $|H_-\rangle$ state.

\section{Readout}
\label{sec:readout}

\begin{figure}
    \centering
    \includegraphics[width=\linewidth]{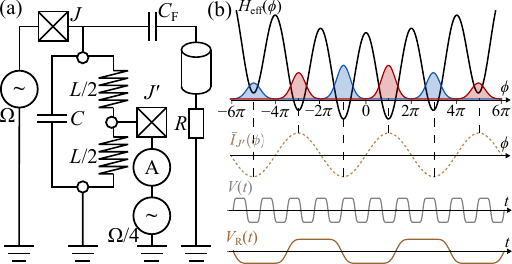}
    \caption{\textbf{Readout.} (a) The readout signal is given by the mean supercurrent $\bar{I}_{J'}$ measured through a weak Josephson junction $J'$ connected to the resonator at $L/2$ of its inductance and driven at $\Omega/4$. (b) This yields an effective $4\pi$-periodic potential so that even wells yield $\bar{I}_{J'} = \bar{I}_0$ while odd wells yield $\bar{I}_{J'} = - \bar{I}_0$.}
    \label{fig:readout}
\end{figure}

The logical state
of the GKP qubit can be read from the mean supercurrent across a weak Josephson junction $J'$ between a node halfway along the inductance of the resonator and a node driven by a second voltage signal, $ V_{\rm R}(t)$ [see Fig.~\ref{fig:readout}(a)]. The voltage tone $ V_{\rm R}(t)$
should have angular frequency $\omega_{\rm R} = \Omega / 4$ and (similar to the main-junction driving tone $V(t)$)
a large voltage amplitude $\max2 e V_{\rm R}(t) \gg \hbar \omega_{\rm R}$, $E_J'$.
The attachment of $J'$ halfway along the inductance means that the Josephson current through $J'$, is proportional to the value of $\sin\hat \phi/2$ in the lab frame.
The large bias amplitude across $J'$ moreover
ensures that the mean current through $J'$ only gets significant contributions near the nodes of $ V_{\rm R}(t)$, which are located at times $t\in \mathbb Z \tilde T$. The mean supercurrent of the biased junction hence is proportional to the mean value of $\sin(\hat \phi/2)$ sampled at integer multiples of $\tilde T$.
As a consequence, the logical $|0\rangle$ and $|1\rangle$ states, whose $\phi$ support at integer multiples of $\tilde T$ is confined near $ 2\pi [\mathbb Z+1/2]$ and $ 2\pi [\mathbb Z-1/2]$, respectively, result in oppositely-valued supercurrents [see Fig.~\ref{fig:readout}(b)]. These can be detected e.g., from the frequency shift of an adjacent transmon due to the induced magnetic field~\cite{danilin_2018,budoyo_2020}.
The readout process can also be applied for initialization. Further details are contained in App.~\ref{app:readout}.

\section{Discussion}
\label{sec:discussion}

In this section, we discuss aspects of experimental implementation, influence of possible corrections beyond our theoretical analysis, scalability concerns, and related work pursuing experimental realization of similar approaches.

\subsection{Experimental implementation}
We now discuss how the key components of the qubit may be implemented experimentally.

{\it Material platform}---%
The superconducting gap of the resonator material, $\Delta$, sets an upper limit for all energy and frequency scales of the qubit, due to the need for avoiding quasiparticle generation---see Fig.~\ref{fig:hierarchy}. We thus expect implementation to be simplest in materials with large superconducting gaps, such as NbN ($\Delta = 3 \, {\rm meV}$) or Pb~\cite{Ruby_2015} ($\Delta =1.5 \,{\rm meV}$). In particular, we expect our simulated parameters in Table~\ref{tab:parameters}, to be compatible with NbN. This configuration involved $\sim 5{\rm GHz}$ resonators, $\sim 150\,{\rm mK}$ temperatures, and voltage amplitudes of $V_{\rm max}\sim 1.5\,{\rm mV}$.
For materials with smaller gaps, such as Al $(\Delta \sim 0.2\,{\rm meV})$~\cite{BaronePaterno1982} the operating frequency, voltage, and temperature must be rescaled correspondingly, and we expect implementation would require $\sim 0.5\,{\rm GHz}$ resonators and $\sim 10\,{\rm mK}$ temperatures.

{\it High-impedance resonator---}Our qubit requires a superconducting resonator with impedance above $h/2e^2\approx 12.9\,{\rm k}\Omega$, corresponding to $\fz\geq2$. As we demonstrated in Sec.~\ref{sec:extensions_to_more_general}, the impedance does not have to closely match $h/2e^2$, but just exceed it (see Fig.~\ref{fig:oblique}). Our simulations indicate that $Q$ factors above $10^5$ are sufficient to observe stabilization of GKP codespace and DQEC.
While the realization of such a high-impedance resonator is challenging, it is within reach of current technology.
For instance, superinductors with impedance of $h/4e^2$ have been demonstrated with NbN~\cite{NiepcePhysRevAppl2019Apr} and quality factors above $10^5$ were achieved \cite{Frasca2023Oct}. Moreover, Al-based hyperinductors with impedances as large as $\sim 30\,{\rm k\Omega}$ have been achieved using air-gapped Josephson junction arrays, kinetic inductors, or geometric inductors~\cite{peruzzo_surpassing_2020,manset_hyperinductance_2025,PechenezhskiyNature2020Sep}.

{\it Josephson junction---}Our qubit requires the high-impedance resonator to be interfaced with a Josephson junction that will be subjected to a large voltage bias.
NbN-based junctions have been demonstrated, e.g., in Refs.~\cite{kammerer2024resonantescapejosephsontunnel,anferov2025millimeterwavesuperconductingqubit,Grimm_2017}.
For these demonstrations, the high-impedance requirement is compatible with the parasitic capacitance of the Josephson junction.
For instance, angular plasma frequencies of $2\pi \times 150{\rm GHz}$ were reported in Refs.~\cite{kammerer2024resonantescapejosephsontunnel,anferov2025millimeterwavesuperconductingqubit}, corresponding to parasitic capacitances of $C_J/E_J\lesssim (0.01 -0.001 ){\rm fF}/h{\rm GHz}$.
For the parameters we considered in our simulations (Table~\ref{tab:parameters}), this would imply a Josephson capacitance of $0.2-0.02 \,{\rm fF}$, and thus only a minor contribution to the total resonator capacitance of $2.5\,{\rm fF}$ (see caption of Table~\ref{tab:parameters}).

{\it Microwave drive---}We expect the drive tones can be achieved with standard microwave control hardware. For instance, the pulse in Fig.~\ref{fig:pulse} that we use in our simulations can be effectively realized with a few microwave tones within the $10-30\,{\rm GHz}$ range. There is significant flexibility in the design of the tone: as we showed in Sec.~\ref{sec:extensions_to_more_general}, the qubit operates as long as the nodes of the voltage tone have a proper spacing, and its amplitude satisfies the hierarchy of Fig.~\ref{fig:hierarchy}.
For instance, the tone can be a single sinusoidal wave, as we demonstrated in Appendix~\ref{app:hex}.
Thus, the tone does not need to be carefully calibrated, and can even tolerate a finite amount of noise, as we show in Appendix~\ref{app:voltage_noise}.

{\it Filtered environment}---The filtered environment can be realized with arrays of dissipative resonators or other filter technologies (see Appendix~\ref{app:chebyshev}). Such filtered environments are, e.g., employed for cat qubits~\cite{krantz_quantum_2019,ferreira_collapse_2020}.

\subsection{Possible limiting factors}
\label{sec:challenges}
We now discuss the possible limiting factors for the qubit. Some of these are accounted for in this work, while others require further investigation.

{\it Finite frequency corrections}---A key limiting factor we foresee
is the finite-frequency corrections to the effective circuit Hamiltonian that we discussed in Sec.~\ref{sec:finite_freq}.
As we discussed there,
finite-frequency corrections
can limit the qubit lifetime by introducing weak terms in the effective Hamiltonian that act nonlocally in phase space and thus directly couple logical sectors. Another way they can limit performance is through rare resonances that kick the resonator out of the codespace via drive-induced state transitions (DUST)~\cite{dumas_measurement-induced_2024}.
Importantly, the finite-frequency corrections above {are} accounted for in our simulations in Sec.~\ref{sec:numerical}.
Based on these simulations, and our analysis in Sec.~\ref{sec:device}, we expect finite-frequency corrections may have an adverse, but small, effect on performance that still allows for significant improvement of coherence time through DQEC.

{\it Parasitic modes---}Another key challenge, {\it not} included in our analysis, is parasitic modes. The oscillating voltage bias through the Josephson junction generates many harmonics, as we see in Fig.~\ref{fig:pulse}(b). These may couple the resonator into additional long-lived {parasitic} modes in the device, which constitute an environment for the qubit. The qubit is technically designed to correct itself against noise from this type of environment. However, for sufficiently strong couplings, the parasitic modes may have a detrimental effect. Understanding the impact of parasitic modes for our proposed device hence remains a key open problem.

{\it Quasiparticle poisoning}---While the qubit is protected against phase-space local noise, it is {\it not} protected against phase-space {non-local} noise, such as quasiparticle tunneling, which shifts $\hat n$ by a discrete value $1/2$. We thus expect quasiparticle poisoning to fundamentally limit lifetime.
Drive-induced quasiparticle generation can be avoided by having the dynamics of the voltage tone to be adiabatic with respect to the gap for quasiparticle generation, as we achieved with the capped voltage tone used in our simulations [see Fig.~\ref{fig:pulse}].
We still expect quasiparticles from other sources to remain a potential bottleneck for the coherence time. The rates of such poisoning events may in general be well controlled, however: for instance poisoning times up to $\sim 20\,{\rm s}$ were recently reported with Pb based devices~\cite{aghaee_interferometric_2024,aghaee202620secondparitylifetime}.
Hence the adverse effects of quasiparticle poisoning could be efficiently mitigated via concatenation in standard quantum error correction codes (see below).

\subsection{Scalability considerations}
There are several aspects of our approach that may be beneficial for scalability---here we outline the main beneficial features we anticipate.

{\it Tolerance for noise and dissipation---}
A key advantage of the qubit is its intrinsic tolerance for noise and dissipation. In particular, the dissipative quantum error correction of the qubit can remove entropy induced by phase space local noise or dissipation without affecting the encoded information. This allows it to operate in the presence of finite noise and dissipation without affecting its lifetime. We have demonstrated the qubit's resilience towards photon loss in our simulations, and also analytically demonstrated resilience towards noise in the driving voltage tone and low-frequency noise (Appendices~\ref{app:low_frequency_noise}~and~\ref{app:voltage_noise}).

{\it Indistinguishability---}A second benefit of our approach is that qubits with our approach are {\it indistinguishable}, meaning they can operate without individual calibration.
For instance, two qubits fabricated with sufficiently small, but finite, deviations of their parameters will lock into the same GKP-stabilization dynamics in response to a shared drive signal. Their individual deviations from the target parameters can be viewed as ultra-low-frequency phase-space local noise; hence the effects by the imperfections of the qubits will be erased by dissipative error correction. This feature could significantly simplify large-scale fabrication and operation.

{\it Tolerance for parameter drift---}A consequence of the feature above is that the qubit remains locked into the regime of GKP stabilization, even if the parameters of the circuit drift over time. Hence our qubit is insensitive to moderate parameter drift, and does not need recalibration. We infer conditions on the parameter drift in our analysis of low-frequency noise in Appendix~\ref{app:low_frequency_noise}.

{\it Large-scale controllability---}Another benefit, also related to the above, is that the qubit is also insensitive to small, but finite, deviations of the driving voltage tone, see Appendix~\ref{app:voltage_noise}. In this sense, our qubits are {\it digitally controllable}. This has the potential to significantly simplify large-scale parallel control. The flexibility in the driving tones also provides a cheap mechanism for implementing gates on the qubits: slowing down one qubit relative to another to eventually lag behind by a quarter cycle effectively achieves the $e^{-i\frac{\pi}{4}\hat \sigma_x}$ gate on one of the qubits (see Sec.~\ref{sec:gates}). Even in situations where the operating frequency windows of two qubits do not overlap, synchronous operations can still be achieved by using the frequency flexibility to make the driving frequencies commensurate.

{\it Low-overhead error correction---}Our proposed approach harnesses dissipation for quantum error correction against phase-space local noise, and thus removes a broad class of errors that are otherwise present in superconducting qubits~\cite{krantz_quantum_2019} without any readout or control overhead. While the lifetime of the qubit may still be fundamentally limited by residual noise sources e.g., from finite-frequency corrections or quasiparticles, reducing the errors to these noise sources may still provide a significant reduction in the total error rate. These residual noise sources can be corrected via concatenation in conventional quantum error correcting codes, a task which may be simplified through the reduced physical qubit error rate.

\subsection{Relationship with other work}

Recently, several works have proposed~\cite{conrad_2021_GKP,KolesnikowarXiv2023Mar,Campagne-Ibarcq2020Aug,Nathan2024May} or demonstrated~\cite{Campagne-Ibarcq2020Aug,fluhmann_encoding_2019,deNeeveNatPhys2022Mar,deneeve2024modularvariablelasercooling,LachanceQuirion2024Apr,larsen_integrated_2025} new approaches for preparing and stabilizing GKP states in circuit-QED, photonics, or trapped-ion platforms.
Besides demonstrated applications for long-coherence qubits \cite{Campagne-Ibarcq2020Aug,Eickbusch2022Dec,SivakNature2023Apr,LachanceQuirion2024Apr}, GKP states have been proposed as displacement sensors \cite{Duivenvoorden2017Jan}, long-distance quantum communication \cite{Fukui2021Aug} and as quantum repeaters \cite{Schmidt2024Apr,AzariICC2024}.
They moreover have multiple uses in quantum information processing, including concatenation with other codes for improved error correction \cite{Fukui2017Nov,GrimsmoPRXQuantum2021Jun,Noh2022Jan}, and interfaces with quantum memories \cite{Dhara2024Jun}.
GKP qubits support natural implementations of Clifford gates \cite{GottesmanPhysRevA2001Jun}, and approaches for magic state generation, and protected non-Clifford gates\cite{Konno2021Oct,SellemarXiv2023Apr,Shaw2024Mar,Matsos2024Sep,obrien_exponentially_2025,nguyen_fault-tolerant_2025,Boudreault_2026}.
The approach we present in this paper may provide a relatively simple means for {dissipatively} generating these exotic and valuable states without need for readout or active feedback control for stabilization.

GKP qubits have been realized in multiple experiments in recent years~\cite{Campagne-Ibarcq2020Aug,SivakNature2023Apr,Brock2024Sep,LachanceQuirion2024Apr}. These experiments involve steering a high-Q resonator mode into the GKP codespace through readout and feedback control of a coupled ancilla superconducting qubit such as a transmon or fluxonium~\cite{ali2026bosonicquantumcontrolweakly,jha2026bosonicerrorcorrectionfluxonium}. Hence, these experiments stabilize GKP qubits via active quantum error correction.
Our approach offers a fundamentally different approach from these experiments by using DQEC, which entirely circumvents the need for readout and feedback control. Furthermore, our approach avoids the need for an ancilla control qubit, that otherwise provides a bottleneck for performance~\cite{Campagne-Ibarcq2020Aug}. Hence, by using DQEC, our approach has the potential to remove key challenges for performance and scalability.

Other proposals for realizing DQEC with GKP qubits have also emerged recently~\cite{Nathan2024May,SellemarXiv2023Apr}:
Ref.~\cite{Nathan2024May} proposes driving a high-impedance resonator with a switchable Josephson junction, to effectively implement sequential activation of GKP stabilizers; Ref.~\cite{SellemarXiv2023Apr} proposes to stabilize the GKP codespace by engineered dissipation using high-bandwidth pulse combs.
While sharing similarities with the present work, our approach provides a path to realization of DQEC which is within experimental reach. Specifically,
the proposals above assume exotic components with no clear path to implementation: a rapid rise-time, high-contrast switchable Josephson junction driven by a wide-bandwidth pulse~\cite{Nathan2024May}, or a wide-bandwidth, high-fidelity pulse comb~\cite{SellemarXiv2023Apr}.
In contrast, our approach
only requires demonstrated (if challenging) components: a high-impedance resonator coupled to a Josephson junction biased by a single, or few, microwave tones. In particular, the pulse in Fig.~\ref{fig:pulse} that we use in our simulations, while superficially similar to the step-like drive from Ref.~\cite{Nathan2024May}, is used to voltage-bias a Josephson junction (a standard feat) rather than to modulate the Josephson coupling through a yet-to-be-defined mechanism, as in Ref.~\cite{Nathan2024May}. Moreover, it can be effectively realized with a few microwave tones within the $10-30\,{\rm GHz}$ range, rather than a sharp pulse with $\sim 100 \,{\rm GHz}$ bandwidths, as required for Ref.~\cite{Nathan2024May}.
Beyond this key difference, our approach directly realizes the GKP stabilizer Hamiltonian, via Floquet engineering, and provides a simple lever of qubit control via modulation of the phase, amplitude, and frequency of the drives.

\section{Conclusion}
In this work, we present a blueprint for a qubit
where quantum error correction of both phase and bit flip errors occurs autonomously through dissipation. The working principle is based on the
Gottesman-Kitaev-Preskill
encoded qubit
which we show can be
realized
in a high-impedance LC resonator through resonant, high-voltage driving through a Josephson junction. In this realization, error weight is proportional to energy; thus the qubit features a restoring force that counteracts bit and phase flips. As a result, simple cooling leads to quantum error correction by relaxing the qubit towards the codespace without corrupting the encoded information. This principle removes the need for feedback control otherwise required in active error correction schemes.

We further demonstrate that the device is most suitably operated with a side-band cooling scheme, raising the operable temperature by an order of magnitude. Beyond error-correction in the steady state, the device exhibits single-qubit Clifford gates with exponential accuracy (deviations are corrected by the dissipative error correction) and magic state preparation above the distillation threshold.

A particular strength of our approach may lie at the interface with other quantum computing platforms. The largely enhanced lifetime by dissipative quantum error correction at minimal hardware overhead makes the device a promising candidate to store quantum information. Moreover, its ability to prepare magic states can supply high-fidelity magic states. In this way, the device may complement fast quantum computing platforms by providing memory and magic, paving the way for algorithms combining fast but short-coherence qubits with long-term quantum memory.
Establishing these interfaces and information pipelines is an interesting direction for future work.

{\it Data Availability. ---} All data and numerical code from which the data has been produced will be available in a Zenodo repository.

{\it Acknowledgements.---}
We gratefully acknowledge insightful discussions and collaboration on related projects with Gil Refael, Liam O'Brien, and Matthew H. Matheny.
We also acknowledge useful discussions with Karsten Flensberg, Anders S\o{}rensen, Jacob Hastrup, Margarita Davydova, Liang Fu, Virgil Baran, Jonathan Conrad, Max Hays, Lev-Arcady Sellem, and Philippe Campagne-Ibarcq.
We thank Thiziri Aissaoui for feedback on the manuscript.
This work made use of resources provided by subMIT at MIT Physics.
M.G. acknowledges support from the German Research
Foundation under the Walter Benjamin program (Grant
Agreement No. 526129603) and by the Air Force Office of Scientific Research under award number FA2386-24-1-4043.
F.N. gratefully acknowledges support from the Carlsberg Foundation, grant CF22-0727.

\bibliography{refs_GKP}

\appendix
\section{Dissipative quantum error correction in the lattice picture}
\label{app:diagonalization}
Here we show that the lattice basis states from Eq.~\eqref{eq:lattice_basis_wf} are eigenstates of $\HGKP$ up to $\mathcal O(\lambda^2\varepsilon_0)$ corrections (Sec.~\ref{seca:diag}), and demonstrate that their exact eigenvalues are identical up to $\mathcal O(e^{-1/\lambda^2})$ corrections. We also provide additional details to the arguments for the protection of the encoded information (Sec.~\ref{seca:prot}), and the emergence of dissipative quantum error correction (Sec.~\ref{seca:dec}).

\subsection{Eigenstates of $\HGKP$}
\label{seca:diag}
To show that the lattice basis states are approximate eigenstates of $\HGKP$, we decompose the state of the system in the lattice state basis via the tensor $\boldsymbol \psi$, with indices $\{\boldsymbol \psi_{ab}^s\}$, such that \begin{equation}
    |\boldsymbol {\psi}\rangle = \sum_{a,b,s}\boldsymbol{\psi}_{ab}^s|\psi_{ab}^s\rangle\end{equation}
The tensor $\boldsymbol \psi_{ab}^s$ can be interpreted as the wavefunction of a particle in the codespace lattice.
Due to the overcompleteness of the basis, the parameterization is not unique.

We now set out to find the evolution of the coefficient tensor $\boldsymbol \psi$ under $\hat H_{\rm GKP}$.
To this end, we note that
\begin{align}
    \hat n|\boldsymbol \psi\rangle = |\boldsymbol n\cdot \boldsymbol \psi\rangle , \qquad \hat \phi |\boldsymbol \psi\rangle = |\boldsymbol \phi\cdot \boldsymbol \psi\rangle
\end{align}
 where $\cdot$ denotes usual tensor contraction, while
$\boldsymbol \phi =\frac{4\pi}{\lambda}\boldsymbol p_y$, $\boldsymbol n=\frac{1}{\lambda}\boldsymbol p_x+\frac{\lambda}{4\pi}\boldsymbol x,
$
with $\boldsymbol x=\frac{1}{\sqrt{2}}(\boldsymbol b_1+\boldsymbol b_1^\dagger)$, $\boldsymbol p_x=\frac{1}{\sqrt{2}i}(\boldsymbol b_1-\boldsymbol b_1^\dagger)$, with $\boldsymbol y$, $\boldsymbol p_y$ defined analogously from $\boldsymbol{b}_2$, and where $(\boldsymbol b_1)^{a'b'}_{ab}=\sqrt{a} \,\delta_{bb'}\delta_{a,a'-1}$ and $(\boldsymbol b_2)^{a'b'}_{ab}=\sqrt{b} \,\delta_{aa'}\delta_{b,b'-1}$ are the matrix representation of two independent bosonic annihilation operators in a 2-mode system.
Here, and henceforth, we shall use bold symbols to indicate matrices and vectors that parameterize operators and states in the codespace lattice basis.
Note that the representation above leaves the index $s$ formally invariant.
More generally, the action of any function of $\hat \phi$ and $\hat n$ $f(\hat \phi,\hat n)$ can be obtained as $f(\hat \phi,\hat n)|\boldsymbol \psi\rangle = |f(\boldsymbol \phi,\boldsymbol n)\cdot \boldsymbol \psi\rangle$.
Note, however, that the overcompleteness of the lattice basis gives us a freedom in representing the action of an operator on these states.
Relevant for us, we also note that $\cos(4\pi \hat n)|\boldsymbol{\psi}\rangle =|\cos(\lambda \boldsymbol y)\cdot \boldsymbol \psi \rangle $, and likewise $\cos(\hat \phi )|\boldsymbol{\psi}\rangle =|\cos(\lambda \boldsymbol x)\cdot \boldsymbol \psi \rangle $~\footnote{This follows by noting that the wavefunction of $\langle \phi|e^{4\pi i \hat n}|\psi_{ab}^s\rangle= \sum_{z=-\infty}^\infty \chi_a(\frac{\phi-\phi_{zs}}{\lambda})\chi_b(\frac{\lambda \phi}{4\pi}+\lambda) (-i)^{a-b} $, and that $\chi_b(\frac{\lambda \phi}{4\pi}+\lambda)=\sum_{b'}[e^{-i\lambda P}]_{bb'} \chi_{b'}(\frac{\lambda \phi}{4\pi}) =\sum_{b'}[e^{-i\lambda X}]_{bb'} (-i)^{b-b'}\chi_{b'}(\frac{\lambda \phi}{4\pi}) $ with $P$ and $X$ the matrix representations of $x$ and $p$ in the Harmonic state basis.}.%
Using the above substitutions, along with $\lambda \equiv ({4\pi \hbar \delta \omega}/{\tilde E_J})^{1/4}$ and $\varepsilon_0=\lambda^2 \tilde E_J$, we find that the Schr\"odinger equation is solved by the state $|\boldsymbol \psi(t)\rangle$ if $\boldsymbol \psi(t)$ satisfies the differential equation
\begin{equation}
    \partial_t \boldsymbol{\psi}(t) = -\frac{i}{\hbar}\boldsymbol H_{\rm GKP}\cdot \boldsymbol \psi(t), \label{eqa:eom}\end{equation}
where
\begin{equation}
    \boldsymbol H_{\rm GKP} =\frac{\varepsilon_0}{2}\Bigg[\boldsymbol p^2_x+ \Big(\boldsymbol p_y-\frac{\lambda^2\boldsymbol x}{4\pi}\Big)^2\Bigg]+\tilde E_J\big[\cos(\lambda\boldsymbol x)+\cos(\lambda \boldsymbol y)\big].
\end{equation}
Hence, the lattice basis representation of the Hamiltonian, $\boldsymbol H_{ \rm GKP}$ is quasilocal, and assigns an energy penalty to lattice sites according to their ``syndrome weight'' $w=a+b$. Specifically, by Taylor expanding in powers of $\lambda$, and using $\varepsilon_0=\lambda^2 \tilde E_J$, it is straightforward to see that
\begin{equation}
    \boldsymbol H_{\rm GKP} = \varepsilon_0(1+\boldsymbol b^\dagger_1\boldsymbol b_1+\boldsymbol b^\dagger_2\boldsymbol b_2)+\mathcal O(\lambda^2\varepsilon_0)
\end{equation}
Thus, to leading order in $\lambda$, $ \boldsymbol H_{\rm GKP} $ is diagonal in the lattice site basis, and assigns an energy penalty to sites according to their syndrome weight. From Eq.~\eqref{eqa:eom} we see that the spectrum of $\boldsymbol{H}_{\rm GKP}$ must coincide with that of $\hat H_{\rm GKP}$; thereby, we conclude that the lattice basis state $|\psi_{ab}^s\rangle$ indeed is an eigenstate of $\hat H_{\rm GKP}$ with eigenvalue $(a+b+1)\varepsilon_0$ up to $\mathcal O(\lambda^
2\varepsilon_0)$ corrections.

Note that the corrections to $\boldsymbol H_{\rm GKP}$ are low-order polynomials in $\boldsymbol x$ and $\boldsymbol y$ of order $\lambda^2 \varepsilon_0 \boldsymbol x^4$, $\lambda^2 \varepsilon_0 \boldsymbol y^4$; hence the corrections to $|\psi_{ab}^s\rangle$ will have support for nearby sites ($a',b'$) on the codespace lattice, with the support decaying exponentially with the distance from site ($a,b$). In particular, the corrections will remain in the codespace up to $\mathcal O(e^
{-\wmax})$ corrections, if $a,b\ll \pi^2/\lambda^2$. By the same arguments, the corrections to the eigenvalues of $\boldsymbol H_{\rm GKP}$ corresponding to $|\psi_{ab}^0\rangle$ and $|\psi_{ab}^1\rangle$ will be identical up to similar $\mathcal O(e^{-1/\lambda^2})$ corrections. Thus, the exact eigenvalues of $\HGKP$ that correspond to the approximate eigenstates $|\psi_{ab}^0\rangle$ and $|\psi_{ab}^1\rangle$ will be identical up to $\mathcal O(e^{-1/\lambda^2})$ corrections.

\subsection{Protection of encoded information}
\label{seca:prot}
The protection of GKP-encoded information can be understood using the lattice picture above.
First recall that the evolution of $\boldsymbol \psi(t)$ generated by Eq.~\eqref{eqa:eom} formally leaves the $s$ index invariant.
Thus, for $i=1,\ldots 3$, \begin{equation}
    {{\tau}}_i(t) \equiv \boldsymbol \psi^*(t)_{ab}^s \sigma^i_{ss'}\boldsymbol \psi_{ab}^{s'}(t)
\end{equation}
is a formal constant of motion~\footnote{This quantity associated with $\boldsymbol \psi$, which technically does not directly correspond to any physical observable, is formally invariant under the evolution generated by $\boldsymbol H$.}.
Importantly, due to the near-orthonormality of the lattice basis states in the codespace [Eq.~\eqref{eq:near_orth}], and their transformation under the logical operators [Eq.~\eqref{eq:yl}], we have
\begin{equation}
    \langle \boldsymbol \psi(t)|\hat \sigma_i|\boldsymbol \psi(t)\rangle ={{\tau}}_i(t) +\mathcal O(e^{-\wmax})
\end{equation}
when $\boldsymbol{\psi}$ has its support confined to $a,b\ll \wmax$.
Thus, by the formal invariance of $ \tau_i(t)$, the encoded information is near invariant if the lattice wavefunctions $\boldsymbol \psi(t)$ remains confined for $a,b\ll \wmax$:
\begin{equation}
    \langle \boldsymbol \psi(t)|\hat \sigma_i|\boldsymbol \psi (t)\rangle =\langle \boldsymbol \psi(0)|\hat \sigma_i|\boldsymbol \psi(0)\rangle+\mathcal O(e^{-\wmax})
\end{equation}

Important for the following, the argument above can be generalized to density matrices: we write the system's density matrix $\hat \rho_{\rm S}(t)=[\boldsymbol \rho_{\rm S}(t)] _{ab;a'b'}^{ss'}(t)|\psi_{ab}^s\rangle \langle \psi_{a'b'}^{s'}|$, with implicit summation over repeated indices used here and throughout.
Provided the support of $\boldsymbol \rho_{\rm S}$ is confined in the codespace $a,b\ll \wmax
$, the considerations above show that the logical state evolution, $\langle \hat \sigma_i(t)\rangle=\Tr[\hat \rho_{\rm S}(t)\hat \sigma_i]$ satisfies
\begin{equation}
     \langle \hat \sigma_i(t)\rangle =\langle \hat \sigma_i(0)\rangle+\mathcal O(e^{-\wmax}),
     \label{eqa:invariance_of_qi}
\end{equation}
analogously to the pure-state case.
\subsection{Dissipative quantum error correction}
\label{seca:dec}

To show that dissipation will stabilize the system in codespace, we consider coupling the system to an environment in thermal equilibrium at some temperature $T_{\rm B}=1/\beta k_{\rm B}\ll \wmax\varepsilon_0$, with $k_{\rm B}$ the Boltzmann constant. We write the coupling term as $\hat H_{\rm int} = \sqrt{\gamma} S(\hat \phi,\hat n)\hat B$, with $\hat B$ an environment observable and $S(\phi,n)$ some function defining the system-observable coupled to the environment. To demonstrate stability, we also consider coupling some noise term to the system, represented by $\hat H_{\rm noise} = \xi(\hat n,\hat \phi) \hat C$, where $\hat C$ is either a time-dependent scalar noise signal or observable of the environment generating the noise and $\hat \xi$ represents the system observable coupled to the noise channel.
With the dissipative environment and noise source included, the Hamiltonian of the combined systems will thus read
$\hat H=\hat H_{\rm GKP}+\xi(\hat \phi,\hat n,t)\hat C+S(\hat \phi,\hat n)\hat B.
$

We now write the density matrix of the combined system-environment configuration as $\hat \rho(t)=\boldsymbol \rho _{abi;a'b'i'}^{ss'}(t)|\psi_{ab}^s\rangle \langle \psi_{a'b'}^{s'}|\otimes |b_i\rangle\langle b_j|$, and $\{|b_i\rangle\}$ an orthonormal basis for the combination of the two environments.
By the considerations of Sec.~\ref{seca:diag}, the equation of motion for the density matrix is solved if $\partial_t\boldsymbol \rho(t)=-\frac{i}{\hbar} [ \boldsymbol H_{\rm SB },\boldsymbol \rho(t)]$, where
\begin{equation}
  \boldsymbol H_{\rm SB }\equiv \boldsymbol H_{\rm GKP}+ \xi(\boldsymbol \phi,\boldsymbol n,t)\boldsymbol C +S(\boldsymbol \phi,\boldsymbol n)\boldsymbol B.
    \label{eqa:rhomat_eom}
\end{equation}
with $\boldsymbol B$ and $\boldsymbol C$ denoting the matrix representations of the operators $\hat B$ and $\hat C$ in the basis $\{|b_i\rangle\}$.
Now, Eq.~\eqref{eqa:invariance_of_qi} implies that the encoded information is invariant in the system if the support of the resonator's reduced density matrix $\boldsymbol \rho_S(t)$ remains confined for lattice sites $a,b\ll \wmax$.
We note that $\boldsymbol \rho_{\rm S}(t)$ describes the reduced density matrix of a particle on the codespace lattice coupled to a dissipative environment with composite-system Hamiltonian given by $\boldsymbol{H}_{\rm SB}$ and system Hamiltonian $\boldsymbol H_{\rm GKP}$.
The latter is dominated by an on-site potential energy term $\varepsilon_0 (\boldsymbol b_1^\dagger \boldsymbol b_1+ \boldsymbol b_2^\dagger \boldsymbol b_2)$, i.e., an on-site potential that assigns an energy penalty $(a+b)\varepsilon_0$ to site $(a,b)$.
In addition, the particle is subject to noise and dissipation through strictly local operators via $\xi(\boldsymbol \phi,\boldsymbol n)$ and $S(\boldsymbol \phi,\boldsymbol n)$.
The system therefore thermalizes to a state with $\boldsymbol \rho_{\rm S} \approx \frac{1}{Z}e^{-\beta_{\rm eff}\varepsilon_0 (\boldsymbol b^\dagger _1 \boldsymbol b_1 + \boldsymbol b^\dagger_2 \boldsymbol b_2)}$ with $Z$ a normalization factor~\cite{BreuerPetruccione}, with exponentially small weight at $a,b\gtrsim \wmax$.
Here $ \beta_{\rm eff}$ denotes the effective inverse temperature of the system determined by the balance of energy influx from the noise, and energy outflux via the dissipative environment. If the coupling to the bath dominates, this temperature is approximately equal to the bath inverse temperature $\beta$. Through canonical perturbation theory, this feature can be established to hold even when accounting for corrections to the Hamiltonian away from the above harmonic approximation and finite system-bath coupling~\cite{thingna_generalized_2012}.
As a result, the leakage rate out of the codespace will follow an Arrhenius law, scaling as $\lambda^2 \varepsilon_0 e^{- \beta_{\rm eff} \varepsilon/\lambda^2}\sim \lambda^2 \varepsilon_0 e^{-\beta_{\rm eff}\tilde E_J}$~\cite{ankerhold_dissipative_1995}.
Thus the lifetime of the qubit remains exponentially large even in the presence of noise, and scales as
\begin{equation}
t_{\rm coherence} \sim \frac{\hbar}{\varepsilon_0} e^{\beta _{\rm eff}\tilde E_J}
\end{equation}
provided the effective inverse temperature $\beta_{\rm eff}$ remains sufficiently large.
Hence the encoded quantum information is dissipatively stabilized against phase-space local noise by coupling low-order polynomials of $\hat \phi$ and $\hat n$ to a thermodynamic environment at temperature $T\ll \tilde E_J/k_{\rm B}$.
\section{Properties of the lattice basis states}
\label{app:lattice_basis}
Here we establish the properties of the lattice basis states that we quote in Sec.~\ref{sec:dec} of the main text:
In Sec.~\ref{seca:charge_wf}, we obtain the wavefunction of $|\psi_{ab}^s\rangle$ as a function of charge, $\hat n$ $\tilde \psi_{ab}^s(n)$; In Sec.~\ref{app:stabilizer_action} we demonstrate that Eqs.~\eqref{eq:s1action}-\eqref{eq:s2action} hold, and thus show that the lattice basis states reside in codespace up to $\mathcal O(e^{-\wmax})$ corrections.
In Sec.~\ref{seca:orth}, we use this to determine the overlap between lattice basis states.
 In Sec.~\ref{seca:logicals}, we obtain the action of the logical operators of the qubit.

\subsection{Wavefunction as a function of charge}
\label{seca:charge_wf}
We obtain $\tilde \psi_{ab}^s(n)$ by Fourier transforming the wavefunction of $\hat \phi$, $\psi_{ab}^s(\phi)$ [defined in Eq.~\eqref{eq:lattice_basis_wf} of the main text]:
\begin{equation}
    \tilde \psi_{ab}^s(n)=\mathcal F[\psi_{ab}^s](n),
    \label{eqa:ft}
\end{equation} where
$
  \mathcal F[f](x)=\frac{1}{\sqrt{2\pi}}\int dy e^{-iyx}f(y)
$
denotes the Fourier transform.
For convenience, we recall that
\begin{equation}
\psi_{ab}^s(\phi)= i^b \sum_{k=-\infty}^\infty \chi_{a}\left(\frac{\phi-\phi_{ks}}{\lambda}\right)\chi_b\left(\frac{\lambda\phi}{4\pi}\right)(-1)^{k}.
\label{eqa:lbwf}
\end{equation}
with $\phi_{ks}\equiv 4\pi k + 2\pi s +\pi$.
In the following, it is convenient to express $\psi_{ab}^s(\phi)$ in terms of the Dirac-comb type function $\mathcal D_s(\phi)=\sum_k\delta(\phi-\phi_{ks})(-1)^k$, with $\delta(x)$ the Dirac delta function. Specifically:
\begin{equation}
\psi_{ab}^s(\phi)= {i^b}\left[\mathcal D_s(\phi)*\chi_a\left(\frac{\phi}{\lambda}\right)\right]\chi_b\left(\frac{\lambda \phi}{4\pi}\right),
\label{eqa:psi_in_terms_of_dirac_comb}
\end{equation}
where $*$ denotes convolution: $f(x)*g(x)=\int_{-\infty}^\infty dy f(x-y)g(y)$.

To evaluate Eq.~\eqref{eqa:ft}, we use the following properties of the Fourier transform: \begin{align} \mathcal F[g](x)&=\frac{1}{a}\mathcal F[f](x/a)\quad {\rm for}\quad g(x)=f(ax)\\ \mathcal F[f*g](x)&=\sqrt{2\pi}\mathcal F[f](x)\mathcal F[g](x)\\ \mathcal F[fg](x)&=\frac{1}{\sqrt{2\pi}}(\mathcal F[f]*\mathcal F[g])(x), \\ \mathcal F[\chi_a](x)&=(-i)^a \chi_a(x).
\end{align}
Using these properties in Eq.~\eqref{eqa:psi_in_terms_of_dirac_comb}, we find
\begin{equation}
\tilde \psi_{ab}^s(n)= {4\pi(-i)^{a}} \Big[\tilde{\mathcal {D}}_s(n)\chi_a(\lambda n)\Big]*\chi_b\left(\frac{4\pi n}{\lambda}\right)
\label{eqa:tildepsi_expr}\end{equation}
where $\tilde {\mathcal D}_s(n)=\mathcal F[\mathcal D_s](n)$.
Explicitly taking this Fourier transform, and using the definition $\phi_{ks}=4\pi k+ 2\pi s +\pi$ along with $-1=e^{i\pi}$, we find \begin{equation}
    \tilde {\mathcal D}_s(n)=\frac{1}{\sqrt{2\pi}}\sum_ke^{-i[4\pi k +2\pi s +\pi] n-i\pi k}.
\end{equation}
Using Poisson's summation formula, we find
\begin{equation}
    \tilde{\mathcal D}_s(n)=\frac{1}{2\sqrt{2\pi}}\sum_\ell\delta(n-\ell/2-1/4)e^{-2\pi i (s+1/2)n} .
\end{equation}
Inserting this into Eq.~\eqref{eqa:tildepsi_expr}, and using $e^{-2\pi i (s+1/2)(\ell/2+1/4)}=(-i)^{2s\ell +\ell+s}e^{-i\pi/4}$,
we find
\begin{widetext}
\begin{equation}
\tilde \psi_{ab}^s(n)= {\sqrt{2\pi}e^{-i\frac{\pi}{4}(2a+1)}}\sum_{\ell=-\infty}^\infty (-i)^{2s\ell+\ell+s}\chi_a\big(\lambda[\ell/2+1/4]\big) \chi_b\left(\frac{4\pi[n-\ell/2-1/4]}{\lambda}\right).
\label{eqa:tildepsi_expr_1}\end{equation}

For $\lambda\ll 1$ and $a,b\ll \wmax\equiv \pi^2/\lambda^2$, $\chi_b\left(\frac{4\pi}{\lambda}(y-x)\right)$ is only nonzero in the $x$-interval $|y-x|\lesssim \lambda $, while $\chi_a(\lambda x)\approx \chi_a(\lambda y)$ in this interval. Hence $\chi_a(\lambda x ) \chi_b\left(\frac{4\pi}{\lambda}(y-x)\right)\approx \chi_a(\lambda y ) \chi_b\left(\frac{4\pi}{\lambda}(y-x)\right)$. Making this approximation above with $x=\ell/2 +1/4$ and $y=n$, we find
\begin{equation}
\tilde \psi_{ab}^s(n)\approx {\sqrt{2\pi}}e^{-i\frac{\pi}{4}(2a+1)}\sum_{\ell=-\infty}^\infty(-i)^{2s\ell+\ell+s}\chi_a\left(\lambda n\right) \chi_b\left(\frac{4\pi [n-\ell/2-1/4]}{\lambda}\right)
\end{equation}
Next, we divide the sum over $\ell$ into even and odd terms, by writing $\ell=2k +s'$:
\begin{equation}
\tilde \psi_{ab}^s(n)\approx e^{-i\frac{\pi}{4}(2a+1)}{\sqrt{2\pi}} \sum_{s'=0}^1 (-i)^{2ss'+s'+s}\sum_{k=-\infty}^\infty(-1)^k\chi_a\left(\lambda n\right) \chi_b\left(\frac{4\pi [n-k - s'/2-1/4]}{\lambda}\right)
\end{equation}
where we used $(-i)^{2s\ell+\ell+s}=(-i)^{2ss'+s'+s}(-1)^k$ under this substitution.
We now recognize $4\pi k +2\pi s'+\pi=\phi_{ks'}$, allowing us to identify the inner sum as $(-i)^a \psi_{ab}^{s'}(4\pi n)$. Thus
\begin{equation}
\tilde \psi_{ab}^s(n)\approx e^{-i\frac{\pi}{4}(4a+1)} {\sqrt{2\pi}}\sum_{s'=0}^1(-i)^{2ss'+s+s'}\psi_{ba}^{s'} (4\pi n) .
\end{equation}
Evaluating the sum over the terms $s'=s$ and $s'=1-s$ and noting $(-i)^{2ss'+s+s'}=(-i)^{|s-s'|}$ for $s,s'\in \{0,1\}$, we find
\begin{equation}
\tilde \psi_{ab}^s(n)\approx e^{-i\frac{\pi}{4}(4a+1)} {\sqrt{2\pi}}\left( \psi^s _{ba}(4\pi n ) -i\psi^{1-s}_{ba}(4\pi n)\right)
\end{equation}
as we sought to show.
\subsection{Action of Stabilizers}
\label{app:stabilizer_action}
We now use the above results to establish Eqs.~\eqref{eq:s1action}-\eqref{eq:s2action} of the main text, that state that $\tilde S_{1,2}|\psi_{ab}^s\rangle=|\psi_{ab}^s\rangle+\mathcal O(e^{-\wmax})$ for $a,b\ll \wmax$;
Here the canonical stabilizer $\tilde S_i$ is defined as $\tilde S_i \equiv \sgn(1-\epsilon-\hat S_i)$, for some positive $\epsilon$ that we pick such that $\epsilon\ll 1$.
For $\tilde S_1$, the result follows immediately from the definition of $|\psi_{ab}^s\rangle$ in Eq.~\eqref{eq:lattice_basis_wf}, since the support of $|\psi_{ab}^s\rangle$ in the $\hat \phi$-interval where $\tilde S_1=-1$ is of order $e^{-\wmax}$.
The result for $\tilde S_2$ follows by similar arguments, using the exact expression for the wavefunction of $\hat n$ in Eq.~\eqref{eqa:tildepsi_expr_1}.

\subsection{Proof of near-orthonormality}
\label{seca:orth}
Here we establish the near-orthonormality of the lattice basis states, quoted in Eq.~\eqref{eq:near_orth} of the main text.
To show this, we evaluate the overlap between two lattice basis states in the $\hat n$ basis: $\langle \psi_{cd}^r|\psi_{ab}^s\rangle =\int_{-\infty}^\infty dn\, \tilde \psi_{cd}^{r*}(n)\psi_{ab}^s (n)$, using the expression for $\tilde \psi_{ab}^s(n)$ in Eq.~\eqref{eqa:tildepsi_expr_1}. By inserting this expression, we find
\begin{align}
        \langle \psi_{ab}^s|\psi_{cd}^r\rangle& = 2\pi \sum_{\ell_1,\ell_2}{i^{c-a+2s\ell_1+\ell_1+s-2r\ell_2-\ell_2-r}}\chi_a\left(\frac{\lambda}{4} [2\ell_1+1]\right)\chi_c\left(\frac{\lambda}{4} [2\ell_2+1]\right) I_{ab}^{\ell_1\ell_2},
        \label{eqa:overlap_1}
        \end{align}
where
\begin{align}
I_{bd}^{\ell_1\ell_2}\equiv
\int_{-\infty}^\infty\!\! dn\,\chi_b\left(\frac{4\pi n-2\pi \ell_1-\pi}{\lambda}\right)\chi_d\left(\frac{4\pi n-2\pi \ell_2-\pi}{\lambda}\right).
\end{align}
Since $\chi_b(x) \sim e^{-x^2/2}$ for $|x|\gg \sqrt{b}$ and $\wmax = \pi^2/\lambda^2$, we have $I_{bd}^{\ell_1\ell_2}= \mathcal O(e^{-\wmax})$ for $\ell_1\neq \ell_2$ and $b,d\ll \wmax$. Furthermore, due to the orthonormality of the functions $\{\chi_a(x)\}$, we have $I_{bd}^{\ell\ell}=\frac{\lambda}{4\pi}\delta_{bd}$. Thus,
\begin{equation}
     I_{bd}^{\ell_1\ell_2}=\frac{\lambda}{4\pi}\delta_{\ell_1\ell_2}\delta_{bd}+\mathcal O(e^{-\wmax})\qquad {\rm for}\quad b,d\ll \wmax.
\end{equation}
Using this in Eq.~\eqref{eqa:overlap_1}, we find
\begin{equation}
        \langle \psi_{ab}^s|\psi_{cd}^r\rangle = \sum_{\ell}\frac{\lambda}{2} i^{c-a+(s-r)(2\ell+1)}\chi_a\left(\frac{\lambda}{4} [2\ell+1]\right)\chi_c\left(\frac{\lambda}{4} [2\ell+1]\right)+\mathcal O(e^{-\wmax}),\quad {\rm for}\quad b,d\ll \wmax.
\end{equation}

We now split the sum above into the terms with even and odd $\ell$ that we reexpress in terms of integrals over Dirac combs:
\begin{align}
    \langle \psi_{ab}^s|\psi_{cd}^r\rangle &=
    \frac{\lambda i^{c-a+r-s}}{2}\int dn[ \mathcal D(n)+(-1)^{r-s}\mathcal D(n+\lambda/2)]\chi_a\left(n\right) \chi_c\left(n\right)+\mathcal O(e^{-\wmax}),\quad {\rm for}\quad b,d\ll \wmax.
    \label{eqa:psioverlap}
\end{align}
where $\mathcal D(n)=\sum_{k}\delta(n-\lambda [k+1/4])$.
Now we use Parseval's identity to write \begin{align}\int dn f(n)g(n)h(n)&=\int d x \mathcal F[f]^*(x) \mathcal F[gh](x)=\frac{1}{\sqrt{2\pi}}\int dx \mathcal F[f]^*(x) (\mathcal F[g](x)*\mathcal F[h](x)).\end{align}
Using this in Eq.~\eqref{eqa:psioverlap} and recalling that $\mathcal F[\chi_a](x)=(-i)^a \chi_a(x)$, while $\mathcal F[g](x)=e^{i yx }\mathcal F[f](x)$ for $g(x)=f(x+y)$, this leads to
\begin{align}
    \langle \psi_{ab}^s|\psi_{cd}^r\rangle &=
    \frac{\lambda i^{2c-2a+r-s}}{2\sqrt{2\pi}}\int dx dx'[1+(-1)^{r-s}e^{-i(x-x')\lambda/2}] \tilde{\mathcal D}(x-x')\chi_a\left(x\right) \chi_c\left(x'
    \right)+\mathcal O(e^{-\wmax}).
\end{align}
\end{widetext}
With $\tilde{\mathcal D}(x)=\mathcal F[\mathcal D](x)=\frac{\sqrt{2\pi}}{\lambda}\sum_{k}\delta(x-\frac{2\pi}{\lambda} k)(-i)^k $.
Inserting this expression for $\tilde {\mathcal D}(x)$ above, and using that, for $a,c\ll \wmax$, $\chi_{a}(x)\chi_c(x+2\pi k/\lambda)=\mathcal O(e^{-\wmax})$ for nonzero integers $k$, we find that only the term with $k=0$ survives, and thus,
\begin{align}
    \langle \psi_{ab}^s|\psi_{cd}^r\rangle &=
    \delta_{ac}\delta_{bd}\delta_{rs}+\mathcal O(e^{-\wmax}).
\end{align}
This was what we wanted to show.

\subsection{Action of logical operators}
\label{seca:logicals}
Here we establish Eqs.~\eqref{eq:zl}-\eqref{eq:yl} of the main text, that give the action of the logical operators on the lattice basis states.

To establish the action of the $\hat \sigma_z$ operator, we note that this operator is diagonal in the eigenbasis of $\hat \phi$. We also note that, for $a,b\ll \wmax$, $\psi_{ab}^s(\phi)$ has all of its support confined in the $\phi$-intervals where $|\phi-\phi_{ks}|\leq \pi$ for some $k\in \mathbb Z$, up to $\mathcal O(e^{-\wmax})$ corrections [see Eq.~\eqref{eq:lattice_basis_wf} of the main text and Fig.~\ref{fig:gkp_states}(a)]. In these intervals $\hat \sigma_z$ takes value $(-1)^s$, and thus, for $a,b\ll \wmax$, $\hat \sigma _z|\psi_{ab}^s\rangle = (-1)^s|\psi_{ab}^s\rangle+\mathcal O(e^{-\wmax})$, establishing Eq.~\eqref{eq:zl}.

To establish the action of the $\hat \sigma_y$ operator on the state $|\psi_{ab}^s\rangle$, we note that this operator is diagonal in the $\hat n$ basis. In Eq.~\eqref{eqa:tildepsi_expr_1}, we obtained an exact expression for the wavefunction in this basis, $\tilde \psi_{ab}^s(n)$. In particular,
\begin{widetext}
\begin{equation}
    (\tilde \psi^0_{ab}(n)\pm i \psi^1_{ab}(n))={\sqrt{2\pi}e^{-i\frac{\pi}{4}(4a+1)} }\sum_{\ell=-\infty}^\infty i^{-\ell}(1\pm (-1)^\ell)\chi_a\big(\lambda[\ell/2+1/4]\big) \chi_b\left(\frac{4\pi[n-\ell/2-1/4]}{\lambda}\right)
\end{equation}
\end{widetext}
For $a,b\ll \wmax$, the $n$ support of $\chi_b\left(\frac{4\pi[n-\ell/2-1/4]}{\lambda}\right)$ is confined to the interval $|n-(\ell/2+1/4)|<1/4$ up to $\mathcal O(e^{-\wmax})$ corrections.
In this interval $\hat \sigma_y$ takes eigenvalue $(-1)^\ell$.
Moreover, $(1\pm (-1)^\ell)$ factor means that only terms where $(-1)^\ell=\pm 1$ contribute, implying
\begin{equation}
    \hat \sigma_y(| \psi^0_{ab}\rangle \pm i |\psi^1_{ab}\rangle )= \pm (| \psi^0_{ab}\rangle \pm i |\psi^1_{ab}\rangle )+\mathcal O(e^{-\wmax}).
\end{equation}
This in turn implies that $\hat \sigma_y|\psi_{ab}^s\rangle = (-i)^s|\psi_{ab}^{1-s}\rangle+\mathcal O(e^{-\wmax})$, establishing Eq.~\eqref{eq:yl}.

\section{Saddle-point approximation of effective Hamiltonian and generalized protocols}

\label{app:SPA}
Here we show how the GKP Hamiltonian in Eq.~\eqref{eq:H_eff_GKP} emerges as the effective Hamiltonian of the circuit,
after making the rotating wave approximation (RWA) and saddle point (or stationary phase) approximation (SPA).

We consider the most general configuration of waveform and circuit impedance in this Appendix. In particular, we consider an arbitrary time-periodic voltage signal $V(t) = V(t+T)$ with period $T=2\pi/\Omega$, whose fundamental frequency is near-commensurate with the resonator frequency $\omc$:
\be
\omc \approx \tilde \Omega ,\quad{\rm where}\quad \tilde \Omega \equiv \frac{p}{q}\Omega.
\ee
where $p,q$ are integers that are relatively prime.
Below, we let $\Phi(t)$ denote the dimensionless phase induced by the oscillating voltage, defined such that
\be
    \partial_t \Phi(t) \equiv \frac{2e}{\hbar} V(t).
\ee
we pick the gauge where $\Phi(t)$ has zero mean, such that $\int_0^T dt \Phi(t)=0$.

To analyze the problem, we go to the co-moving phase-space frame rotating with angular frequency $\tilde \Omega$, as described in Sec.~\ref{sec:device} of the main text, using the rotating frame transformation generated by $H_{\rm r} = \frac{\hbar\tilde \Omega}{2}({\hat \phi^2}/{2\pi \fz }+2\pi \fz \hat n^2)$. In this frame, the Hamiltonian is given by
\be
\tilde H(t) = \hat H_{\rm d}
    -E_J \cos\big[\hat \phi \cos \tilde{\Omega}t+ 2\pi \fz \hat n \sin \tilde{\Omega}t - \Phi(t)\big]\!
\label{eqa:tilde_h_def}
\ee
where
$\hat H_{\rm d}\equiv \frac{\hbar \delta \omega}{2} \big( \frac{\hat \phi^2}{2\pi \fz}+2\pi \fz \hat n^2 \big)$ and $\delta \omega = \omc - \tilde \Omega$.
Note that the Hamiltonian $\tilde H(t)$ is time-periodic with {\it extended} period $qT$, $\tilde H(t) = \tilde H(t+qT)$.
Our goal is to calculate the effective, or Floquet, Hamiltonian generated by $\tilde H(t)$, $\Heff$.

We first make the RWA, which is valid given our assumption $\partial_t \Phi(t) = 2e V(t)/\hbar\gg E_J, \delta \omega$. The RWA approximates $\Heff$ by the time-average of $\tilde H(t)$~\cite{Kuwahara_2016} over its full (extended) period:
\be
\Heff \approx \HRWA,\quad {\rm where}\quad \HRWA \equiv \int_0^{ qT }\frac{dt}{ qT} \tilde H(t) .
\label{eqa:HRWA_def}
\ee
We focus on computing the time-average of the second term in Eq.~\eqref{eqa:tilde_h_def}, since $H_{\rm d}$ is time-independent.
For simplicity, we thus set $\delta \omega =0$ below, unless noted otherwise.
We now note
\be
\HRWA = -\int_0^{qT}\frac{dt}{2qT} E_J e^{-i[\hat \phi\cos\tilde \Omega t - 2\pi\fz \hat n\sin\tilde \Omega t- \Phi (t)]} + h.c.
\label{eqa:hrwa_dimless-def}
\ee

To compute the integral above, we now use SPA, which is valid when the characteristic scale of $\partial_t \Phi = 2e V(t)/\hbar$ is much larger than $\tilde \Omega$---something we have already assumed to ensure the RWA is valid.
In this limit, the integral in Eq.~\eqref{eqa:HRWA_def} only gets contributions from the points $\{t_i\} \subset [0,qT)$ where $\tilde \Phi$ is stationary. These points coincide with the nodes of the voltage signal, i.e., $\{t_i\} = \{t\in [0,qT)|V(t) = 0\}$.
Using the SPA we then obtain
\be
\HRWA \approx -\sum_{i=1}^N \frac{E_Je^{-i[\hat \phi\cos\tilde \Omega t_i - 2\pi\fz \hat n\sin\tilde \Omega t_i- \Phi (t_i)-\frac{\pi}{4}\mu_i]} }{2qT\sqrt{ |\partial_t ^2 \Phi(t_i)|/2\pi}} + h.c.
\label{eqa:hrwa_dimless}
\ee
where
$
\mu_i \equiv \sgn[\partial_t ^2 \Phi(t_i)].
$
Using $\partial_t \Phi(t) = 2eV(t)/\hbar $, this leads to
\be
\HRWA\! \approx\! \sum_{i=1}^N \!\frac{E_J\cos[\hat\phi \cos\tilde \Omega t_i \! - \! 2\pi\fz \hat n\sin\tilde \Omega t_i\!-\! \Phi (t_i)\!-\!\frac{\pi}{4}\mu_i] }{-qT\sqrt{2 e |\partial_t V(t_i)|/h}} .
\label{eqa:hrwa_general}
\ee
where the sum runs over all the nodes of $V(t)$ within the interval $[0,qT)$, $t_1,\ldots t_N$.
This result is the most general expression for the effective Hamiltonian of the circuit, in principle valid for any periodic waveform $V(t)$ in the regime where $eV(t)\gg \hbar \tilde \Omega$ away from the nodes of $V(t)$.

We now consider two special cases in which the square-GKP and hexagonal-GKP stabilizer Hamiltonians emerge.

\subsection{Emergence of square GKP stabilizer Hamiltonian}
We can obtain the square GKP stabilizer Hamiltonian by driving at $\Omega \approx 2\omc$, i.e., $p=1$, $q=2$ with a waveform $V(t)$ that only consists of odd-harmonic sine waves,
\be
V(t) = \sum_{k=0}^\infty c_k \sin([2k+1]\Omega t)
\label{eqa:odd_harmonic_sine_drive}
\ee
for some coefficients $\{c_k\}$.
This choice of waveform ensures that $V(t)=-V(-t)$ and $V(t) = -V(t+T/2)$, implying the nodes of $V(t)$ lie at $t \in \frac{T}{4}\mathbb Z$.
We moreover require that $\{c_k\}$ are chosen such that these instances are the {\it only} nodes of $V(t)$.
For instance, the waveform can be a pure sine wave ($c_0 = V_0$ and $c_k=0$ for $k\geq 1$), or take the capped-tone shape in Fig.~\ref{fig:pulse} of the main text.

With the choice above, the nodes of the voltage signal within the extended period $2T$ occur at the four instances $\{t_i \} =\{ 0, \frac{1}{2}T, T,\frac{3}{2} T\}$, corresponding to $\{\tilde \Omega t_i\} =\{ 0, \frac{\pi}{2}, \pi, \frac{3\pi}{2}\}$.
Without loss of generality, we pick the time origin such that $\partial_t V(0) \geq 0$.
As a result $\mu_i = (-1)^{i+1}$, while $|\partial_t V(t_i)|=|\partial_t V(0)|$ for $i=1,\ldots 4$.
Finally, with the gauge choice that $\Phi(t)$ has zero time-average, $\Phi(t_i)=(-1)^{i} \thmax$, where
\be
\thmax \equiv \int_{0}^{T/4} dt \frac{2eV(t)}{\hbar} .
\ee
Inserting these results in Eq.~\eqref{eqa:hrwa_general}, using $q=2$, and explicitly writing out all $4$ terms, we obtain
\begin{widetext}
\begin{align}
\HRWA &\approx \frac{-E_J \left(\cos\left[\hat\phi + \thmax-\frac{\pi}{4}\right] +\cos\left[-\hat\phi+ \thmax-\frac{\pi}{4}\right]+ \cos\left[2\pi \fz \hat n- \thmax+\frac{\pi}{4}\right] +\cos\left[-2\pi \fz \hat n -\thmax+\frac{\pi}{4}\right]\right)}{2T\sqrt{2 e |\partial_t V(0)|/h}}.
\end{align}
\end{widetext}
Using $\cos(a+b)= \cos(a)\cos(b)+\sin(a)\sin(b)$, we obtain
\begin{align}
\HRWA &\approx -\frac{ E_J \cos(\thmax-\pi/4) }{T\sqrt{2 e |\partial_t V(0)|/h}}\left(\cos\hat\phi + \cos 2\pi \fz \hat n\right)
\end{align}
with $\fz \in \mathbb Z$ we thus recover the stabilizer Hamiltonian for a square-lattice GKP qudit with $\fz$-fold degeneracy.
In the presence of nonzero detuning, an extra term $H_{\rm d}$ is added to $\HRWA$.
This is the result we quoted in the main text.

\subsection{Emergence of hexagonal GKP stabilizer Hamiltonian}
\label{app:hex}
We now show how the hexagonal GKP code can be generated by an odd-harmonic sine drive similarly to the square-lattice Hamiltonian [Eq.~\eqref{eqa:odd_harmonic_sine_drive}] (or a higher-order featuring odd sines), but with $\Omega \approx 3\omc$ or $\Omega \approx 3\omc/2$---i.e., with $q=3$ and $p=1,2$.
For simplicity we focus on the case $\Omega \approx 3\omc$ ($q=3,p=1$).
In this case, the waveform $V(t)$ has $6$ nodes within the extended period $3T$, namely at $\tilde \Omega t = z/6$ for $z=0,1,2,3,4,5$, corresponding to $\tilde \Omega t = 2\pi z /6$.

Explicitly summing over all these $6$ nodes in Eq.~\eqref{eqa:hrwa_general}, we obtain
\begin{align}
\HRWA \!\approx \tilde E_{J}^{(3)}\!\sum_{z = 1}^{3}\!\cos\!\left( \hat\phi c_z - 2\pi \fz \hat n s_z+(-1)^z\left[ \thmax+\frac{\pi}{4}\right]\right) \notag \\ + \tilde E_{J}^{(3)}\sum_{z = 1}^{3}\cos\left(-\hat\phi c_z + 2\pi \fz \hat n s_z -(-1)^z\left[ \thmax-\frac{\pi}{4}\right]\right)
\end{align}
where $\tilde E_J^{(3)}=\frac{-E_J}{3T\sqrt{2e |\partial_t V(0)|/h}} $
$,
c_z \equiv \cos\left(\frac{2\pi z}{6}\right)$, and $s_z \equiv \sin\left(\frac{2\pi z}{6}\right)$.
Here we paired up the terms $i$ and $i+3$ (for $i=1,2,3$) from the sum in Eq.~\eqref{eqa:hrwa_general} to form the summand above.
Since $\cos(x) = \cos(-x)$, the two terms in each summand are identical, leading to
\be
\HRWA \approx\tilde E_J^{(3)} \sum_{z = 1}^{3}\cos( \hat\phi c_z - 2\pi \fz \hat n s_z+(-1)^z[ \thmax-\pi/4]).%
\ee
When $\fz = \frac{d}{\sin 2\pi/3} = \frac{2d}{\sqrt{3}}$ for integer $d$, all $3$ terms in the sum above commute.
In this case, $\HRWA$ is the hexagonal GKP stabilizer Hamiltonian that protects a $d$-dimensional qudit in its low-energy subspace.
Note that the effective confining potential is not an oscillatory function of $\thmax$, in contrast to the square-lattice implementation we considered in Sec.~\ref{sec:device} of the main text. Instead, modulation of $\thmax$ will only cause a displacement of the centers of the stabilized GKP states in phase space.

\begin{figure}
    \centering
    \includegraphics[width=0.7\linewidth]{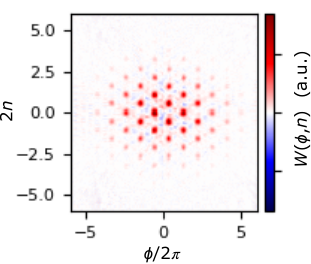}
    \caption{
    Numerically simulated Wigner function of a steady-state density matrix for driving at $\Omega \approx 3 \omc$ after time-evolution of $t = 2000 \times 2\pi \omc^{-1}$ starting from a random mixed state. This density matrix corresponds to a mixed state in the hexagonal GKP encoding. System parameters: Josephson energy $E_J = 2.25 \hbar \omc$, dimensionless impedance $\fz = 4/\sqrt{3}$,
    phase drive $\Phi(t) = \frac{2e}{\hbar}A \cos \Omega t$ with $A = 40.5\pi$ and $\Omega = 3\omc - \delta \omega$ with detuning $\delta \omega = 0.004 \omc$, bath temperature $T_{\rm B} = 0.016 \hbar \omc/k_{\rm B}$. These calculations used a low-pass filter with bandwidth $\Lambda_B = 0.01 \omc$, spectral density confined to $\leq \hbar \tilde \Omega/2$, and dimensionless coupling strength $\kappa = 10^{6}$.}
    \label{fig:app-Wigner-q3}
\end{figure}

We verify that the hexagonal states indeed emerge by calculating numerically the Wigner function of the steady-state density matrix after initialization from a random mixed state when driving close to thrice the resonance frequency with suitable system parameters, see Fig.~\ref{fig:app-Wigner-q3} for the result and details of the simulation.

\section{$S$ gate protocol}
\label{app:S_T_gates}

Here we provide further details on how a protected phase, or $S$, gate, $e^{-i\frac{\pi}{4}\hat \sigma_z}$ might be generated with the circuit with the square-GKP encoding, using the phase-revival mechanism described in Ref.~\cite{Nathan2024May}.
We use the odd-harmonic sine waveform used in the main text and discussed in Sec.~\ref{app:pulse}.
To implement an $S$ gate, we first shift the GKP lattice by $\phi \to \phi + \pi$ by adiabatically increasing the amplitude of an extra tone with angular frequency $\tilde \Omega$, $V_{\rm s}$, from zero to $\pi \hbar \tilde \Omega/2e$.
This results in a shift of the phase signal $\Phi(t)$ by $\Delta \Phi(t) = \frac{2eV_{\rm s}}{\hbar \tilde\Omega} \cos \tilde \Omega t$, resulting in the effective Hamiltonian being given by
\be
\hat H_{\rm RWA}\!\approx \!- \frac{E_J\cos(\thmax-\pi/4)}{T\sqrt{2e|\partial_t V(0)|/h}}[\cos(\hat \phi-2eV_{\rm s}/\hbar \Omega)+\cos4\pi \hat n].
\ee
This result follows from direct insertion of the new phase signal $\Phi(t)$ into Eq.~\eqref{eqa:hrwa_general} under the assumption that $V_{\rm s}$ is only weakly affecting $\partial_t V(t)$ at the nodes of the voltage oscillation (a simplifying, but not essential assumption).
Under the adiabatic ramp of $V_{\rm s}$, the wells in the $\cos\hat \phi$-potential of the effective Hamiltonian, $\hat H_{\rm RWA}$ get shifted adiabatically from $2\pi \mathbb Z+\pi $ to $2\pi \mathbb Z +\pi + 2eV_{\rm s}/\hbar \Omega $, until they are located at $\hat \phi \in 2\pi \mathbb Z$.

After the adiabatic shift of the well centers, we abruptly change the main driving tone to a new tone with frequency $\Omega' \approx \Omega/2$, i.e., with approximately doubled period $T'\approx 2T$. The amplitude of the period-doubled driving pulse should be adjusted {so that $\thmax=-3\pi/4$} and ideally have the same nodal slope $\partial_t V(0)$ as the original driving pulse.
After the period-doubling of the signal, there are just two nodes of $V(t)$ in the extended period (which now coincides with $T$), namely $t = 0$ and $t=T/2$.
Using the formula in Eq.~\eqref{eqa:hrwa_general}, the approximate effective Hamiltonian of the system becomes
\be
\Heff \approx -\tilde E_J \cos\hat\phi +\frac{\hbar \delta \omega'}{2} \left( \frac{\hat \phi^2}{4\pi }+4\pi \hat n^2 \right),
\ee
where $\delta\omega'\equiv \Omega'-\omc$ is the detuning of the frequency of the period-doubled signal from $\omc$.
To arrive at the above result, we used $\thmax = -3\pi/4$, $\tilde E_J = \frac{2 E_J}{T\sqrt{2e|\partial_t V(0)|/h}}$ \footnote{The factor of $2$ in $\tilde E_J$ arises because the two terms in the sum in Eq.~\eqref{eqa:hrwa_general} for $p=q=1$ are identical.}.
The detuning of the period-doubled pulse, $\delta \omega'$ can be distinct from the usual driving pulse, and will play a crucial role.

The effective Hamiltonian above describes a free particle with mass $1/4\pi \delta \omega '$ in a potential consisting of a parabola with curvature $\delta \omega '/4\pi$ and a cosine potential of amplitude $\tilde E_J$.
When $\tilde E_J \gg \hbar \delta \omega '$, the low-energy eigenstates of $\Heff$ are confined within the individual wells; the spectra and eigenstates of each well are near-identical, except for an overall energy shift for each well, which for the well at $\hat \phi = 2\pi k$ is given by $\frac{ \pi \hbar \delta \omega'}{2} k^2$.
A GKP state with expectation value $\langle\hat \sigma_z\rangle =\pm 1$ is a coherent superposition of states with support in even ($+$) or odd ($-$) wells, whose constrictions to individual wells are near-identical.
Under evolution by $\Heff$ above for a duration of $\Delta t = 1 / \delta \omega '$, the component of the state in well $k$ will have acquired a phase factor $e^{-i\frac{\pi}{2} k^2}$ relative to the component in well zero.
Since $k^2 \mod 4 = 0$ for even $k$ and $k^2 \mod 4 =1$ for odd $k$, components in odd wells will thus have acquired a phase factor $i$ relative to their counterparts in even wells after the time span $\Delta t$.
This is equivalent to the action of an $S$ gate.
Reverting to the original driving tone after this time-span will again activate the stabilizer term $\cos(4\pi \hat n)$ in the effective Hamiltonian, stabilizing the state after the gate.
A similar mechanism is described in Ref.~\cite{Nathan2024May}, which shows that such a gate is expected to be exponentially robust against control noise.

\section{Readout protocol}
\label{app:readout}
Here we provide additional details on the operating principle of the readout protocol.

To recap, the readout is achieved by (1) connecting a secondary, weak Josephson junction $J'$ at a new circuit node located halfway along the inductor, with Josephson energy $E_{J'}$, and (2) applying a separate voltage tone $V_{\rm R}(t)$ across $J'$ with a quadrupled period (angular frequency $\Omega_{\rm R} = \Omega/4$). The circuit diagram is shown in Fig.~\ref{fig:readout}(a) in the main text. The new node is henceforth referred to as the readout node. In the following, we show that the time-averaged supercurrent $\bar I$ across $J'$ is positive for even and negative for odd wells, thereby distinguishing the two logical states.

We first seek to express the mean supercurrent in the secondary junction in terms of the resonator state evolution $\hat \rho(t)$.
We assume that the capacitance at the readout node is small, and that the secondary junction inductance $L_{J'} = \varphi_0^2/E_{J'}$ remains small relative to that of the main inductor, $L$, such that (a) the phase at the readout node remains relatively unperturbed by the secondary Josephson junction, and (b) the spectrum of modes in the resonator remains relatively unperturbed by the secondary Josephson junction, allowing all but the fundamental mode to be frozen out of the dynamics.
In this case, the phase at the readout node is given by $\hat \phi/2$, due to its location halfway along the inductor.
The supercurrent through the secondary Josephson junction will then be given by
\be
I(t) = \frac{2e E_{J'}}{\hbar}\sin\Big[\frac{\hat \phi}{2}-\Phi_{\rm R}(t)\Big]
\label{eq:supercurrent_expression}
\ee
with $\Phi_{\rm R}(t)$ the phase-oscillation induced by $V_{\rm R}(t)$.

We can find $\bar I$ via
\be
\bar I = \lim_{N\to \infty}\frac{1}{N} \sum_n \bar I_n
\ee
where
\be
\bar I_n \equiv \int_{4nT}^{(n+1)4T}\!\!\!\frac{dt}{4T}\, \Tr[\hat \rho(t) \hat I(t)].
\ee

Using the expression in Eq.~\eqref{eq:supercurrent_expression}, we find
\be
\bar I_n =\frac{2e E_{J'}}{ \hbar }\int_{4nT}^{4(n+1)T}\!\!\!\frac{dt}{4T}\, \Im \Tr[\hat \rho(t)e^{-i\hat \phi/2}]e^{-i\Phi_{\rm R}(t)}
\ee
In the limit of rapid phase oscillation, $\partial_t \Phi_{\rm R}(t) \gg \partial_t \Tr[\hat \rho(t)e^{-i\hat \phi/2}]$, we can approximate the integral above using the stationary phase approximation, as we did to obtain the effective Hamiltonian in Sec.~\ref{app:SPA}.
Assuming an odd-harmonic sine waveform, $\Phi_{\rm R}(t)$ has stationary points at the voltage nodes, $t_0=4nT$ and $t_1=4(n+1/2)T$.
The stationary-phase approximation then yields
\be
\bar I_n \approx \frac{e E_{J'}}{2 T \hbar}\Im\sum_{k=0}^1 \frac{e^{- i\Phi_{\rm R} (t_k)-i\frac{\pi}{4}\mu_k} }{\sqrt{ |\partial_t^2 \Phi_{\rm R}(t_k)|/2\pi}} \Tr[\hat \rho(t_k)e^{-i\hat \phi/2}]
\ee
with $\mu_k = \sgn(\partial_t^2 \Phi_{\rm R}(t_k))$.
For odd-harmonic sine waveform, we have $\Phi_{\rm R}(t_k) = (-1) ^k \Phi_{{\rm R}, 0}$, and $\mu_k = (-1)^k$ for $k=0,1$, with $ \Phi_{{\rm R}, 0}$ the phase oscillation amplitude induced by $V_{\rm R}(t)$.
Using these results along with $\partial_t^2 \Phi_{\rm R}(t) = 2e\partial_t V_{\rm R}(t)/\hbar$, we find
\be
\bar I_n \approx \frac{e E_{J'}}{2T \hbar \sqrt{ \varphi_0 |\partial_t V_{\rm R}(0)|}}\Im\sum_{k=0}^1 \Tr[\hat \rho(t_k)e^{-i\hat\phi/2}]e^{- i(-1)^k[\Phi_{{\rm R}, 0}-\frac{\pi}{4}]}
\ee
Neglecting the effect of the weak Josephson junction $J'$, the time evolution of the resonator density matrix is given by
\be
\hat \rho(t)\approx \hat R_{\tilde \Omega t}^\dagger e^{-i\Heff t/\hbar} \hat \rho(0)e^{i\Heff t/\hbar }\hat R_{\tilde \Omega t}
\ee
where $\hat R_\theta$ rotates phase space by an angle $\theta$, and $\tilde \Omega = \pi/T$ is the angular frequency of the rotating frame.
Since the strength of the GKP potential by assumption is much weaker than $2\hbar \omc $, $\tilde{E}_J \ll 2 \hbar \omc$, the evolution in the rotating frame is near-stationary over a single period, $e^{-i\hat H_{\rm GKP }T} \approx 1$, implying
\begin{equation}
   \hat \rho(2(2n+1)T) \approx \hat \rho(4nT).
\end{equation}
Hence
\be
\bar I_n \approx \frac{e E_J'}{2T\hbar}\Im \sum_{k=0}^1 \frac{e^{- i(-1)^k[\Phi_{{\rm R},0}-\frac{\pi}{4}]} }{\sqrt{ 2e |\partial_t V_{\rm R}(0)|/h}} \Tr[\hat \rho(4nT)e^{-i\hat \phi/2}]
\ee
Evaluating the sum, we find
\be
\bar I_n \approx I_0 \Tr[\hat \rho(4nT)\sin(\hat \phi/2)]
\ee
where
\be
I_0 \equiv \frac{e E_{J'} \cos(\thmax-\frac{\pi}{4})}{ T \hbar \sqrt{\varphi_0 |\partial_t V_{\rm R}(0)|}}
\ee
Thus, the mean supercurrent is given by
\be
\bar I \approx \bar I _ 0\lim_{N\to \infty}\sum_{n=1}^N \frac{1}{N}\Tr[\hat \rho(4nT)\sin (\hat \phi/2)]
\ee

We now consider the possible measurement outcomes:
eigenstates with support in even wells ---i.e., $\hat \phi \in 4\pi \mathbb Z + \pi$ (such as $|0\rangle$)---will result in $\bar I \approx \bar{I}_0$, while eigenstates with support in odd wells (such as $|1\rangle$)---i.e., $\phi \in 4\pi \mathbb Z - \pi$---will result in $-\bar I \approx - \bar{I}_0$; the logical state is thus measured by the protocol.

The readout protocol can be used for initialization: initializing the system in a random state, the stabilization protocol will take the system into the code subspace.
Subsequently measuring the state with the protocol above will collapse it into a logical $|0\rangle$ or $|1\rangle$ state, depending on the measurement outcome.

\section{Modeling dissipative dynamics}
\label{app:dissipative-dynamics}
In this section, we describe how the driven-dissipative dynamics of the circuit may be modeled with the universal Lindblad equation from Ref.~\cite{Nathan_2019,NathanPRB2020_ULE}. In particular, we show how this dynamics results in the time-independent effective master equation in Eq.~\eqref{eq:Lindblad_eq_HighFrequency} of the main text in the regime where
\be
\delta \omega, \tilde E_J /\hbar \ll \tilde \Omega
\label{eqa:condition}
\ee
with $\tilde \Omega= p\Omega/q$ for relatively prime integers $p,q$ such that $\tilde \Omega \approx \omc$, and $\tilde E_J = \frac{E_J \cos\left(\Phi(0)-\pi/4\right)}{T \sqrt{\fz e |\partial_t V(0)| / h}}$ denoting the barrier height in the Josephson potential of the effective Hamiltonian [Eq.~\eqref{eq:H_eff_GKP}].
For concreteness, we focus on the odd-harmonic sine waveform with $\fz=2$, $p/q= 1/2$, while noting that our analysis can be applied to more impedances and driving tones.

To recap,
the circuit can be analyzed as a periodically driven system with time-dependent Hamiltonian
\be
 \hat H(t)=
\frac{\hbar \omc}{2}\left(\frac{\hat \phi^2}{2\pi \fz}+2\pi \fz \hat n^2\right)
-E_J \cos(\hat \phi+\Phi(t)),
\ee
We couple the circuit to an external bath via
\be
\hat H_{\rm RC} = \frac{2e \hat n \hat Q_ {\rm B}(t)}{C_ {\rm B}},
\ee
where $\hat Q_{\rm B}(t)$ denotes the fluctuating charge on the bath-side of the coupler, and $C_{\rm B}$ the capacitance of the coupler.
We model $\hat Q_{\rm B}(t)$ as a quantum noise variable with equilibrium PSD,
\be
J(\omega)\equiv \frac{4e^2}{2\pi C_{\rm B}^2}\int dt \langle \hat Q_B(t)\hat Q_{\rm B}(0)\rangle e^{i\omega t},
\ee
For now, we let $J(\omega)$ be arbitrary.

In the Markovian regime, where the characteristic rate of system-bath coupling is much weaker than the characteristic inverse correlation time of the bath the time-evolution of the reduced density matrix of the system evolves according to the Lindblad equation~\footnote{Here we ignored the Lamb shift that weakly renormalizes the Hamiltonian}
\be
\partial_t \hat \rho(t) = -\frac{i}{\hbar}[H(t),\hat \rho(t)] + \hat L(t)\hat \rho(t)\hat L^\dagger(t) - \frac{1}{2} \{\hat L^\dagger(t)\hat L(t),\hat \rho(t)\}
\ee
with time-dependent jump operator
\be
\hat L (t) = \int ds g(t-s) \hat U(t,s) \hat n \hat U(s,t)
\ee
where $\hat U(t,s)\equiv \hat U(t)\hat U^\dagger(s)$ with $\hat U(t) \equiv \mathcal T e^{-i\int _{0}^t dt' \hat H(t')/\hbar }$ the lab frame evolution operator, and
\be
g(t) = \int d\omega e^{-i\omega t}\sqrt{2\pi J(\hbar \omega)}
\ee
denotes the {jump correlator}~\cite{NathanPRB2020_ULE}.

The jump operator and master equation above in principle can be solved numerically.
However, we can obtain a simpler and illuminating expression for the dynamics by transforming to the comoving frame discussed in Sec.~\ref{sec:device} of the main text, and using Floquet theory.
The discussion below is devoted to this analysis, by first transforming the master equation to the comoving frame (Sec.~\ref{seca:cmf}), and subsequently, formulating the dynamics there in terms of the effective Hamiltonian $\Heff$ [Eq.~\eqref{eq:heff}] (Sec.~\ref{seca:floquet_thm}).

\subsection{Going to the rotating frame}
\label{seca:cmf}

We first go to the comoving frame introduced in Sec.~\ref{sec:device} of the main text.
In this frame, the evolution operator is given by $\tilde U(t) =\hat U^\dagger_{\rm r}(t)\hat U(t)$,
where \be
\hat U_{\rm r}(t)\equiv e^{-i\hat H_{\rm r}t/\hbar},\quad \hat H_{\rm r} = \frac{\hbar\tilde \Omega}{2}\left(\frac{\hat \phi^2}{2\pi \fz }+2\pi \fz \hat n^2\right)
\ee
generates a phase space rotation by angle $2 \pi t/\tilde T$, and in particular satisfies $\hat U_{\rm r}(t)=\hat U_{\rm r}(t+ \tilde T)$, and $\hat U_{\rm r}(k\tilde T)=1$ for $k\in \mathbb Z$.
The time-evolution in the rotating frame is generated by $\tilde H(t)$ [Eq.~\eqref{eq:tildehdef}], such that $\partial_t \tilde U(t) = -\frac{i}{\hbar}\tilde H(t)\tilde U(t)$.
The state of the system in this frame,
\be
\tilde \rho(t) \equiv \hat U_{\rm r}^\dagger(t) \hat \rho(t)\hat U_{\rm r}(t),
\ee
evolves according to the master equation
\be
\partial_t \tilde \rho(t) = -\frac{i}{\hbar}[\tilde H(t),\tilde \rho(t)] + \tilde L(t)\tilde \rho(t)\tilde L^\dagger(t) - \frac{1}{2} \{\tilde L^\dagger(t)\tilde L(t),\tilde \rho(t)\}
\ee
where
$
\tilde L(t) = \hat U_{\rm r}^\dagger(t)\hat L(t)\hat U_{\rm r}(t)
$.
Using that $\hat U_{\rm r}(t)$ generates a phase space rotation by angle $\tilde \Omega t$, such that $\hat U_{\rm r}^\dagger(t)\hat n \hat U_{\rm r}(t) =\hat n\cos\tilde \Omega t + \frac{\hat \phi}{4\pi } \sin\tilde \Omega t$, we thus find
\be
\tilde L (t) = \int ds g(t-s) \tilde U(t,s) \left[ \hat n\cos\tilde \Omega t + \frac{\hat \phi}{2\pi \fz} \sin\tilde \Omega t\right]\tilde U(s,t)
\label{eqa:rfjo}\ee
with $\tilde U(t,s)\equiv \tilde U(t)\tilde U^\dagger(s)$.

\subsection{Invoking the Floquet theorem}
\label{seca:floquet_thm}
We now use the Floquet theorem to express the evolution in terms of the effective Hamiltonian $\Heff$ introduced in the main text [see text above Eq.~\eqref{eq:H_eff_GKP}].
Noting that $\tilde H(t) = \tilde H(t+\tilde T)$ with $\tilde T= 2T$, we can invoke the Floquet theorem~\cite{floquet_sur_1883,oka_floquet_2019}.
This theorem states that the time-evolution generated by $\tilde H(t)$ of any state $|\Psi(t)\rangle$, can be expressed in terms of a complete orthonormal basis of time-periodic {\it Floquet states}, $\{|\psi_n(t)\rangle = |\psi_n(t+\tilde T)\rangle\}$ and associated {\it quasienergies} $\{\varepsilon_n\}$, as $|\Psi(t)\rangle = \sum_n c_ne^{-i\varepsilon_n t/\hbar}|\psi_n(t)\rangle $. Equivalently, the time-evolution operator of the system, $\tilde U(t)$, can be parameterized via
\be
\tilde U(t) = \cM(t)e^{-i\Heff t/\hbar },
\label{eqa:micromotionplusheff}
\ee
where $\cM(t)=\sum_n|\psi_n(t)\rangle\langle \psi_n|$ denotes the unitary {\it micromotion operator} and $\Heff=\sum_n \varepsilon_n |\psi_n\rangle\langle \psi_n|$ denotes the {\it effective Hamiltonian} of the system, described in the main text.
Note that $\cM(t)=\cM (t+\tilde T)$ and $\cM(0)=\cM(n\tilde T)=1$, for $n\in \mathbb Z$.

Importantly,
there is a gauge freedom in choosing the quasienergies, since Eq.~\eqref{eqa:micromotionplusheff} only defines each quasienergy $\varepsilon_n$ up to a multiple of $2\pi \hbar/\tilde T$: a shift of $\varepsilon_n $ by $2\pi \hbar/\tilde T$ and corresponding right-multiplication of the micromotion operator $\cM(t)$ by the $\tilde T$-periodic operator $|\psi_n\rangle\langle \psi_n|e^{2\pi i t /\tilde T}$,
will result in the same time evolution operator $\tilde U(t)$.
In the high frequency regime, a particular choice of quasienergy is more natural, however: here, a particular choice of branch cut causes $\Heff$ to coincide with the time-averaged Hamiltonian $\HRWA$ in the limit $\tilde T\to 0$.
In our analysis, we use this choice, such that $\Heff \approx \HRWA$ and $\cM (t) \approx 1$~\cite{Schon_2013,Zhang_2017,NathanPhysRevRes2020Dec}; the convergence can be proven in the high-frequency limit~\cite{eckardt_high-frequency_2015}.
In the regime $\tilde E_J/\hbar, \delta \omega \ll 1/\tilde T$ that we consider, the above convention for quasienergies is ensured by fixing $\varepsilon_n\in [-\pi \hbar/\tilde T,\pi \hbar/\tilde T]$. This is because the eigenvalues of $\HRWA$ will lie in this interval, except for those of high-eigenvalue eigenstates far displaced in phase space and photon number, which will not play a role for the dynamics [see discussion under Eq.~\eqref{eq:app-X-JumpMatrixElements} below].

\subsection{Dissipative dynamics in Floquet frame}
Having laid out the preliminaries of Floquet theory, we now finally transform to the {\it Floquet frame}~\cite{MoriAnnuRevCMP2023}, i.e., the frame comoving with $U_{\rm r}(t)\cM(t)$.
Specifically, we consider the evolution of the density matrix
\be
\hat \rho_{\rm F}(t)=\cM^\dagger(t)\tilde \rho(t)\cM(t),
\ee
which coincides with $\tilde \rho(t)$, and hence also $\hat \rho(t)$, for $t\in \mathbb Z \tilde T$.
Direct computation shows that the equation of motion for $ \hat \rho_{\rm F}$ reads
\begin{align}
\partial_t \hat \rho_{\rm F}(t) = &-\frac{i}{\hbar}[\Heff, \hat \rho_{\rm F}(t)]\label{eqa:ule_1}
\\ &+ \hat L_{\rm F}(t)\hat \rho_{\rm F}(t) \hat L_{\rm F}^\dagger(t) - \frac{1}{2}\{ \hat L_{\rm F}^\dagger(t) \hat L_{\rm F}(t),\hat \rho_{\rm F}(t)\}.\notag
\end{align}
where $\hat L_{\rm F}(t) \equiv \cM^\dagger(t)\tilde L(t) \cM(t)$.
Expanding $\hat L(t)$ in terms of its Fourier series and using Eqs.~\eqref{eqa:rfjo}-\eqref{eqa:micromotionplusheff}, we find~\cite{Phd_thesis,Nathan_2019,NathanPRB2020_ULE}
\be
\hat L_{\rm F}(t) =\sum_z e^{-i\tilde \Omega z t} \hat L_z\label{eqa:lprime_coefficient_expanstion}\ee
where
\be
\hat L_z = \sum_{ab}\sqrt{2\pi J(\omega_{ab}+z\tilde \Omega)}|\psi_a\rangle\langle \psi_a|\bar n_{z}| \psi_b\rangle\langle \psi_b|
\ee
and
\begin{equation}
    \bar n_{z} \equiv \frac{1}{\tilde T}\int_0^{\tilde T} dt \cM^{\dagger}(t)\left[\hat n\cos\tilde \Omega t + \frac{\hat \phi}{2\pi \fz} \sin\tilde \Omega t\right] \cM(t) e^{iz\tilde \Omega t},
    \label{eq:app-X-JumpMatrixElements}
\end{equation}
with $\omega_{ab}\equiv(\varepsilon_a-\varepsilon_b)/\hbar$.

While Eq.~\eqref{eqa:ule_1} in principle can be solved numerically, we can make a further simplifying rotating wave approximation that cancels out the cross terms from coefficients $L_z$ and $L_{z'}^{\dagger}$ for $z\neq z'$ that arises by insertion of Eq.~\eqref{eqa:lprime_coefficient_expanstion} into Eq.~\eqref{eqa:ule_1}.
As a first step, we expand Eq.~\eqref{eqa:ule_1} in terms of the coefficients $\{\hat L_z\}$, obtaining
\begin{align}
&\partial_t \hat \rho_{\rm F}(t) = -\frac{i}{\hbar}[\Heff, \hat \rho_{\rm F}(t)]\label{eqa:ule_1point5}\\ &+ \sum_{z_i=-\infty}^\infty e^{-i(z_1-z_2)\tilde \Omega t} \left[\hat L_{z_1} \hat \rho_{\rm F}(t) \hat L_{z_2}^\dagger - \frac{1}{2}\{ \hat L_{z_2}^\dagger\hat L_{z_1},\hat \rho_{\rm F}(t)\}\notag
\right].\end{align}
Next, we note that states in the code-subspace coincide with the low-lying eigenstates of the GKP Hamiltonian, and hence have quasienergies in $[-\tilde E_ J,\tilde E_J]$.
Moving away from the code-subspace towards the high-energy limit, the energy spectrum gradually turns into an evenly spaced spectrum of harmonic oscillator eigenstates confined near LC photon number $n_{\rm ph} = E/\hbar \delta \omega$ for energy $E$.
At the same time, the eigenstates of the GKP Hamiltonian (having energy of order $\lesssim \tilde E_J$) are confined near photon number $n_{\rm ph}\lesssim \tilde E_J/\hbar \delta \omega$.
When $\hbar \tilde \Omega \gg \tilde E_J$, the phase space (photon number) support of states in the code subspace and states with $E>\hbar \tilde \Omega$ is thus highly disjoint, and $\frac{1}{\tilde T}\int_0^{\tilde T} dt \langle \psi_a(t)|\hat n|\psi_b(t)\rangle e^{iz\tilde \Omega t} = 0$ for these states.
As long as $\hat \rho(t)$ maintains its support in the code subspace, we can therefore set $\langle \psi_a|\bar n_{z}| \psi_b\rangle = 0$ up to exponentially small corrections for all states where $|\varepsilon_a-\varepsilon_b| \gtrsim \hbar \tilde\Omega$.
In the limit $\hbar \tilde \Omega \gg \tilde E_J$, cross-terms in Eq.~\eqref{eqa:ule_1point5}, where $z_1\neq z_2$, are thus highly oscillatory relative to all other frequency scales in the equation, and hence can be eliminated via a rotating wave approximation. This results in the {\it time-independent} master equation in [Eq.~\eqref{eq:Lindblad_eq_HighFrequency}] of the main text,
\be
\partial_t \hat \rho_{\rm F} = -\frac{i}{\hbar}[\Heff, \hat \rho_{\rm F}]+ \sum_{z=-\infty}^\infty \left[\hat L_z\hat \rho _{\rm F}\hat L_z^\dagger - \frac{1}{2}\{ \hat L_z^\dagger\hat L_z,\hat \rho_{\rm F}\}\right].
\label{eqa:ule_2}\ee
Since $\hat \rho_{\rm F}(t) = \hat \rho(t)$ for $t \in \mathbb Z \tilde T$, the time-independent master equation above describes the stroboscopic dynamics of $\hat \rho(t)$.
This was the result we sought to establish.

\subsection{Consequence of photon loss}
\label{seca:photon_loss}
We finally show that a capacitively coupled bosonic environment with a smooth and broad-range spectral density, $J_{\rm pl}(E)$, results in a jump operator proportional to the resonator annihilation operator $a_0 =\sqrt{\pi \fz} (n-i\phi/2\pi \fz)$.
To analyze this situation, we can repeat the arguments leading down to Eq.~\eqref{eq:app-X-JumpMatrixElements} to analyze the consequences of this noise source.
The result of this analysis shows that the additional coupling to a bosonic environment leads to an additional Lindblad-form term in Eq.~\eqref{eqa:ule_1} generated by a $\tilde T$-periodic jump operator $L'_{\rm pl}(t)$ whose Fourier coefficients $\{L'_{{\rm pl},z}\}$ can be found from Eq.~\eqref{eq:app-X-JumpMatrixElements} with $J(E)$ replaced by $J_{\rm pl}(E)$.
When $J(E)$ is broad-range and smooth on the energy scale $\tilde E_J$, $J(z\tilde \Omega + \varepsilon_a -\varepsilon_b) \approx J(z\tilde \Omega)$ for quasienergies $\varepsilon_a,\varepsilon_b$ in the range of the effective potential from the GKP Hamiltonian (which is of order $\tilde E_J$).
As a result,
\be
L'_{{\rm pl},z} \approx \sum_z \sqrt{2\pi J(z\tilde \Omega)} \bar n_z.
\ee
Next, we recall that $\cM(t) \approx 1$.
Thus,
\be
\bar n_z \approx \frac{\delta_{z,1}}{2}\left(\hat n-i\hat \phi/2\pi \fz\right)+ \frac{\delta_{z,-1}}{2}\left(\hat n+i\hat \phi/2\pi \fz\right)
\ee
We recognize $\frac{1}{2}\left(\hat n-i\hat \phi/2\pi \fz\right)= \frac{1}{\sqrt{4\pi \fz}} \hat a$, with $\hat a$ the annihilation operator of the cavity mode.
Moreover, in the regime $k_{\rm B}T_{\rm B} \ll \hbar \tilde \Omega $, we have $J(z\tilde \Omega)\approx 0$ for $z<0$. As a result
\be
L'_{{\rm pl},z} \approx \sqrt{2\pi J(\tilde \Omega)} a_0.
\ee
We identify $2\pi J(\tilde \Omega)$ as the photon loss rate associated with the dissipative channel.
This result demonstrates a capacitively coupled bath with smooth spectrum and temperature smaller than $\hbar \tilde \Omega/k_{\rm B}$ will result in a jump operator proportional to the resonator annihilation operator. This was what we wanted to show.

\section{Capped voltage tone}
\label{app:pulse}

To reduce the maximal voltage across the Josephson junction during the drive, we cap the voltage by considering a drive of the form
\begin{equation}
    V(t) = \frac{V_0}{A_0} \left[ \sin \Omega t + \sum_{n=1}^N A_n \sin\big( [2n+1] \Omega t \big)\right]
\end{equation}
where the coefficients $A_n$ are chosen such that the $2n$'th derivative vanishes at $\Omega t = \pi/2$ while the slope at the voltage nodes $dV/dt(\Omega t = 0) = V_0 \Omega$ is kept equal to the slope of a sinusoidal drive.

For the voltage pulse used in the main text, we used $N = 6$. The coefficients are explicitly
\begin{align*}
    A_1 & = 1/4, \quad A_2 = 1/12 \\
    A_3 & = 1/42, \quad A_4 = 1/198 \\
    A_5 & = 1/1452, \quad A_6 = 1/22308
\end{align*}
and the factor $A_0 = 1024/429$ so that $\max V(t) = \frac{1024}{3003} \, V_0 \approx 0.340992 \, V_0$. The voltage cap is chosen by comparing lifetime calculations with different caps, and then choosing the minimal voltage cap where the lifetime is not significantly reduced.

\section{Realization of filter}
\label{app:chebyshev}

{Here we demonstrate that a suitable filtered environment can be realized via a single lossy LC resonator or an array of these. Specifically, we consider implementing the filtered environment as a series of $N$ capacitively coupled LC resonators, where the first resonator is coupled to the qubit, and the last resonator is coupled to an ohmic environment at temperature $T_{\rm B}$. Here $N$ may take value $1$, in which case the environment is implemented via a single resonator.

For the array of $N$ coupled resonators described above, the spectral density of the quadrature coupled to the system is given by \cite{Sete2015BandpassPurcell,Cattaneo2021EngineeringDissipation,Thorbeck2024PassiveLeakageReset}}
\begin{equation}
    S_N(\omega)=\frac{\gamma}{U_N(x(\omega))^2+\gamma^2 U_{N-1}(x(\omega))^2},
    \label{eq:app_PSD_chebyshev}
\end{equation}
where $x(\omega )=\frac{|\omega|-\omega_0}{\Lambda_{\rm f}},$
 \(U_n(x)\) denotes the Chebyshev polynomial of the second kind, $\Lambda_f$ is controlled by the capacitive coupling strength, $\omega_0$ denotes the resonator frequency, and \(\gamma\) is proportional to the coupling of the last oscillator to the ohmic environment.
The bath power spectral density that emerges from the spectral density above is given by~\cite{gardiner_quantum_2004}
\[
J_N(\omega)=S_N(\omega)\,\Bigl[\Theta(\omega)+2n_B( |\omega|)\Bigr]\,\hbar |\omega|,
\]
with
\(
n_B(\omega)=\frac{1}{e^{\hbar \omega/(k_{\rm B} T_{\rm B})}-1}.
\)
The center and width of the energy filter can thus be controlled by the array parameters $\gamma$, $\omega_0$, and $\Lambda_{\rm f}$.
For a single lossy resonator (\(N=1\)), the spectral density reduces to a Lorentzian centered at \(\omega_0\), with width proportional to \(\gamma\,\Lambda_f\).

\begin{figure*}
    \centering
    \includegraphics[width=0.8\linewidth]{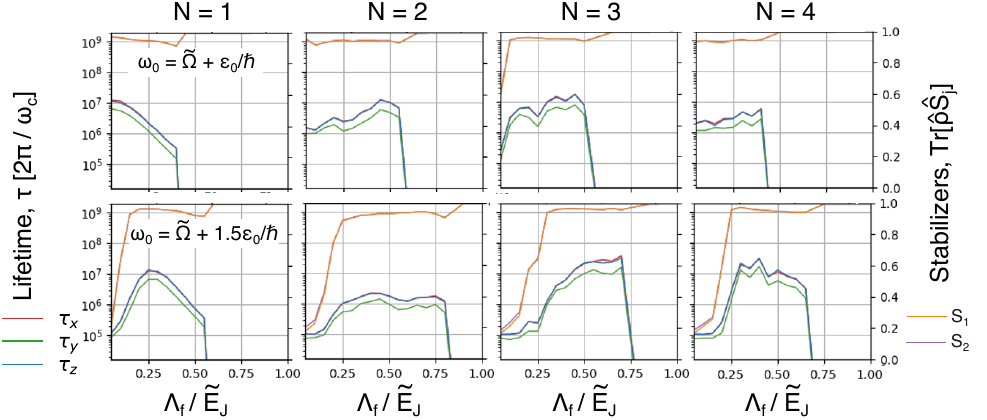}
    \caption{\textbf{Qubit performance with Lorentzian ($N = 1$) or Chebyshev ($N \geq 2$) band pass filter.} Qubit information lifetimes $\tau_x, \tau_y, \tau_z$ and steady-state stabilizer eigenvalues ${\rm Tr}[\hat \rho \hat S_j]$ for band pass filters consisting of $N = 1,2,3,4$ (columns 1-4) coupled resonators centered at $\omega_0 = \tilde \Omega + \varepsilon_0/\hbar $ (top row) and $\omega_0 = \tilde \Omega + 1.5\varepsilon_0/\hbar $ (bottom row).
    Each panel shows the lifetimes and stabilizer expectation values as a function of filter bandwidth $\Lambda_{\rm f}$. The power spectral density of the filtered bath is defined in Eq.~\eqref{eq:app_PSD_chebyshev}. Here, we use a filter coupling strength of $\kappa = 0.001$ and filter temperature $k_B T_B = 0$. The damping parameter in the Chebyshev filter PSD is set to $\gamma = 1$. All remaining parameters are as in Tab.~\ref{tab:parameters}.}
\label{fig:app_chebyshev_filter}
\end{figure*}

Fig.~\ref{fig:app_chebyshev_filter} shows numerical results from simulations of the circuit using the Filter PSD above, using the parameters from Table~\ref{tab:parameters} and waveform in Fig.~\ref{fig:pulse}. We vary the number of coupled resonators $N$ and filter width $\Lambda_{\rm f}$, and probe two values of the frequency center $\omega_0$, namely $\omega_{\rm 0} = {\tilde{\Omega}}+ \varepsilon_0/\hbar $ and $\omega_{0} = {\tilde{\Omega}} + 1.5 \varepsilon_0/\hbar $, where $ \varepsilon_0 = \sqrt{4 \pi \tilde{E}_J \hbar \delta \omega} $ is the predicted level spacing of the GKP states---see Sec.~\ref{sec:dec} and Fig.~\ref{fig:2}.
Centering the filter bandwidth at $ {\tilde{\Omega}} + \varepsilon_0/\hbar$ optimizes the side-band cooling mechanisms driven by photon-loss at energy of a drive photon plus a relaxation in the GKP subspace between two adjacent levels. A slightly higher frequency center at ${\tilde{\Omega}} + 1.5 \varepsilon_0/\hbar $ allows also relaxation processes from further separated excited states in the GKP subspace given sufficient filter width $\Lambda_f$.

The numerical results indicate that a too large filter width is detrimental for qubit performance, consistent with our analysis in Sec.~\ref{sec:lindblad}: with wider filter windows, $J(\omega)$ acquires support below $\tilde \Omega$, allowing side-band heating processes, which cause transitions between GKP states that increase their quasienergy, and hence their effective syndrome weight, as defined in Sec.~\ref{sec:dec} (see also Fig.~\ref{fig:gkp_states}). This will ultimately take the system out of codespace.

We see in Fig.~\ref{fig:app_chebyshev_filter} that the side-band cooling mechanism can perform well and reach a lifetime of around $5 \times 10^7 \, \omc^{-1}$ already with a single resonator in the environment ($N=1$).
The lifetime is further improved with a Chebyshev filter of $N = 3$ coupled lossy resonators, reaching up to around $ 10^8 \, \omc^{-1}$. We attribute the shorter lifetime of the Lorentzian ($N = 1$) filter to its heavy tails, which decay only as $J(\omega)\sim (\omega - \omega_0)^{-2}$, which allows side-band heating transitions. Instead, for the Chebyshev filter, including more resonators strongly suppresses the tails as $J(\omega)\sim (\omega - \omega_0)^{-2 N}$ which reduces the sideband heating.

\section{Consequences of voltage bias noise}
\label{app:voltage_noise}
Here we show that noise in the voltage bias signal $V(t)$ will not adversely affect qubit performance when its amplitude and frequency range fall below a threshold we identify below [Eq.~\eqref{eqa:voltage_noise_conditions}].
We focus in this Appendix on the implementation of the square-GKP qubit in Eq.~\eqref{eq:no_oscillation} where a modulation of the phase bias does not cause a modulation of the amplitude of the stabilizer terms in the Hamiltonian:
\begin{align}
    \hat H_{\rm eff}\approx&\frac{\hbar \delta \omega}{2} \left( \frac{\hat \phi^2}{4\pi }+4\pi \hat n^2 \right) + \tilde E_0 \cos(\hat \phi-\Phi(t_0)-\pi/4)\notag \\ &+\tilde E_1\cos(4\pi \hat n-\Phi(t_1)+\pi/4),\label{eqa:no_oscillation}
\end{align}
As we will show below, this implementation offers the greatest resilience towards noise and drift of bias. We note that our results apply to the more general case considered in Eq.~\eqref{eqa:HRWA_def}.

We model voltage bias noise as a shift in the voltage tone, $V(t)\to V_{\rm tar}(t)+\xi_V(t)$, with $V_{\rm tar}(t)$ the target voltage bias tone.
The voltage noise in turn induces a noise in the phase bias, $\delta \Phi(t)=\frac{2e}{\hbar}\int_0^t dt' \xi_V(t')$. In terms of this quantity, the Hamiltonian in the presence of voltage bias noise reads
\begin{equation}
    \hat H(t) =\frac{\hbar\omc}{2}\Big[\frac{\hat \phi^2}{4\pi }+4\pi \hat n^2 \Big] - E_J \cos[\hat \phi-
\Phi(t)-\delta \Phi(t)].
\end{equation}

To infer the consequences of the voltage bias noise, we note that, even with noise, the Hamiltonian above can still be analyzed using the approach of Appendix~\ref{app:SPA} to obtain an effective Hamiltonian for the circuit. In particular, the arguments leading to the effective Hamiltonian in Eq.~\eqref{eqa:no_oscillation} still hold for the above, provided $|\xi_V(t)|\ll V_0$, and that the angular frequency range of $\xi_V(t)$, $\omega_\xi$, is much smaller than $eV_{\rm max}/h$.
Under these conditions, within a given period of $V_{\rm tar}$ bias the location and slope of the nodes of $V(t)$ are approximately unmodified.
In this case, the analysis from Appendix~\ref{app:SPA} shows that the effective Hamiltonian in period $k$ is given by Eq.~\eqref{eqa:no_oscillation} after the substitution $\Phi(t_i)\to \Phi(t_i)+\delta \Phi(k\tilde T+t_i)$, which shifts the center of the stabilized GKP states in phase space.
Hence, the voltage noise leads to an effective drift of $\Phi(t_i)$ in Eq.~\eqref{eqa:no_oscillation} by $\delta \Phi(z\tilde T+t_i)$.
When the drift of $\delta \Phi(t)$ is much smaller than $2\pi$ on the timescale for relaxation into codespace, $t_{\rm stab}$, the resonator state will be relaxed with respect to the instantaneous effective Hamiltonian, and stabilization will persist, up to a displacement in phase space.
Thus, we expect the operation of the protocol is unaffected by the voltage bias noise as long as the change of $\delta \Phi(t)$ over the timescale $t_{\rm stab}$ is much smaller than $2\pi$ , and the two conditions above are satisfied.
This leads to the following 3 conditions.
\begin{align}
    |\xi_V(t)|&\ll V_{\rm max},\notag \\ \hbar \omega_\xi /2e&\ll V_{\rm max}\label{eqa:voltage_noise_conditions} \\ |\delta \Phi(t)-\delta \Phi(t')| &\ll2\pi\quad{\rm for}\quad|t-t'|\lesssim t_{\rm stab}\notag
\end{align}
We thus expect the protocol to tolerate voltage bias noise as long as its amplitude and frequency scales remain much smaller than $V_{\rm max}$, and the induced noise in the phase bias changes slowly on the stabilization timescale $t_{\rm stab}$.

\section{Consequences of low-frequency flux- and charge- noise}
\label{app:low_frequency_noise}
Here we analyze the effects of low-frequency charge and flux noise, such as $1/f$ noise or parameter drift.
As above, we focus for simplicity on the square-GKP qubit implementation also considered in the main text, while noting our results apply to the more general case considered in Appendix~\ref{app:SPA}.

We model the low-frequency flux and charge noise through the classical noise signals $ \delta n(t)$ and $\delta \phi(t)$ that couple capacitively and inductively to the circuit. Specifically, these noise signals modify the Hamiltonian such that, up to a scalar shift,
\begin{align}
\hat H(t) =&\frac{\hbar \omc}{2}\Big[\frac{(\hat \phi-\delta \phi(t))^2}{4\pi }+4\pi [\hat n-\delta n(t)]^2 \Big]\notag \\ &- E_J \cos[\hat \phi-
\Phi(t)] ,
\end{align}
We consider low-frequency noise where the frequency range for the fluctuations $\delta \phi(t)$ and $\delta n(t)$ are smaller than the resonator frequency $\omc$; residual spectral support at higher frequencies can be treated as a separate channel~\cite{ATA_paper}, and be analyzed numerically via quantum master equation methods, as we do in the main text.

We now show that low-frequency charge and flux noise is equivalent to a perturbation in the voltage bias, which we analyzed in the previous Appendix (Appendix~\ref{app:voltage_noise}).
To see this, we consider the evolution in a comoving frame resulting from the time-dependent displacement $e^{i\hat D(t)}$ with $\hat D(t) = a(t)\hat \phi +b(t)\hat n$ and $a(t)$, $b(t)$ two scalar functions we identify below; specifically, we consider the evolution of the density matrix $\hat \rho_{\rm F}(t)=e^{i \hat D(t)}\hat \rho(t)e^{-i\hat D(t)}$, with $\hat \rho(t)$ the lab frame density matrix.
It is straightforward to verify that, up to a scalar shift, the comoving frame density matrix $\hat \rho_{\rm F}(t)$ evolves with the Hamiltonian
\begin{equation}
\hat H'(t)=e^{i\hat D(t)}\hat H(t) e^{-i\hat D(t)}-\hbar \partial_t \hat D(t),
\end{equation}
up to a scalar shift, where we used that $e^{i \hat D(t)}\partial_t \hat D(t) e^{-i \hat D(t)}=\partial_t \hat D(t)+s(t)$, with $s(t)$ a scalar.
Writing out the expression explicitly, we find, again up to a scalar shift,
\begin{widetext}
\begin{equation}
\hat H'(t) =\frac{\hbar \omc}{8\pi}\big[\hat \phi+b(t)-\delta \phi(t)\big]^2+2\pi \hbar\omc \big[\hat n-a(t)-\delta n(t)\big]^2
-\hbar \dot a(t)\hat \phi -\hbar \dot b(t)\hat n - E_J \cos[\hat \phi
+b(t)+\Phi(t)] ,
\end{equation}
Rearranging the terms linear in $\hat \phi$ and $\hat n$, and again neglecting scalar offsets, we find
\begin{equation}
\hat H'(t) =\frac{\hbar\omc}{2}\Big[\frac{\hat \phi^2}{4\pi }+4\pi \hat n^2 \Big]-\left(\hbar \dot a(t)-\frac{\hbar\omc}{4\pi}[b(t)-\delta \phi(t)]\right)\hat \phi -\big(\hbar \dot b(t)+4\pi \hbar\omc [a(t)+\delta n(t)]\big) \hat n - E_J \cos[\hat \phi
+b(t)+\Phi(t)] ,
\end{equation}
\end{widetext}
We see that the two middle terms vanish if $a(t)$ and $b(t)$ satisfy the differential equation
\begin{equation}
   \hbar \begin{bmatrix}
        \dot a \\ \dot b
    \end{bmatrix} = \hbar\omc\begin{bmatrix}
        b/4\pi\\ -4\pi a
    \end{bmatrix}-\hbar\omc\begin{bmatrix}
        \delta \phi /4\pi \\ 4\pi \delta n \end{bmatrix}\label{eqa:aeom}
\end{equation}
We therefore pick $a(t)$ and $b(t)$ to satisfy the above equations.
With this choice, we thus find
\begin{equation}
\hat H'(t) =\frac{\hbar\omc}{2}\Big[\frac{\hat \phi^2}{4\pi }+4\pi \hat n^2 \Big]- E_J \cos[\hat \phi+b(t)+\Phi(t)] .
\end{equation}
Hence the evolution in the presence of low-frequency flux and charge noise is equivalent to a voltage noise with phase noise $\delta \Phi(t)=b(t)$, which we analyzed in Appendix~\ref{app:voltage_noise}.

We now consider the properties of the flux noise signal $\delta \Phi(t)=b(t)$.
To this end, we rewrite the equation for $a(t)$ and $b(t)$ in Eq.~\eqref{eqa:aeom} in frequency domain, which yields
\begin{equation}
   -i\hbar \omega \begin{bmatrix}
         a(\omega) \\ b(\omega)
    \end{bmatrix} = \hbar\omc DJ\left( \begin{bmatrix} a(\omega) \\ b(\omega)
    \end{bmatrix}+ \begin{bmatrix}
        -\delta n(\omega) \\ \delta \phi (\omega) \end{bmatrix}\right)
\end{equation}
with $$J=\begin{bmatrix}
    0 & 1 \\- 1 &0\end{bmatrix}, \qquad D=\begin{bmatrix}
        (4\pi)^{-1} & 0 \\ 0 & 4\pi
    \end{bmatrix}.$$
The solution to this equation is given by
\begin{equation}
\begin{bmatrix}
    a(\omega)\\ b(\omega)
\end{bmatrix}=\left(1-i\frac{\omega}{\omc}JD^{-1} \right)^{-1}\begin{bmatrix}
   - \delta n(\omega)\\ \delta \phi(\omega)
\end{bmatrix}.
\end{equation}
Since we assume the frequency support of $\delta \phi(t)$ and $\delta n(t)$ to be confined at frequencies smaller than $\omc$, the solution above is well-defined. In particular
$
    b(\omega)\approx \delta \phi(\omega)+ \frac{4\pi i \omega}{\omc}\delta n(\omega)
$ implying
\begin{equation}
    b(t) \approx \delta \phi(t)-\frac{\pi}{\omc}\delta \dot n(t)
\end{equation}
Thus, exploiting that we assumed the spectral support of the noise is confined below $\omc$, and using the conditions in Eqs.~\eqref{eqa:voltage_noise_conditions}, the low-frequency noise will not affect the protocol if
\begin{align}
    |\delta \dot n(t)|&\ll \omc \\ |\delta \phi(t)-\delta \phi(t')| &\ll2\pi\quad{\rm for}\quad|t-t'|\lesssim t_{\rm stab}\notag
\end{align}
where $t_{\rm stab}$ denotes the timescale for relaxation into the GKP codespace.%
Hence the protocol is not adversely affected by low-frequency flux and charge noise, provided that the change in the offset charge $\delta n(t)$ over one driving period is much smaller than $1$, and that the change of the offset phase over the stabilization time $t_{\rm stab}$ remains much smaller than $2\pi$.

\section{Numerical implementation}
\label{app:numerics}
In all of our simulations, we evaluate the master equation
\eqref{eqa:ule_2} numerically in a finite Hilbert space by discretizing $\hat \phi$ to a finite phase grid of $d$ points
$\phi_k\in[-\phi_{\max},\phi_{\max})$ with spacing $\Delta\phi=2\phi_{\rm max}/d$.
The cutoff $\phi_{\max}$ and inverse spacing $2\pi /\Delta \phi$ determine the region of phase space faithfully covered by our simulation. We pick these to be much larger than the inverse of the dimensionless GKP squeezing parameter $\lambda \equiv \left(\frac{2\pi \fz \hbar \delta \omega}{\tilde E_J}\right)^{1/4},$ which controls the extent of the stabilized GKP states in phase space, and is much smaller than $1$ in the regime we simulate (see Sec.~\ref{sec:dec}).
Specifically, we pick
\begin{equation}
    \phi_{\rm max} = 8\pi \fz \lambda^{-2},\quad {\Delta \phi}
    \approx\frac{\pi }{4}\lambda^{2}.
\end{equation}
with $\approx$ indicating $\Delta \phi$ should be the divisor of $2\phi_{\rm max}$ closest to $\frac{\pi }{4}\lambda^{2}$.
In our basis, $\hat \phi$ is a diagonal matrix with diagonal entries $\{\phi_k\}$.

When representing the charge operator $\hat n$, we are restricted by the fact that off-diagonal matrix elements in our basis represent translations of $\hat \phi$ by an integer multiple of $\Delta \phi$---i.e., by an integer power of $e^{i\Delta \phi \hat n}$. To faithfully cover as large an interval of phase space as possible, we therefore represent the charge operator $\hat n$ by the banded matrix corresponding to the $32$nd order central finite difference representation of $\partial_\phi$, i.e., the matrix representation of $\hat n^{(32)}_{\rm disc}\equiv \frac{1}{\Delta \phi}\sum_{a=1}^{16} s_a \sin(a\Delta \phi \hat n) $, where the coefficients $\{s_a \}$ are fixed by requiring $\sum_{a=1}^{16} s_a \sin(ax) = x + O(x^{33})$. This ensures $\hat n_{\rm disc}^{(32)} = \hat n+\mathcal O\big[(\hat n \Delta \phi )^{33}\big]$.
The residual correction of order $(\Delta \phi\hat n)^{33}$ will then give a negligible error in the region of phase space we simulate, where $|\hat n|\lesssim 1/\lambda$.
Crucially, the finite-difference operator $\hat n_{\rm disc}^{(32)}$ can be exactly represented in the discrete $\hat \phi$ basis we use, and has matrix elements
\begin{equation}
  [\hat n_{\rm disc}]_{j,j\pm a} =\left\{\begin{array}{ll} \mp\frac{\ii\,s_{a}}{2\Delta\phi}\,\quad &{\rm for}\quad |a|\leq 16\\0 &{\rm for}\quad |a|>16
  \end{array}\right. .
  \label{eqa:bands}
\end{equation}

The logical $\hat\sigma_z$ operator is diagonal in the discrete $\phi$ basis, with elements $\{\Xi(\phi_k/2\pi)\}$ on the diagonal. We represent $\hat \sigma_y$ by acting on the state with the discretized $\hat \sigma_z$ operator after Fourier transforming the wavefunction of the system---an operation that maps $\hat \phi$ to $2\pi \fz \hat n$, and hence transforms $\hat \sigma_y$ into $\hat \sigma_z$.

We compute the effective Hamiltonian $\Heff$ and jump operators $\hat L_z$ that parameterize the master equation in Eq.~\eqref{eqa:ule_2} using the time-evolution operator $\tilde U(t)$ generated by $\tilde H(t)$.
This enables the calculation of $\Heff$ via $\Heff=\ii\hbar\log\tilde U(\tilde T)/\tilde T$ and the jump operator via Eq.~\eqref{eq:jump_operator}. For the latter, we use that the micromotion operator
$\cM(t)=\tilde U(t) {e^{i \Heff t / \hbar}}$.
To calculate $\tilde U(t)$, we
discretize the extended driving period $t\in [0,\tilde T]$ into a grid of spacing $\Delta t$, $(t_1,\ldots t_N)$, with $t_n=(n-1/2)\Delta t$, such that
$ \tilde U(t_n) \simeq \prod_{j=n}^{1}
   e^{-\ii \tilde H(t_j)\Delta t/\hbar},
$
We fix $\Delta t$ by
requiring that the drive phase advance by at most a quarter cycle per step,
$ \max_t\big|\Phi(t+\Delta t)-\Phi(t)\big| \;\le\; \pi/2 ,
  \label{eqa:Ncrit}
$
At the default parameters of Table~\ref{tab:parameters} this gives
$N\approx1700$ steps per extended period. The criterion is verified
a posteriori against the realised samples of $\Phi(t)$.

We solve the master equation in Eq.~\eqref{eqa:ule_2} via the stochastic Schr\"odinger equation~\cite{carmichael_open_1993,dalibard_wave-function_1992}, as detailed in the main text.

\end{document}